\documentclass[aps,twocolumn,showpacs,superscriptaddress,groupedaddress,nofootinbib]{revtex4} 
\usepackage{graphicx}
\usepackage[sort&compress]{natbib}
\usepackage{subfigure}
\usepackage{amsmath}
\usepackage{amssymb}
\usepackage{amsfonts}
\usepackage{rotating}
\usepackage{cancel}
\usepackage{lipsum}
\usepackage{mathtools}
\usepackage{color}
\usepackage{bbm}
\usepackage{dsfont}
\usepackage{bbold}
\usepackage{multirow}
\usepackage{ulem}

\begin{document}
\arraycolsep1.5pt

\newcommand{\Ima}{\textrm{Im}}
\newcommand{\Rea}{\textrm{Re}}
\newcommand{\mev}{\textrm{ MeV}}
\newcommand{\be}{\begin{equation}}
\newcommand{\ee}{\end{equation}}
\newcommand{\ba}{\begin{eqnarray}}
\newcommand{\ea}{\end{eqnarray}}
\newcommand{\gev}{\textrm{ GeV}}
\newcommand{\nn}{{\nonumber}}
\newcommand{\cm}{\textcolor{blue}}
\newcommand{\cor}[2]{{\color{red}\sout{#1}} {\bf\color{blue} #2}}
\newcommand{\dtres}{d^{\hspace{0.1mm} 3}\hspace{-0.5mm}}
\def\pmat#1{\begin{pmatrix}#1\end{pmatrix}}

\def\del{\partial}

\title{Light- and strange-quark mass dependence of the $\rho(770)$ meson revisited}

\author{R. Molina}
\email{raqumoli@ucm.es}
\affiliation{Departamento de F\'{\i}sica Te\'orica and IFIC,
Centro Mixto Universidad de Valencia-CSIC,
Institutos de Investigaci\'on de Paterna, Aptdo. 22085, 46071 Valencia, Spain}
\affiliation{
Institute of Physics of the University of Sao Paulo, Rua do Mat\~ao, 1371 -Butant\~a, S\~ao Paulo -SP, 05508-090}
\affiliation{Universidad Complutense de Madrid, Departamento de Fisica Teorica II, \& IPARCOS, Plaza Ciencias, 1, 28040 Madrid, Spain}
\author{J.~Ruiz~de~Elvira}
\email{elvira@itp.unibe.ch}
\affiliation{Albert Einstein Center for Fundamental Physics, Institute for Theoretical Physics,
University of Bern, Sidlerstrasse 5, 3012 Bern, Switzerland }

\date{\today}
\begin{abstract}
Recent lattice data on $\pi\pi$-scattering phase shifts in the vector-isovector channel, pseudoscalar meson masses and decay constants for strange-quark masses smaller or equal to the physical value allow us to study the strangeness dependence of these observables for the first time. We perform a global analysis on two kind of lattice trajectories depending on whether the sum of quark masses or the strange-quark mass is kept fixed to the physical point. The quark mass dependence of these observables is extracted from unitarized coupled-channel one-loop Chiral Perturbation Theory.
This analysis guides new predictions on the $\rho(770)$ meson properties over trajectories where the strange-quark mass is lighter than the physical mass, as well as on the SU(3) symmetric line.
As a result, the light- and strange-quark mass dependence of the $\rho(770)$ meson parameters are discussed and precise values of the Low Energy Constants present in unitarized one-loop Chiral Perturbation Theory are given.
Finally, the current discrepancy between two- and three-flavor lattice results for the $\rho(770)$ meson is studied.

\end{abstract}

\pacs
{
13.75.Lb  
11.30.Rd, 
12.38.Gc, 
14.40.-n 
}
\maketitle
\section{Introduction}

The $\rho(770)$ meson is the lightest vector meson in the hadron spectrum and one of the most studied hadrons in the literature. It is one of the best examples of a $q\bar{q}$ resonance well described within the quark model. Its phase shift fits well into a simple Breit-Wigner (BW) parameterization up to small corrections~\cite{Pisut:1968zza,Lafferty:1993sx} and it is usually considered as the prototype of narrow resonance in the light-quark sector.
It also dominates the $\pi\pi$ scattering amplitude in the $I=J=1$ decaying almost exclusively to two pions~\cite{Tanabashi:2018oca} channel below 1 GeV,\footnote{Where $I$ and $J$ refer to isospin and angular momentum, respectively.} decaying almost exclusively to two pions~\cite{Tanabashi:2018oca}.
The $\rho$-meson mass and width are well known from experiment;  the Particle Data Group (PDG) quotes for their BW values $M=775.1(3)$ and $\Gamma=149.1(8)$, respectively~\cite{Tanabashi:2018oca}\footnote{These values correspond to the average value for the charged meson seen in $\tau$ decays and $e^+e^-$ collisions.} From the theory side, the most precise determination of its pole parameters comes from the Roy-equation analysis of $\pi\pi$ scattering~\cite{Ananthanarayan:2000ht,Colangelo:2001df,GarciaMartin:2011cn,GarciaMartin:2011jx,Pelaez:2019eqa}.
The contribution of the $\rho(770)$ is also important for the hadronic total cross section $\sigma(e^+e^-\to \textrm{hadrons})$~\cite{Aubert:2009ad,Babusci:2012rp,Ablikim:2015orh},
which explains applications that go well beyond low-energy meson physics, ranging from the hadronic-vacuum polarization and  
the light-by-light contributions to the anomalous magnetic moment of the muon (see, for instance,~\cite{Eidelman:1995ny,Jegerlehner:2009ry,Colangelo:2017qdm,Colangelo:2018mtw}) to the electromagnetic and tensor-nucleon form factors~\cite{Belushkin:2006qa,Lorenz:2014yda,Hoferichter:2016duk,Hoferichter:2018zwu}.
Furthermore, it also plays a crucial role in the analysis of heavy meson decays~\cite{Kang:2013jaa,Niecknig:2015ija} and in the restoration of chiral symmetry at high temperatures~\cite{Pisarski:1995xu,Harada:2000kb,Rapp:2009yu,GomezNicola:2012uc,Nicola:2013vma,Nicola:2016jlj,GomezNicola:2017bhm,Nicola:2018vug,Nicola:2019ohb}.

Although, the $\rho$-meson properties fit well within the naive quark-model picture, its nature in terms of QCD degrees of freedom is still under discussion~\cite{Hu:2016shf}. Nevertheless, at low energies QCD becomes non perturbative, what hinders the study of hadron composition in terms of the fundamental QCD degrees of freedom. LatticeQCD simulations attempt to tackle this problem, however, several challenges are met when dealing with hadron scattering processes \cite{Briceno:2017max,Mohler:2015zsa}.
In this regard, the $m_q$ and $1/N_c$ expansions~\cite{tHooft:1973alw,Witten:1979kh,Cohen:2014vta}\footnote{$m_q$ and $N_c$ stand for the quark mass and number of colors, respectively.} provide model independent predictions to identify different kinds or hadrons.
These parameters can be used to study whether the response of resonance properties to a change on $N_c$ or $m_q$ compares well with the behavior expected for different QCD configurations. 
For instance, by studying the $N_c$ dependence of the $\rho$-meson properties, it was found that it also has a small non-$q\bar{q}$ component~\cite{Pelaez:2006nj,RuizdeElvira:2010cs,Guo:2012ym,Guo:2012yt,Ledwig:2014cla}. In addition, the analysis of the quark-mass dependence of the $\rho(770)$ parameters by means of the generalization of the Feynman-Hellmann theorem for resonances suggests that it requires non-negligible corrections beyond the quark model~\cite{RuizdeElvira:2017aet}.
In this way, the extraction of the light- and strange-quark mass dependence of the $\rho$-meson parameters from LatticeQCD simulations provides a powerful tool to confront quark-model predictions.

At low energies, Chiral Perturbation Theory (ChPT)~\cite{Weinberg:1978kz,Gasser:1983yg,Gasser:1984gg} is the Effective Field Theory (EFT) that controls the quark-mass dependence of hadronic observables.
ChPT encodes the interactions of the pseudo-Goldstone bosons of the spontaneous chiral symmetry breaking, and hence, it is capable to describe the quark-mass dependence of the light-pseudoscalar meson masses and decay constants at low energies, being these completely inherited from QCD~\cite{Gasser:1983yg,Gasser:1984gg}.
Nevertheless, ChPT is constructed as an expansion in quark masses and momenta and hence it is only valid below a certain scale.
Therefore, ChPT does not provide direct information about resonance properties.
On the contrary,  unitarized  Chiral Perturbation Theories (UChPTs)~\cite{Truong:1988zp,Dobado:1989qm,Dobado:1992ha,Dobado:1996ps,Nieves:1998hp,Oller:1998hw,Nieves:1999bx,GomezNicola:2001as} are based on imposing exact unitarity while matching ChPT at low energies. Thus,  the region of validity of the chiral expansion is extended, allowing one to generate poles on unphysical Riemann sheets in the complex-energy plane and to access resonance properties.
In particular, the Inverse Amplitude Method (IAM)~\cite{Truong:1988zp,Dobado:1989qm,Dobado:1992ha,Dobado:1996ps} generates the $\rho(770)$ resonance from $\pi\pi$ scattering and provides a tool to study the light- and strange-quark mass dependence of the $\rho$-meson properties, while reproducing the chiral series at low energies. Thus, in this work we utilize the IAM to investigate the quark mass dependence of the $\rho$-meson pole parameters, such as its mass, width and couplings to the $\pi\pi$ and $K\bar{K}$ channels. The analysis of these properties requires the determination of the Low Energy Constants (LECs) involved in the pseudoscalar meson masses, decay constants and meson-meson scattering.  A  chiral trajectory specifies the way in which the light- and strange-quark masses vary. In UChPT, the behavior of resonance properties over chiral trajectories is controlled by chiral symmetry and unitarity. It is desired to determine LECs which provide a full description of the resonance properties on chiral trajectories where $m_s$ and/or $m_{u,d}$ vary. These kind of predictions of an EFT can be tested by lattice QCD simulations. 

LatticeQCD (LQCD) is the only known tool to extract non-perturbative information from QCD. It is the instrument to determine the low-energy parameters of the chiral Lagrangian that govern the quark mass dependence of resonance properties, hence, rendering evidence of the EFT predictions.   In recent simulations, lattice data on $I=J=1$ $\pi\pi$ scattering have been extracted for several pion masses for two light flavors ($N_f=2$)~\cite{Aoki:2007rd,Gockeler:2008kc,Feng:2010es,Lang:2011mn,Pelissier:2012pi,Bali:2015gji,Guo:2016zos,Erben:2019nmx,Fischer:2020fvl} and including also the strange quark ($N_f=2+1$)~\cite{Wilson:2015dqa,Dudek:2012xn,Bulava:2016mks,Feng:2014gba,Alexandrou:2017mpi,Fu:2016,metivet}. See also~\cite{Werner:2019hxc,Miller:2020xhy} for recent $N_f=2+1+1$ simulations. Surprisingly, results for the $\rho$ mass in $N_f=2$ simulations are at odds with experimental predictions. Namely, the $N_f=2$ simulation with the lightest pion mass $m_\pi\simeq 150$ MeV by the RQCD Collaboration~\cite{Bali:2015gji} predicts a $\rho$-meson mass around $60$ MeV below the physical value~\cite{Bali:2015gji}. Other $N_f=2$ simulations also show disagreement with the closest pion-mass result for $N_f=2+1$.
For example, the $N_f=2$ GWU simulation at $m_\pi\simeq 226$ MeV~\cite{Guo:2016zos} gives a $\rho$-meson mass around $45$ MeV lighter than the $N_f=2+1$ Hadron Spectrum (HadSpec) outcome of the simulation for $m_\pi\simeq 236$ MeV~\cite{Wilson:2015dqa}. It has been argued in recent analyses~\cite{Guo:2016zos,Hu:2016shf} with the UChPT model of~\cite{Oller:1998hw}, that this difference can be explained through the effect of $K\bar{K}$ loops in the $\pi\pi\to K\bar{K}\to \pi\pi$ reaction, where the kaon is off-shell. This effect has been shown to be consistent among the $N_f=2$ simulations~\cite{Hu:2016shf}. Moreover, while the error ellipses of $N_f=2$ lattice data analyses do overlap, hence showing consistency among the simulations, the same cannot be stated for the $N_f=2+1$ results, where one finds inconsistencies among lattice simulations~\cite{Hu:2016shf}.
 
The light-quark mass dependence on decay constants has also been studied in LQCD simulations in~\cite{Bruno:2016plf,Blum:2014tka,Bazavov:2010hj,Bazavov:2009bb,Aubin:2008ie,Noaki:2009sk,Aoki:2008sm,Baron:2010bv}. However, almost no attention has been paid in the past to their strange--quark-mass dependence, nor of the $\rho(770)$ phase shift. Most of these simulations have been performed with a strange-quark mass kept fixed at the physical point, $m_s=m_{s}^0$.\footnote{The superscript ``0'' stands for physical point from now on.} A reflection of this can be found in the Flag Review~\cite{Aoki:2019cca}; the averaged LEC values from  different fits to decay constants with ChPT do not represent a global analysis of data and they do not track other trajectories rather than those with roughly $m_s=m_{s}^0$.
In addition, these LECs do not describe the meson-meson interaction at the energies where the $\rho$ meson begins to resonate.
The only exception up to recently  was a  simulation of the pion decay constant done by MILC over the $m_s=0.6\,m_{s}^0$ trajectory ~\cite{Bazavov:2010hj}. Thus, in spite of the great advance of lattice simulations, data out of the chiral trajectory $m_s=m_{s}^0$ are still scarce, even though the response of hadron properties to different chiral trajectories could elucidate their strangeness nature, in particular, and dynamical nature, in general. A larger amount of highly precise data on a variety of chiral trajectories are necessary to shed light on the composition of hadrons.

Recently, the CLS Collaboration generated ensembles on different chiral trajectories with $\text{Tr}{\cal M}=C$, (where $\text{Tr}{\cal M}=m_u+m_d+m_s$  and $C$ denotes a constant), in large volumes~\cite{Andersen:2018mau,Bruno:2016plf}. Moreover, this constant varies a little with the inverse gauge coupling, $\beta$, of the simulation, which characterizes the set of ensembles generated. These trajectories are of particular interest since the hadron response along the trajectory will manifest as a consequence of both, variations in the light and strange quarks. These recent lattice simulations motivate the present analysis by investigating them in combination with the simulations over $m_s=m_s^0$ trajectories. This provides a good ground to study the strange-quark mass dependence of decay constants and the $\rho(770)$ phase shift, which we intend to do here. Of course, new LQCD simulations over trajectories with larger variations on these constants or for different values of the strange-quark mass in the $m_s=k$ trajectories would improve the analysis presented here.

The study of hadron properties ($\rho$ meson) we conduct here needs to emphasize the role of pseudoscalar decay constants, which are strongly connected to the coupling of vector mesons to pions. This is supported by the assumption of dominance of vector mesons in the pion-photon coupling, the so-called Vector Meson Dominance (VMD)~\cite{sakurai}, which connects the size of the pion decay constant ($f_\pi$) and the $\rho\to\pi\pi$ coupling ($g_{\rho\pi\pi}$) in the EFT~\cite{Birse:1996hd}.
In this context, the large experimentally observed decay width of the $\rho(770)$ meson is directly connected to its coupling to two pions, which explains why the $\rho$-meson phase shift and $\rho$-meson properties are tightly related to the size of $f_\pi$. 
In this sense, these two observables should always be determined together in lattice simulations. Beyond that, the quark mass dependence of  pseudoscalar decay constants fixes the chiral trajectories in the lattice and hence, they can be used to set the lattice scale by letting them go to the physical point.

The analysis we perform here will be useful to further check the KSFR relation~\cite{Riazuddin:1966sw}, which under VMD states that $g_{\rho\pi\pi}=m_\rho/\sqrt{2}f_\pi$ in the SU(3) limit where $m_u=m_d=m_s$. While, it is common that in previous lattice/experimental data analyses of $\rho$-meson phase shift the $\rho$-meson mass increases monotonically with $m_\pi$, so that the KSFR relation is fulfilled~\cite{Hanhart:2008mx,Pelaez:2010fj,Nebreda:2010wv,Guo:2016zos,Hu:2017wli}, this behavior was not observed in the recent data of~\cite{Andersen:2018mau}. Whether this is a consequence of the lightness of the strange-quark mass used in these simulations or not will also be checked in the present analysis.

In conclusion, we analyze here the lattice data on $\rho$-meson phase shifts in $N_f=2+1$ of~\cite{Dudek:2012xn,Wilson:2015dqa,Bulava:2016mks,Andersen:2018mau},
 in combination with decay constant lattice data from~\cite{Bruno:2016plf,Blum:2014tka,Bazavov:2010hj,Bazavov:2009bb,Aubin:2008ie}. Moreover, the lattice data of~\cite{Feng:2014gba,Alexandrou:2017mpi,Noaki:2009sk,Aoki:2008sm} are also considered in separated analyses.
 
 Let us make some initial remarks. Experimental phase-shift data on $\pi\pi\to\pi\pi$ scattering in the $I=J=1$ channel were successfully reproduced using the IAM ~\cite{GomezNicola:2001as,Hanhart:2008mx,Pelaez:2010fj,Nebreda:2010wv}. Here, the LECs are extracted by performing a fit to lattice phase-shift data instead, taking into account the covariance matrix for energy levels, similarly as in~\cite{Hu:2017wli}. The main differences with the work of~\cite{Hu:2017wli} are:
 \begin{itemize}
  \item[1. ] A global fit to lattice data on two distinct chiral trajectories, $\text{Tr}{\cal M}=C$ and $m_s=k$, is done instead of considering trajectories only over $m_s=m_s^0$ simulations.\footnote{Here, $k=m_s^0$ or $0.6m_s^0$, where only data on $f_\pi$ are included in the latter~\cite{Bazavov:2010hj}.} As mentioned previously, this includes data from~\cite{Andersen:2018mau,Bruno:2016plf} for $\text{Tr}{\cal M}=C$ and from~\cite{Dudek:2012xn,Wilson:2015dqa,Bulava:2016mks,Blum:2014tka,Bazavov:2010hj,Bazavov:2009bb,Aubin:2008ie} for $m_s=k$.
  \item[2. ] We perform a simultaneous fit of phase shift and decay constant lattice data. Note that in~\cite{Hu:2017wli}, only phase-shift data were analyzed, while the quark mass behavior of pseudoscalar decay constants was fixed with the LECs obtained in the fit done in~\cite{Nebreda:2010wv}, which only included lattice data on $m_s=m_s^0$.
  \item[3. ] The theoretical framework used here is the IAM in coupled channels~\cite{GomezNicola:2001as}  instead of the simplified UChPT model considered in~\cite{Oller:1998hw}, which was taken into account in~\cite{Hu:2017wli}. This is, we include here one-loop diagrams not just in the $s$ channel, but also in the $t$ and $u$ channels, hence, consistently with chiral symmetry at low energies.
  \item[4. ] For the CLS data in~\cite{Andersen:2018mau,Bruno:2016plf} the systematic error in the lattice spacing is taken into account in the final fit by using the bootstrap method, assuming that the lattice spacing is normally distributed with the standard deviation associated to the lattice error. 
 \end{itemize}

 Although in the present analysis we only include $N_f=2+1$ lattice data, this work can be considered as complementary to the previous $N_f=2$ and $N_f=2+1$ analyses done in~\cite{Hanhart:2008mx,Pelaez:2010fj,Guo:2016zos,Hu:2016shf} and~\cite{Nebreda:2010wv,Hu:2017wli}, respectively, or to the recent two-loop $N_f=2$ IAM study in~\cite{Niehus:2020gmf}.
 If the strange-quark mass has no effect on the $\rho$-meson properties extracted from the lattice simulations, then, the disagreement among $N_f=2$ and $N_f=2+1$ lattice results will be due to the scale setting or other finite volume effects, such as the lattice spacing or the box size.
 Thus, we study in detail in which particular quark-mass regime the $\rho$-meson properties in the simulation are sensitive to both, the strange- and light $u$-, $d$-quark masses. 

This paper is organized as follows. In section~\ref{sec:theory} we explain the formalism considered. In section~\ref{sec:chpt}, we show the results of a global analysis on decay constants over several chiral trajectories. Section~\ref{sec:iama} provides the results of the combined fit both to phase shift and decay constant lattice data. In particular, we first analyze in~\ref{sec:ms} lattice data over $m_s=k$ trajectories, while the same analyses for the $\text{Tr}{\cal M}=C$ data are shown in section~\ref{sec:trm}.  Following this, we present our final results on a global fit on both trajectories in section~\ref{sec:gl}.  Finally, the main conclusions are presented in section~\ref{sec:con}.

\section{Theoretical framework}\label{sec:theory}

\subsection{Chiral Perturbation Theory}

At low energies QCD interactions become non-perturbative and EFTs provide the proper framework to perform systematic calculations. 
The basic premise of EFTs is that the dynamics at low energies (or large distances) do not depend on the details of the dynamics at high energies (or short distances). 
As a result, low-energy hadron physics can be described using an effective Lagrangian containing only a few degrees of freedom, hence, ignoring those present at higher energy scales.

Chiral perturbation theory is the low-energy EFT of QCD. It is  built as the most general expansion in terms of derivatives and quark masses~\cite{Gasser:1983yg,Gasser:1984gg} compatible with QCD symmetries, 
which relevant degrees of freedom at low energies are the pseudo Nambu--Goldstone bosons (NGB) of the chiral symmetry spontaneous breakdown, i.e.,  pion, kaon and eta mesons.

At leading order (LO) in this expansion, the chiral Lagrangian reads
\begin{equation}\label{L2chiraldef}
\mathcal{L}_2=\frac{f_0^2}{4}\left\langle \partial_\mu U(x)^\dagger \partial^\mu U(x)+\chi^\dagger U(x)+\chi U(x)^\dagger\right\rangle,
\end{equation}
where $f_0$ coincides with the pion decay constant in the chiral limit and $\chi= 2B_0 {\mathcal M}$, 
with $B_0$ a constant to be related with the quark condensate and
${\cal M}={\rm diag}\left(m_{ud}, m_{ud}, m_s\right)$ is the three-flavor quark-mass matrix, where exact isospin symmetry $m_{ud}=\frac{m_u+m_d}{2}$ is assumed.
The matrix $U(x)=\exp(\frac{i \sqrt 2 \phi(x)}{f_0})$ collects the contribution of pions, kaons and etas, with  
\begin{align}
  \phi(x)=&
  \left({\begin{array}{ccc}
        \tfrac{\pi^0(x)}{\sqrt 2}+\tfrac{1}{\sqrt{6}}\eta(x) & \pi^+(x) & K^+(x) \\
        \pi^-(x) & -\tfrac{\pi^0(x)}{\sqrt{2}}+\tfrac{1}{\sqrt{6}}\eta(x) & K^0(x) \\
        K^-(x) & \bar K^0(x) & -\tfrac{2}{\sqrt{6}}\eta(x)\\
      \end{array}}\right). \nonumber
\end{align}

By expanding the LO chiral Lagrangian in powers of $f_0$, one can identify the mass field terms obtained with the pseudo NGB fields, which yields a relation between meson and quark masses
\begin{align}
M^2_{0\pi}=&2\,m_{ud} B_0\, ,\nonumber\\
M^2_{0K}=&(m_{ud}+m_s) B_0\nonumber\, ,\\
M^2_{0\eta}=&\frac{2}{3}(m_{ud}+2\,m_s)B_0\, .\label{eq:ml}
\end{align}

The constant $B_0$ is related with the quark condensate value in the chiral limit, 
\begin{equation}
\Sigma_0=-\langle 0 \vert\bar{q}q \vert 0\rangle_0=B_0 f^2_0,\label{eq:b0}
\end{equation}
with $q\in\{u,d,s\}$, leading to the well known Gell--Mann--Oakes-Renner formula  $2m_{ud}\Sigma_0=M_{0\pi}^2f_0^2$~\cite{GellMann:1968rz}, 
i.e., even though both $m_{ud}$ and $\Sigma_0$ are scale dependent quantities, and hence, they are not observables, 
their product is scale independent.

At higher orders, all terms in the Lagrangian come multiplied by LECs, which contain information about higher energy scales.
In addition, they absorb the divergences which appear in the chiral expansion, so that,  the theory is renormalizable order by order. Unfortunately, the LECs cannot be determined perturbatively from QCD. While the LECs which multiply energy-dependent terms can be extracted quite well from dispersion theory~\cite{Bijnens:2011tb,Bijnens:2014lea,Hoferichter:2015tha,Siemens:2016jwj}, 
Lattice QCD provides in principle a model independent way to determine the values of LECs which fix the quark mass dependence~\cite{Leutwyler:2015jga,Aoki:2019cca}. 

The NLO Lagrangian was first derived in~\cite{Gasser:1983yg} for two flavors.  The effect of the strange quark was studied in~\cite{Gasser:1984gg}. 
Omitting field tensor and vacuum terms, the SU(3) NLO ChPT Lagrangian reads  
\begin{widetext}
\begin{align}\label{L4}
\mathcal{L}_4=&L_1\left\langle D^\mu U^\dagger D_\mu U\right\rangle^2+L_2\left\langle D^\mu U^\dagger D^\nu U\right\rangle \left\langle D_\mu U^\dagger D_\nu U\right\rangle+L_3\left\langle D^\mu U^\dagger D_\mu D^\nu U^\dagger D_\nu \right\rangle\nonumber\\
& + L_4\left\langle D^\mu U^\dagger D_\mu U\right\rangle\left\langle\chi^\dagger U+ \chi U^\dagger\right\rangle +L_5 \left\langle D^\mu U^\dagger D_\mu U(\chi^\dagger U+U^\dagger\chi)\right\rangle+L_6\left\langle\chi^\dagger U +\chi U^\dagger\right\rangle^2\nonumber\\
&+L_7\left\langle\chi^\dagger U -\chi U^\dagger\right\rangle^2+L_8\left\langle \chi^\dagger U\chi^\dagger U + \chi U^\dagger \chi U^\dagger\right\rangle.
\end{align} 
\end{widetext}
In Eq.~\eqref{L4}, $L_1$, $L_2$ and $L_3$ multiply massless terms and hence they also contribute in the chiral limit. 
$L_4$ and $L_5$ accompany terms depending linearly on the quark masses and they contribute to the renormalization of the NGB wave functions and decay constants.
Lastly, $L_6$, $L_7$ and $L_8$ come together with quadratic terms  on the quark mass. These only contribute to the renormalization of the NGB masses and have a minor role in the determination of the $\rho(770)$ meson properties. 

One-loop correction to the pion, kaon and eta NGB  masses read~\cite{Gasser:1984gg}
\begin{align}
m_\pi^2=& M_{0\,\pi}^2\left[1+\mu_\pi-\frac{\mu_\eta}{3}+\frac{16 M_{0\,K}^2}{f_0^2}\left(2L_6^r-L_4^r\right)\right.\label{eq:pimass}\\
&+\left.\frac{8 M_{0\,\pi}^2}{f_0^2}\left(2L_6^r+2L_8^r-L_4^r-L_5^r\right)\right]\,,\nonumber
\end{align}
\begin{align}
m^2_K=& M^2_{0\,K}\left[1+\frac{2\mu_\eta}{3}+\frac{8 M_{0\,\pi}^2}{f_0^2}\left(2L_6^r-L_4^r\right)\right.\nonumber\\
&+\left.\frac{8 M_{0\,K}^2}{f_0^2}\left(4L_6^r+2L_8^r-2L_4^r-L_5^r\right)\right]\,, \label{eq:kmass}
\end{align}
\begin{align}
m^2_\eta= &M^2_{0\,\eta} \left[1+2\mu_K-\frac{4}{3}\mu_\eta+\frac{8M^2_{0\,\eta}}{f_0^2}(2L_8^r-L_5^r)\right.\nonumber\\
&+\left.\frac{8}{f_0^2}(2 M^2_{0\,K}+M^2_{0\,\pi})(2L_6^r-L_4^r)\right]\nonumber\\
&+ M^2_{0\,\pi}\left[-\mu_\pi+\frac{2}{3}\mu_K+\frac{1}{3}\mu_\eta\right]+\nonumber\\
&\frac{128}{9f_0^2}(M^2_{0\,K}-M^2_{0\,\pi})^2(3L_7+L_8^r)\,,\label{eq:etamass}
\end{align}
with
\begin{equation}\label{eq:tadpole}
\mu_P=\frac{M_{0\, P}^2}{32 \pi^2 f_0^2}\log\frac{M_{0 \,P}^2}{\mu^2},\qquad P=\pi,K,\eta\,.
\end{equation}
The superscript $r$ denotes renormalized LECs, which carry the dependence on the regularization scale $\mu$~\cite{Gasser:1984gg}. 
This scale dependence cancels exactly in the calculation of any observable.
In the following, we will identify the physical NGB masses with the one-loop ChPT prediction above. Nevertheless, note that the quark mass dependence is always expressed in terms of the leading order NGB masses.

In addition, while at LO the NGB decay constant $f_0$ is independent of the quark mass,  
one-loop corrections in the pseudoscalar decay constants lead to
\begin{align}
f_\pi=& f_0\left[1-2\mu_\pi-\mu_K+\frac{4 M_{0\,\pi}^2}{f_0^2}\left(L_4^r+L_5^r\right)+\frac{8 M_{0\,K}^2}{f_0^2}L_4^r\right],\label{eq:fpis}\\ 
f_K=&f_0\left[1-\frac{3\mu_\pi}{4}-\frac{3\mu_K}{2}-\frac{3\mu_\eta}{4}+\frac{4M_{0\,\pi}^2}{f_0^2}L_4^r\right.\nonumber\\
&+\left.\frac{4M_{0\,K}^2}{f_0^2}\left(2L_4^r+L_5^r\right)\right],\label{eq:fk}\\
f_\eta=&f_0\left[1-3\mu_K+ \frac{4L_4^r}{f_0^2}\left(M_{0\,\pi}^2+2M_{0\,K}^2\right) +\frac{4M_{0\,\eta}^2}{f_0^2}L_5^r\right]\,,
\label{eq:feta}
\end{align}
which are also identified with the physical quantities. 

\subsection{Meson-meson scattering in ChPT}

The scattering of NGB meson is computed in ChPT as an expansion in momenta and meson masses.
Denoting as ${\cal A}^{I}(s,t,u)$ the scattering amplitude of the NGB process $a\to b$ with defined isospin $I$, one has the generic form 
\begin{equation}
{\cal A}^I(s,t,u)={\cal A}^I_{2}(s,t,u)+{\cal A}^I_4(s,t,u)+\dots,
\end{equation}
where $s$, $t$ and $u$ are the usual Mandelstam variables and ${\cal A}^I_k= {\cal O}(p^k)$, where $p$ means either  meson momenta or masses. 
The LO amplitude ${\cal A}_2$ is obtained at tree level from the ${\cal L}_2$ Lagrangian. 
The NLO contribution contains one-loop diagrams from ${\cal L}_2$ plus tree-level contribution from ${\cal L}_4$ involving LECs. 

\begin{figure}
  \includegraphics[width=0.4\textwidth]{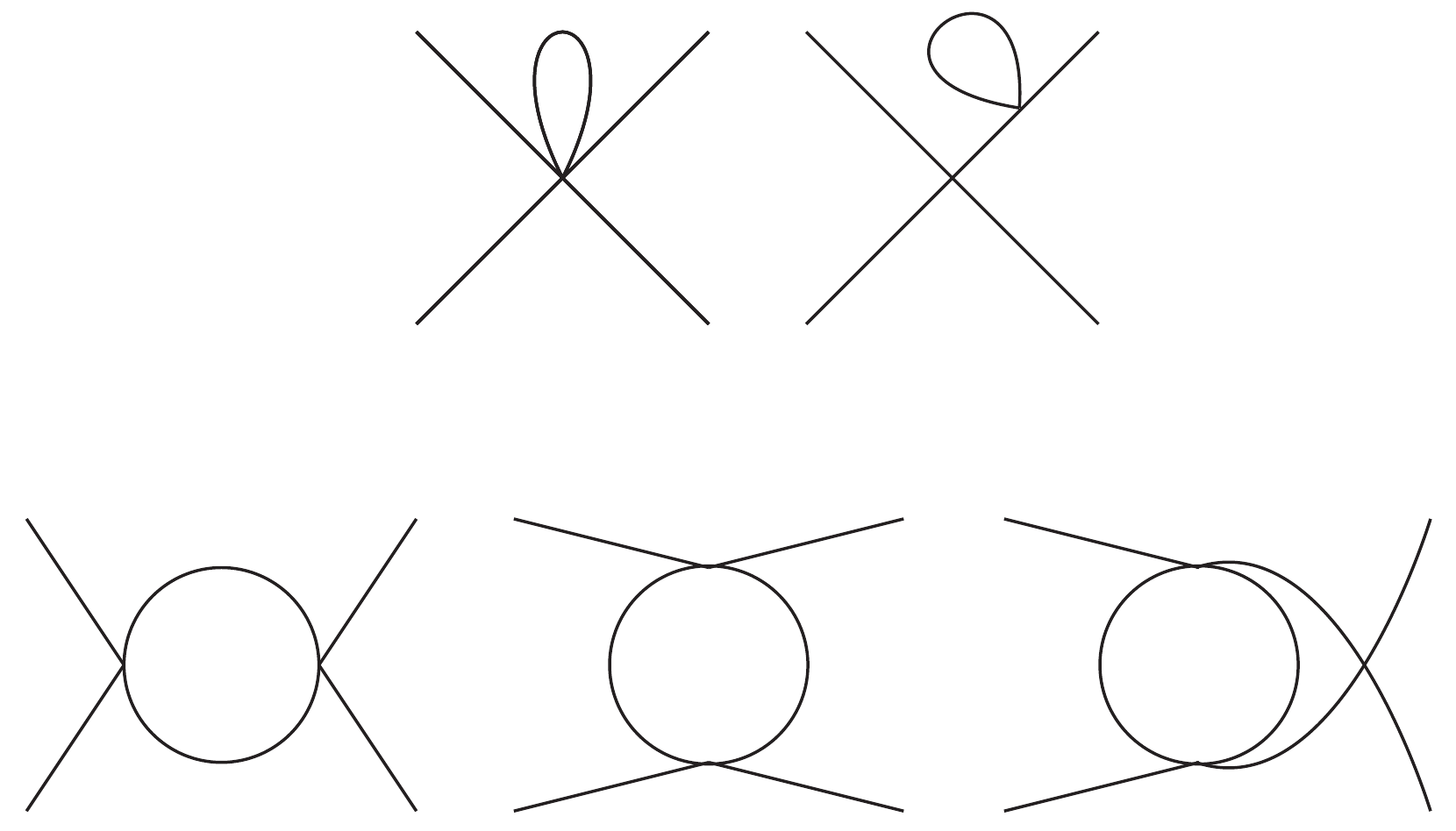}  
  \caption{Generic one-loop diagrams entering in meson-meson scattering. Top diagrams correspond to tadpoles, while bottom diagrams represent loops in the $s$, $t$, and $u$ channels.}\label{fig:diagrams}
\end{figure}
The $\pi\pi\rightarrow\pi\pi$ scattering amplitude at one-loop order in ChPT was computed first in~\cite{Gasser:1983yg} in a two-flavor formalism and in~\cite{Gasser:1984gg} for three flavors. The $\pi K\rightarrow \pi K$ and $\pi\eta\rightarrow\pi\eta$  scattering  amplitudes were evaluated in~\cite{Bernard:1990kx,Bernard:1990kw}~and~\cite{Bernard:1991xb}, respectively. 
The one-loop expressions for the SU(3) pseudo NGB reactions used here can be found  in~\cite{GomezNicola:2001as}. 
The SU(2) and SU(3) two-loop $\pi\pi$ scattering amplitudes were obtained in~\cite{Knecht:1995tr,Bijnens:1995yn} and ~\cite{Bijnens:2004eu}, respectively.
The two-loop $\pi K\rightarrow\pi K$ amplitude was determined in~\cite{Bijnens:2004bu}. Recently, first three-loop calculations have been explored in~\cite{Bijnens:2018lez}. 

Using the normalization conventions given in~\cite{GomezNicola:2010tb,RuizdeElvira:2018hsv}, the $s$-channel partial-wave projection of the amplitude is defined as
\begin{equation}\label{eq:pw}
t^{IJ}(s)=\frac{1}{32\pi N}\int\limits_{-1}^1{dx\,P_J(x) {\cal A}^I\left(s,t(s,x),u(s,x)\right)},
\end{equation}
where $N$ is a normalization factor equal to $2$ if all the particles are identical and $1$ otherwise.  
The Mandelstam variables $t(s,x)$ and $u(s,x)$ are defined by the kinematics of the corresponding $a\to b$ process and $x=\cos \theta$, being $\theta$ the scattering angle in the center-of-mass frame. 

  
Being an expansion in momenta and masses, it is clear that ChPT cannot satisfy unitarity,
which in the elastic case implies the relation
\begin{equation}\label{eq:elasticuni}
  \Ima \, t_{IJ}(s)=\sigma(s)\vert t_{IJ}(s)\vert^2\;\Rightarrow\; \vert t_{IJ}\vert <1/\sigma(s),
\end{equation}
where $\sigma(s)=2 q(s)/\sqrt s$ and $q$ is the momentum in the center-of-mass frame.
In the following, we only consider the $I = J = 1$ channel and the superscript index $IJ$ will be suppressed to ease the notation.
Nevertheless, ChPT satisfies elastic unitarity perturbatively.
For instance, defining as 
\begin{equation}
t(s)=t_2(s)+t_4(s)+\cdots,
\end{equation}  
the chiral series of the $I=J=1$ $\pi\pi$ partial-wave amplitude,
with $t_2(s)$ and $t_4(s)$ the tree-level and one-loop ChPT partial-wave amplitudes, in the elastic case one finds the relations
\begin{align}\label{eq:unichpt}
\Ima\,t_2(s)=&\,0,\nonumber\\
\Ima\,t_4(s)=&\,\sigma(s)\vert t_2(s)\vert^2,\nonumber\\
\cdots&,
\end{align}
which implies that the unitarity bound in Eq.~\eqref{eq:elasticuni} is increasingly violated in ChPT at larger energy values.
In practice, it implies that the chiral series is limited to scattering momenta around 200 MeV above threshold.
Furthermore, the ChPT series does not converge equally well in all parts of the low-energy region.
This is particularly evident in the scalar-isoscalar channel where strong pion-pion rescattering effects slow the convergence of the chiral series~\cite{Meissner:1990kz}. 
Finally, at increasingly large momenta, several partial-waves become resonant. Resonances are non-perturbative effects and, as such, they cannot be reproduced within the ChPT power expansion.
Furthermore, they usually saturate the unitarity bound in Eq.~\eqref{eq:elasticuni}, which implies that elastic unitarity can be  violated in the resonance region.

\subsection{Unitarity and analyticity}\label{sec:uni}

Below the four-pion production threshold, located at $s=16 m_\pi^2$, $\pi\pi$ scattering is purely elastic and, consequently, it can be described in terms of its phase shift.
Above this energy, there are possible intermediate processes such as $2\pi\to n\,\pi$, with $n=4,6,\dots$ or $\pi\pi\to \bar K K,\,\eta\eta,\dots$, which, in principle, have to be taken into account. In our case of interest, the $P$-wave $\pi\pi$-scattering partial wave, inelasticities are completely negligible below the $K\bar K$ threshold and very small below 1.4 GeV~\cite{Protopopescu:1973sh,Hyams:1973zf,Grayer:1974cr,Estabrooks:1974vu,GarciaMartin:2011cn,Pelaez:2019eqa,Perez:2015pea}. Thus, in this work elastic scattering is assumed to occur below the $K \bar K$ threshold and above only the $\pi\pi$ and $K\bar K$ channels are considered.

The unitarity condition for the $S$-matrix, $SS^\dagger=\mathbb 1$, implies that, for two-coupled channels, it can be parameterized in terms of only three independent parameters.
It is customary to choose them as the $\pi\pi \to \pi\pi$ and $K\bar K\to K \bar K$ phase shifts, denoted as as $\delta_1$ and $\delta_2$, respectively, and the inelasticity $\eta$. 
Thus, the S-matrix is expressed as
\begin{equation}
S=\pmat{\eta\,e^{2\,i\,\delta_1}& i\sqrt{1-\eta^2}\,e^{i\,(\delta_1+\delta_2)}\\
i\sqrt{1-\eta^2}\,e^{i\,(\delta_1+\delta_2)}& \eta\,e^{2\,i\,\delta_2}}\,.
\label{eq:sma}
\end{equation}

The $T$-matrix elements $t_{ij}$ of the scattering amplitude are related to $S$-matrix elements as,
\begin{equation}
S_{ij}=\delta_{ij}+2\, i\sqrt{\sigma_i\sigma_j}\,t_{ij}
\label{eq:sel}
\end{equation}
with
\begin{align}\label{eq:sigma}
\sigma_i = \left\{
     \begin{array}{crr}
      \sqrt{1-4\,m_i^2/s}&\qquad \sqrt{s}>2\,m_i\,&\\
      0 &  \qquad\mathrm{else} &\\
     \end{array}
   \right. 
\end{align}
and $i,j=1,2$.
The relation between the $S$- and $T$-matrix, Eq.~\eqref{eq:sel}, allows one to derive the following unitarity condition for the $T$-matrix elements
\begin{align}\label{eq:unicoupledel}
\Ima\,t_{11}=&\,\sigma_1\, \vert t_{11}\vert^2+\sigma_2\, \vert t_{12}\vert^2,\nonumber\\
\Ima\,t_{12}=&\,\sigma_1\, t_{11}\,t_{12}^*+\sigma_2\,t_{12}\,t_{22}^* \nonumber,\\
\Ima\,t_{22}=&\,\sigma_1\, \vert t_{12}\vert^2+\sigma_2\, \vert t_{22}\vert^2,
\end{align}
or
\begin{equation}\label{eq:unicoupled}
\Ima\, T= T\,\Sigma\,T^*, 
\end{equation}
 in matrix form, being
\begin{equation}
T=\pmat{t_{11}& t_{12}\\
t_{12}& t_{22}}\,,\quad \Sigma=\pmat{\sigma_{1}& 0\\
0& \sigma_2}\, .
\label{eq:tma}
\end{equation}
Eq. (\ref{eq:unicoupled}) implies the coupled-channel unitarity relation
\begin{equation}\label{eq:uni-coupled}
\Ima\, T^{-1}=-\Sigma\,
\end{equation}
is fulfilled. The phase space definition, Eq.~\eqref{eq:sigma}, ensures that  in the elastic case, i.e., below the $K\bar K$ threshold, elastic unitarity is satisfied. In the one channel case, Eq. (\ref{eq:uni-coupled}) simplifies to
\begin{equation}\label{eq:unielasticim}
\Ima \,1/ t_{11}(s)=-\sigma_1(s)\ .
\end{equation}

The unitarity conditions in Eqs.~\eqref{eq:uni-coupled}~and~\eqref{eq:unielasticim} imply that  the inverse of the imaginary part of an scattering amplitude in the physical region is completely fixed by unitarity. The strong relation between unitarity and resonances has motivated the development of several ChPT inspired methods based on imposing exact unitarity.
Some of them are the so-called $K$-matrix method~\cite{Gupta:1977pd} and the chiral unitarity approach. The latter was considered first in~\cite{Oller:1997ti,Oller:1998hw} to describe $\pi\pi$ and $K\bar K$ scattering in the scalar-isoscalar channel, leading to fairly precise determinations of the $f_0(500)$ and $f_0(980)$ resonance properties. There are also more involved unitarization methods. For example, the Bethe-Salpeter (BS) equations were solved for $\pi\pi$ scattering in~\cite{Nieves:1998hp,Nieves:1999bx}, both in the on-shell and off-shell schemes, while the N/D method was employed in~\cite{Oller:1998zr} providing also results for the rest of lightest scalars, namely the $\kappa(700)$ and $a_0(980)$.
However, none of them generates the $\rho(770)$ pole in the $\pi\pi$ scattering $P$ wave.

The energy-dependence of an scattering amplitude is also strongly constrained by analyticity. 
Analyticity is based on the Mandelstam hypothesis~\cite{Mandelstam:1959bc}, i.e., the assumption that an scattering amplitude is represented by a complex function 
that presents no further singularities than those required by general principles such as unitarity and crossing symmetry.
In this way, poles in the real axis are associated with bound states (absent in low-energy meson-meson scattering) and production thresholds give rise to cuts.
Cuts are a consequence of the unitarity condition given in Eq.~\eqref{eq:unicoupled}, which, together with the Schwartz-reflection principle, imply that an scattering amplitude must have a cut
where unitarity demands its imaginary part to be non-zero. It occurs due to both, direct and crossed channels, leading to a right- (RHC) and left-hand cut (LHC), respectively.

Once analyticity is established, Cauchy's integral formula allows one to construct a representation that relates the amplitude at an arbitrary point in the complex plane to an integral
over its imaginary part along the right- and left-hand cuts, the so called dispersion relations. 
The convergence of the dispersive integral often requires subtractions, which introduce a certain number of a priori undetermined constants.  
The Froissart--Martin bound~\cite{Froissart:1961ux,Martin:1962rt} guarantees that at most two subtractions are needed to ensure the convergence at infinity,
but one subtraction is enough for the $\pi\pi$ scattering amplitude in the vector-isovector channel.
Thus, a once-subtracted dispersion relation for $I=J=1$ $\pi\pi$ scattering reads 
\begin{align}\label{eq:dronce}
t(s)=&\,t(0)+\frac{s}{\pi}\int\limits_{4m_\pi^2}^{\infty}{d\,s^\prime \frac{\Ima\, t(s')}{s^{\prime}\left(s^\prime-s-i\,\epsilon\right)}}\nonumber\\&
+\frac{s}{\pi}\int\limits_{-\infty}^{0}{d\,s^\prime \frac{\Ima\, t(s')}{s^{\prime}\left(s^\prime-s-i\,\epsilon\right)}}, 
\end{align}
where the first and second integrals stand for the RHC and LHC contributions, respectively. 
The subtraction constants involve the evaluation of the amplitude at $s=0$, so that, they can be pinned down by matching to ChPT in the regime where the chiral expansion is expected to show better convergence properties. 
However, while the value of $\Ima\,t(s)$ in the physical RHC is constrained from unitarity, the LHC contribution is in principle unknown.
On the one hand, most UChPT methods differ in the way the LHC is treated. While the $K$-matrix and chiral unitarity approach models simply neglect the LHC contribution, the BS and N/D methods approximate it with ChPT. 
On the other hand, Roy--Steiner equations~\cite{Roy:1971tc,Hite:1973pm} solve this problem exactly using crossing symmetry. 
They provide a representation involving only the physical region, but which, at the same time, intertwines all partial-waves with different isospin and angular momentum.
Although, Roy--Steiner-equation solutions allow for high-precision descriptions of different scattering processes at low energies~\cite{Ananthanarayan:2000ht,Colangelo:2001df,GarciaMartin:2011cn,Buettiker:2003pp,Hoferichter:2015hva},  
and provide the proper framework to extract resonance pole parameters~\cite{Caprini:2005zr,DescotesGenon:2006uk,GarciaMartin:2011jx,Masjuan:2014psa,Caprini:2016uxy,Pelaez:2016klv}, or to evaluate an scattering amplitude in an unphysical region~\cite{Hoferichter:2015dsa,Hoferichter:2016ocj,RuizdeElvira:2017stg}, their analysis requires experimental information for the high-energy contribution and higher partial waves. 
Thus, they are in principle inappropriate for the analysis of lattice data at different quark masses. 
In this article, we follow the IAM, which will be outlined in the next section~\ref{sec:iam}.

\subsection{Elastic Inverse Amplitude Method}\label{sec:iam}

The Inverse Amplitude Method exploits the relation between a dispersion relation for the inverse of an scattering amplitude and the ChPT amplitude at a given order. 
At NLO in the chiral expansion, taking into account that ChPT amplitudes grow as $s^2$ when $s\to\infty$, one needs three subtractions to ensure the convergence at high energies.
Thus, a thrice-subtracted dispersion relation for a elastic ChPT $\pi\pi$-scattering partial wave reads
\begin{align}\label{eq:t2-t4}
t_2(s)=&t_2(0)+t_2^\prime(0)s,\nonumber\\
t_4(s)=&t_4(0)+t_4^\prime(0)\,s+t^{\prime\prime}_4(0)\,s^2+\frac{s^3}{\pi}\int\limits_{4m_\pi^2}^{\infty}{ds^\prime\frac{\sigma(s) t_2(s)^2}{s^{\prime\,3}(s^\prime-s-i\epsilon)}}\nonumber\\
&+\frac{s^3}{\pi}\int\limits_{-\infty}^{0}{ds^\prime\frac{\Ima\,t_4(s)}{s^{\prime\,3}(s^\prime-s-i\epsilon)}},
\end{align}
where we have used Eq.~\eqref{eq:unichpt} to fix the absorptive part of $t_4(s)$ in the physical region.
Note that Eq.~\eqref{eq:t2-t4} is strongly related to a thrice subtracted dispersion relation for the function $g(s)=t_2(s)^2/t(s)$,~
\begin{align}\label{eq:Gdr}
g(s)=&g(0)+g^\prime(0)\,s+g^{\prime\prime}(0)s^2-\frac{s^3}{\pi}\int\limits_{4m_\pi^2}^{\infty}{ds^\prime\frac{\sigma(s) t_2(s)^2}{s^{\prime\,3}(s^\prime-s-i\epsilon)}}\nonumber\\
&+\frac{s^3}{\pi}\int\limits_{-\infty}^{0}{ds^\prime\frac{\Ima\,g(s)}{s^{\prime\,3}(s^\prime-s-i\epsilon)}},
\end{align}
so that in an elastic approximation the RHC contribution coincides exactly with that of $-t_4(s)$.
The subtraction constants require the evaluation of the scattering amplitude and its derivatives at $s=0$, the kinematic region where ChPT provides a reliable description. 
Thus, using ChPT at NLO one gets
\begin{equation}
\begin{aligned}
 & g(0)\simeq   t_2(0)-t_4(0),\\
 & g'(0)\simeq  t_2'(0)-t'_4(0),\\
 & g''(0)\simeq  -t''_4(0).
\end{aligned}\label{eq:SCelastic}
\end{equation}

Being the RHC exactly fixed from unitarity, and once the subtraction constants are estimated using ChPT, the only remaining unknown information in Eq.~\eqref{eq:Gdr} is the LHC. 
The left-hand cut might indeed play a relevant role below threshold, but it is expected that its contribution should be less important as one moves into the physical region.
Thus, for a qualitative description it is sufficient to approximate the left-hand cut using ChPT. At NLO, one finds
\begin{equation}\label{eq:LHCChPT}
\Ima\,g(s) \simeq\, t_2(s)^2 \Ima\,\frac{1}{t_2(s)+t_4(s)}\simeq - \Ima\, t_4(s).
\end{equation}

Inserting Eqs.~\eqref{eq:SCelastic}~and~\eqref{eq:LHCChPT} in Eq.~\ref{eq:Gdr} one obtains
\begin{equation}
t(s)^{\mathrm{IAM}}=\frac{t_2(s)^2}{ t_2(s)-t_4(s)}\label{eq:iam},
\end{equation}
which stands for the well-known equation of the IAM method.
The IAM was derived first in~\cite{Truong:1988zp,Dobado:1989qm} using only unitarity for $\pi\pi$ scattering.
Its derivation from a dispersion relation and application thereafter to $\pi K$ scattering was investigated in~\cite{Dobado:1992ha,Dobado:1996ps}, 
whereas the remaining  IAM meson-meson scattering processes were studied in~\cite{GomezNicola:2001as} to one loop.
The two-loop version of the IAM was derived in~\cite{Nieves:2001de} and its generalization to include the effect of Adler zeros was obtained in~\cite{GomezNicola:2007qj}.

The IAM provides a simple algebraic equation that ensures elastic unitarity while at low energies reproduces the chiral expansion.
This fact implies that the IAM can be used to describe the resonance region below 1 GeV, i.e., well beyond the applicability range of ChPT.  
Furthermore, it is based on a dispersion relation, hence, its use in the complex plane is justified, providing a simple tool to study resonance properties. 
The main difference between the IAM and the on-shell BS or N/D method, is that, in the IAM only the absorptive part of the left-hand contribution is expanded at low energies. 
It implies that the left-hand cut energy dependence is still controlled by a dispersion relation instead of being fully given by ChPT. 
In addition, the IAM generates not only scalar but also vector resonances~\cite{Oller:1998hw}, without involving new additional parameters rather than the ChPT LECs. 
Hence, it reproduces at low energies the quark mass dependence predicted by ChPT.

Nevertheless, it has also several caveats. While the RHC is solved exactly using elastic unitarity, the LHC is approximated using ChPT. 
The direct consequence of this fact is that the IAM breaks crossing symmetry.
Besides, while the IAM provides higher order ChPT contributions needed to fulfill unitarity, some of the leading order logarithms from higher-order loop graphs appear with the wrong coefficients~\cite{Gasser:1990bv}.

In addition, it is worth mentioning that the IAM describes experimental data, including resonance pole parameters, of meson-meson scattering in the region below 1 GeV only within a 10\%-15\% accuracy~\cite{GomezNicola:2001as,Pelaez:2003dy}. This small difference highlights the relevance of the LHC in the physical region below 1 GeV. 

Clearly, leaving the LECs as free parameters to be adjusted to data instead of being fixed to the ChPT values improves the description of the experimental data.
Indeed, $\pi\pi$ and $\pi K$ scattering experimental data were described in~\cite{GomezNicola:2001as,Nebreda:2010wv} using the IAM with LEC values compatible with pure ChPT determinations.
Small LECs changes are indeed expected since the IAM includes contributions that go beyond the pure chiral expansion at a given order.    
However, it is important to remark that while ChPT is a natural theory in the sense that its predictions are linear in LECs changes, the IAM as well as other UChPT models are strongly dependent on precise LECs determinations.
Small changes on the LEC values might produce large effects on the phase-shift and pole parameter predictions.

Finally, let us remark that the dispersive derivation of the IAM only constrains its energy dependence, and hence, it is not clear whether it provides the correct quark-mass dependence.
While the IAM reproduces the ChPT series at low energies, thus, ensuring that it provides the quark-mass dependence predicted from QCD close to the chiral limit,
it also introduces higher-order contributions that spoil the chiral series at higher energies and for heavier quark masses.
Thus, high quality lattice data for different light- and strange-quark masses are key to ensure that the chiral extrapolation performed within the IAM is well consistent with QCD.

\subsection{Coupled channel formalism}\label{sec:iamcc}

The generalization of the inverse amplitude method to coupled channels should be in principle straightforward if one assumes the factorization of the RHC and LHC contribution for the different channels involved.
In this case, we can define the matrix version of the function $g(s)$ in Eq.~\eqref{eq:Gdr} as $G(s)=T_2(s) T(s)^{-1} T_2(s)$,
where $T_k$ stands for the ${\cal O}(p^k)$ $I = J = 1$ ChPT matrix (see Eq.~\eqref{eq:tma}).
Similarly as in Eq.~\eqref{eq:Gdr}, a thrice-subtracted dispersion relation for $G(s)$ reads
\begin{align}
G(s)=&G(0)+G'(0)s+G''(0)s^2-\frac{s^3}{\pi}\int\limits_{s_{th}}^{\infty}{ds^\prime\,\frac{T_2(s)\Sigma(s) T_2(s)}{s^{\prime\,3}(s^\prime-s-i\epsilon)}}\nonumber\\
&+\frac{s^3}{\pi}\int\limits_{-\infty}^{s_L}{ds^\prime\,\frac{\Ima\,G(s)}{s^{\prime\,3}(s^\prime-s-i\epsilon)}},
 \end{align}
where $s_{th}$ and $s_L$ stand for the corresponding right- and left-hand cut branching points, respectively.
The numerator of the RHC contribution corresponds to the matrix version of Eq.~\eqref{eq:unichpt}, i.e., 
\begin{equation}\label{eq:chptcoupleduni}
\Ima\, T_4(s)=T_2(s)\Sigma(s)\, T_2(s),
\end{equation}  
and hence, the right-hand cut of G(s) coincides with that of the matrix $-T_4(s)$. 
The subtraction constants can be evaluated using ChPT.  By means of expanding $T^{-1}$ as
\begin{equation}\label{eq:Tinvmatrixexp}
T^{-1} \simeq \left(T_2+T_4+\cdots\right)^{-1} \simeq T_2^{-1}\left(\mathbb{1}-T_4\,T_2^{-1}+\cdots\right), 
\end{equation}
one recovers the equivalent version of Eq.~\eqref{eq:SCelastic} in matrix form. However, the problem now is the evaluation of the left-hand cut. 
Although the RHC branching point $s_{th}=4m_\pi^2$ is common for all the elements of the T-matrix, the LHCs of the various channels do differ.
Namely, while the $\pi\pi$ scattering LHC starts at $s=0$, the LHC  for the $K\bar K\to K\bar K$ partial wave opens at $s=4m_K^2-4m_\pi^2$.
In this way, proceeding as we did for the elastic IAM, i.e., taking the perturbative expansion in Eq.~\eqref{eq:Tinvmatrixexp} for the absorptive part of $G(s)$ along the LHC,
one is indeed mixing the LHCs of all T-matrix elements. 
This translates into a violation of the factorization hypothesis, which produces spurious left-hand cuts breaking unitarity~\cite{Iagolnitzer:1973fq,Badalian:1981xj,Guerrero:1998ei,GomezNicola:2001as,Ledwig:2014cla}.
As a summary, the analogous of Eq.~\eqref{eq:iam} cannot be derived in coupled-channels using a dispersion relation. 

Alternatively, one can still exploit unitarity  in order to derive a coupled channel version of Eq.~\eqref{eq:iam} valid in the real axis.
Taking into account Eq.~\eqref{eq:uni-coupled}, $\mathrm{Im}\,T^{-1}=-\Sigma$, one can write
\begin{equation}\label{eq:iamcc}
T=\left[\Rea\, T^{-1}-i \Sigma\right]^{-1}.
\end{equation}
Now, $\Rea\, T^{-1}$ can be approximated once more with ChPT. Using  Eq.~\eqref{eq:Tinvmatrixexp} one gets
 \begin{align}
   T\simeq& \,T_2\left[T_2-\Rea\,T_4-i T_2\Sigma T_2\right]^{-1}T_2\nonumber\\
   =&\,T_2\left[T_2-T_4\right]^{-1}T_2
  \label{eq:iamde}
 \end{align}
which provides the IAM coupled channel unitarization formula. Note that to derive Eq.~\eqref{eq:iamde} we have used Eq.~\eqref{eq:chptcoupleduni}.
Nevertheless, it is important to note that Eq.~\eqref{eq:iamde} is only justified in the real axis where the ChPT coupled channel unitarity relation~\eqref{eq:chptcoupleduni} is fulfilled. 

At this point, it is also important to discuss at which energy the couple-channel formalism should be taken into account. Given the phase-space definition in Eqs.~\eqref{eq:sigma} and~\eqref{eq:tma}, the unitarity relation in Eq.~\eqref{eq:unicoupled} acquires dimension two only when one crosses the $K\bar{K}$ production threshold. Thus, Eq.~\eqref{eq:iamde} should be used only above the $K\bar K$ threshold, i.e., when its dimension coincides with the number of states accessible and the coupled-channel unitarity relation in Eq.~\eqref{eq:uni-coupled} is fulfilled. Below this energy one should consider the one-dimensional IAM equation.
Thus, this procedure yields a discontinuity at $4m_K^2$, instead of a single continuous function. 
Alternatively, one can include the $K\bar K$ channel for all energies. This provides a continuous function but it  again introduces spurious left-hand cuts, leading to a violation of unitarity. 
Nevertheless, these violations are in general small, around 2\%-5\%~\cite{Guerrero:1998ei,GomezNicola:2001as}.
In this paper we consider the second approach for Eq.~\eqref{eq:iamde}, but in order to reduce the effect of spurious cuts, we introduce an extra term in the $\chi^2$ of our fit to lattice data, which penalizes unitarity violations of the S-matrix by some factor, as explained in~Sect.~\ref{sec:fitting}.

Eq.~\eqref{eq:iamde} was used in~\cite{GomezNicola:2001as} to study all possible amplitudes for meson-meson scattering
leading to a fairly good description of all available experimental data below $ 1.2$ GeV with reasonable LEC values.
These amplitudes were analytically continued to the complex plane in order to look for poles associated to the lightest scalar and vector resonances~\cite{Pelaez:2003xd,Pelaez:2003dy}, with determinations compatible with experimental values within uncertainties.
This result suggests that the role of spurious LHCs which prevent the dispersive derivation of the coupled-channel IAM formula are also small.
Furthermore, we have explicitly checked that by removing the $t$- and $u$-channel loop functions (Fig.~\ref{fig:diagrams}) in the $\pi\pi\to \bar KK$ and $\bar KK\to\bar KK$ ChPT amplitudes that generate the spurious cuts, the mass and width of the $\rho$-meson obtained in the global Fit IV (see Sect.~\ref{sec:gl}) change less than 1 and 6 MeV, respectively, i.e., within the uncertainties quoted. Nevertheless, the effect of the $t$ and $u$ channels in the $\pi\pi$ amplitude lead to a shift of 6 and 15 MeV for the mass and width of the $\rho$-meson in Fit IV, respectively (without readjusting the LECs). 

To conclude, Eq.~\eqref{eq:iamde} is the tool we use to analyze lattice scattering data in the $\rho(770)$ channel.
The explicit expressions for the elements of the $T_2$ and $T_4$ for $\pi\pi\to\pi\pi$, $\pi\pi\to K\bar{K}$ and $K\bar{K}\to K\bar{K}$ are given in the appendix of~\cite{GomezNicola:2001as}.

\subsection{Resonances}

Resonances are formally defined as poles lying on unphysical Riemann sheets. 
An unphysical Riemann sheet is reached when the physical right-hand cut is crossed continuously from the upper-half plane to the lower-half plane above a given production threshold.  
In the elastic scattering case, there are only two Riemann sheets, the physical and unphysical one, which are called, first and second sheet, respectively. 
These two Riemann sheets must coincide in the real axis,
\begin{equation}
S^I\left(s+i\epsilon\right)=S^{II}\left(s-i\epsilon\right) .
\end{equation}
In addition, the scattering amplitude on the first Riemann-sheet satisfies the Schwartz reflection principle, 
i.e., $S\left(s + i\epsilon\right) = S^*(s-i\epsilon)$, which together with unitarity, $SS^*=\mathbb 1$, yields the relation
\begin{equation}\label{eq:SIIRS}
S^{II}\left(s-i\epsilon\right)=S^{I}\left(s-i\epsilon\right)^{-1}\,.
\end{equation}
The analytic continuation of Eq.~(\ref{eq:SIIRS}) into the complex plane implies that a pole on the second Riemann sheet corresponds to a zero in the physical one.
By means of Eq.~\eqref{eq:sel} one can translate this relation to the $T$-matrix elements, leading to
\begin{equation}\label{eq:IIRS}
t^{II}(s)=\frac{t^I(s)}{1+2\,i\,\sigma(s)\,t^I(s)}\, .
\end{equation}
where $\sigma(s)=\sqrt{1-4m^2/s}$, and its determination is chosen as  $\sigma(s^*)=-\sigma(s)^*$, to ensure the Schwartz reflection symmetry.

When further channels are opened, more unphysical Riemann sheets can be defined by continuing the square momenta of the intermediate states over the different thresholds. 
Thus, there are $2^n$ Riemann-sheets for a given number $n$ of opened channels. 
The generalization  of Eq.~\eqref{eq:IIRS} in a coupled-channel formalism is straightforward
\begin{equation}\label{eq:IIRS2}
  T^{(n)}(s)=T(s)\left(\mathbb{1}+2\,i\,\Sigma(s)^{(n)}\,T(s)\right)^{-1}\,,
\end{equation}
where $\Sigma^{(n)}$ is a diagonal matrix containing the phase space factors of those channels that have been crossed continuously.
In particular, for the $\pi\pi$ and $K\bar K$ $I=J=1$ coupled-channel case, we will have four different Riemann sheets defined as
\begin{align}
\Sigma^{II}=\pmat{\sigma_{\pi}& 0\\ 0& 0}, \quad \Sigma^{III}=\pmat{\sigma_{\pi}& 0\\  0& \sigma_{K}},\quad \Sigma^{IV}=\pmat{0& 0\\  0& \sigma_{K}},\nonumber\\
\label{eq:sigmaRS}
\end{align}
where $\sigma_\pi=\sqrt{1-4m_\pi^2/s}$ and $\sigma_K=\sqrt{1-4m_K^2}$ are the phase space factors of the $\pi\pi$ an $K\bar K$ channels, respectively.

Therefore, a pole in the $T$ matrix corresponds to a zero of the determinant of the matrix inside the brackets of Eq. \eqref{eq:IIRS2}, which is denoted by $\sqrt{s_\mathrm{pole}}=E_0=(M-i\,\Gamma/2)$, 
where M and $\Gamma$ stand for the mass and width of the resonance, respectively.

In addition, the dynamics of a resonance is strongly related to its coupling to a given channel, which is defined from the pole residue as
\begin{equation}
g_ig_j=-16\pi \lim\limits_{s\to s_\mathrm{pole}}{\left(s-s_\mathrm{pole}\right)t_{ij}(s)(2J+1)/(2p(s))^{2J}}, 
\label{eq:iamco}
\end{equation}
where $p(s)$ stands for the center-of-mass-system momentum of the corresponding process.

\subsection{Formalism in the finite volume}\label{sec:finitevol}
The L\"uscher's approach~\cite{Luscher:1986pf,Luscher:1990ux} allows one to relate the measured discrete value of the energy in a finite volume to the scattering phase shift at the same energy in the continuum. The volume-dependence of the discrete spectrum of the lattice QCD gives the energy dependence of the scattering phase shift.
This method, originally derived for a single scattering process was soon extended to coupled channels for potential scattering~\cite{Liu:2005kr}, non-relativistic effective theories~\cite{Bernard:2008ax,Lage:2009zv} and  to relativistic scattering~\cite{Hansen:2012tf,Briceno:2012yi,Li:2012bi,Guo:2012hv}.
Extensions of the L\"uscher formalism to three-particle systems under certain conditions are also available, see for instance~\cite{Polejaeva:2012ut,Hansen:2014eka,Briceno:2017tce,Mai:2017bge,Doring:2018xxx,Hansen:2019nir,Blanton:2019igq,Pang:2019dfe,Briceno:2019muc,Romero-Lopez:2019qrt,Hansen:2020zhy} and references therein.

The L\"uscher's approach is based on the analysis of the dominant power-law volume dependence that enters
through the momentum sums in a BS equation, where all quantities are written in terms of non-perturbative correlation functions.
In order to extract this dependence one assumes that the BS kernel, which accounts for the LHC and subtraction constant contributions in Eq.~\eqref{eq:dronce} and only involves a exponentially suppressed dependence on the volume~\cite{Luscher:1986pf}, coincides for large volumes with its infinite-volume form.
In this way, the difference between finite- and infinite-volume integrals entering on the BS equations only depends on on-shell values of the two-particle integrand leading to the the quantization condition\footnote{Actually, in its relativist extension, L\"uscher's formulation neglects the volume dependence of the propagator dressing function or, equivalently, the real part of the two-particle propagator, which might lead to significant corrections for small volumes~\cite{Chen:2012rp}.}
\begin{equation}
\det \left[ i\,T+ {\cal F}^{-1} \right]=0,
\end{equation}
where $T$ is the scattering amplitude in the continuum and $\mathcal F$ is a matrix that contains sums of the generalized Zeta functions subduced into the relevant finite volume little groups~\cite{Hansen:2012tf,Briceno:2012yi}.

L\"uscher's method was subsequently rederived in~\cite{Doring:2011vk,Doring:2012eu} by discretizing the $s$-channel loop functions which appear in the IAM coupled-channel equation of Eq.~\eqref{eq:iamde} and neglecting the $t$- and $u$-channel contributions. The discretization of the $t$ and $u$ channels has been discussed in~\cite{Albaladejo:2012jr,Albaladejo:2013bra}. In the latter, the exponentially suppressed volume dependence of the LHC contribution was explicitly taken into account, concluding that the  LHC volume dependence is numerically negligible for lattice sizes $L > 2m_\pi^{-1}$ while for lattice volumes $m^{-1}_\pi<L<2\,m^{-1}_\pi$, it only affects noticeably the first energy level. 
Furthermore, note that neglecting the volume dependence of the LHC contribution in the finite volume is by no means equivalent to ignoring the LHC in the continuum; lattice energy levels are non-perturbative quantities and, as such, they include all physical effects, both from the RHC and LHC contributions. The same cannot be stated for the dispersive formalism defined in Sect.~\ref{sec:iam}~and~\ref{sec:iamcc} since one explicitly factorizes the RHC and LHC contributions. 
However, to extract information from the energy levels and connect them with the T-matrix in the continuum one does need a generalized L\"uscher method including all physical effects, which might become particularly difficult, for example, in the case of multi-channel and intermediate states of three or more particles.   

In principle, one could use the formalism in~\cite{Albaladejo:2012jr,Albaladejo:2013bra} to evaluate the energy levels and fit them to the lattice data.
Nevertheless, in order to avoid the discretization of loops we follow here the method used in~\cite{Hu:2017wli}.
Namely, we fit the  phase shift values extracted from the lattice using L\"uscher's method, while the eigenenergies are reconstructed by means of a Taylor expansion taking into account the correlation between energy $E_n$ and phase shift $\delta(E_n)$, as well as the covariance matrix of eigenenergies provided by the lattice. This method is explained in the  subsection below.

\subsection{Fitting procedure}\label{sec:fitting}

The low energy constants of SU(3) Chiral Perturbation Theory to one loop are extracted from fits to lattice phase-shift data in the $I=J=1$ channel together with pseudoscalar meson decay constants and masses.
This includes the $N_f=2+1$  phase-shift data of~\cite{Andersen:2018mau,Dudek:2012xn,Wilson:2015dqa,Bulava:2016mks,Feng:2014gba,Alexandrou:2017mpi} together with data from~\cite{Bruno:2016plf,Blum:2014tka,Bazavov:2010hj,Bazavov:2009bb,Aubin:2008ie,Noaki:2009sk,Aoki:2008sm} for decay constants.

We analyze lattice simulations on two different chiral trajectories, where either the sum of the three-lightest quarks or the strange-quark mass is fixed to the physical point, i.e., $\text{Tr}{\cal M}=C$ or $m_s=k$, respectively.
The corresponding tree-level pseudoscalar meson masses relations are
\begin{equation}
m_{0K}^2=-\frac{1}{2}m^2_{0\pi}+C\,B_0\ ,\label{eq:trm}
\end{equation}
for $\text{Tr}{\cal M}=C$ and
\begin{equation}
m_{0K}^2=+\frac{1}{2}m^2_{0\pi}+k\,B_0\ ,\label{eq:msp}
\end{equation}
for $m_s=k$, with $k=m_{s}^0$ or $0.6\,m_{s}^0$. 

As a result from a combined analysis of data on these two kind of trajectories, in Sect.~\ref{sec:gl} we also show predictions for $\rho$-meson phase shifts, pseudoscalar meson decay constants and masses in other trajectories where the strange-quark mass is fixed to values smaller than the physical one, $m_s=k$ with $k<m_s^0$, on the SU(3) symmetric trajectory, $m_s=m_{ud}$, i.e.,
\begin{equation}
m_{0k}^2=m^2_{0\pi}\ ,
\end{equation}
and for trajectories where the light-quark mass is kept fixed at the physical point $m_{ud}=m_{ud}^0$, i.e.,
\begin{align}
m^2_{0\pi}=&m^{02}_{0\pi},\nonumber\\
m^2_{0K}=&m^2_{0K,\mathrm{phys}}+(m_s-m_{s}^0)\,B_0\ .
\label{eq:muphys}
\end{align}

We employ one-loop ChPT for the analysis of pseudoscalar meson masses and decay constants, see Sect.~\ref{sec:chpt}, in combination with the coupled-channel IAM discussed in Sect.~\ref{sec:iamcc} for the $\rho$-meson phase shifts. The fitting parameters are the LECs entering into our expressions, i.e., $L_i$, with $i=\{3,\dots,8\}$, $L_{12}=2\,L_1-L_2$, and the parameters which fix the chiral trajectories in Eqs.~\eqref{eq:trm} and~\eqref{eq:msp}, $C\,B_0$ and  $k\,B_0$.
The chiral scale $\mu$ is fixed to $770$ MeV and the pion decay constant in the chiral limit $f_0$ is set to $80$ MeV.
We fixed $f_0$ because its inclusion as a new fitting parameter did not entail any substantial reduction of the $\chi^2$.
In the following we describe the contributions to the $\chi^2$.

Meson-meson scattering in the lattice translates into discrete energies which are correlated. 
In order to take into account those the following function is minimized,

\begin{equation}
 \chi^2_E=(\vec{E}-\vec{{\cal E}})^T {\mathcal C}^{-1}(\vec{E}-\vec{{\cal E}})\ ,\label{eq:chi2}
\end{equation}
where $\vec{{\cal E}}$ is  the vector of eigenenergies measured on the lattice, ${\cal C}$ their covariance matrix and $\vec{E}$ the corresponding energies of the fit function.

Nevertheless, we do not fit directly lattice energy levels but phase shifts extracted using the L\"uscher formula. 
In order to take into account the energy correlations, we follow the method considered in~\cite{Hu:2017wli}.
This is, for each energy level , $E^i$, a Taylor expansion of both, the phase shift extracted from the lattice, $\delta_L$, and the one evaluated in the IAM, $\delta_{\text{IAM}}$, is performed around the energy given by the lattice simulation, ${\cal E}^i$.
If one assumes that both $\delta_L$ and $\delta_{\text{IAM}}$ coincide exactly at $E^i$, at leading order, one finds 
\begin{eqnarray}
 E^i={\cal E}^i+\frac{\delta_L({\cal E}^i)-\delta_{\text{IAM}}({\cal E}^i)}{\delta_{\text{IAM}}'({\cal E}^i)-\delta_{L}'({\cal E}^i)}\, ,
\end{eqnarray}
which provides a direct way to evaluate $\chi_E^2$ in Eq.~\eqref{eq:chi2} in terms of phase shift values.
The minimization of Eq.~\eqref{eq:chi2} allows one to avoid dealing with the generalized Zeta functions encoded in the L\"uscher quantization condition.
This makes the fitting procedure considerably faster. Furthermore, using a UChPT model without a LHC in~\cite{Hu:2017wli} or a two-loop version of the IAM in two flavors~\cite{Niehus:2020gmf}, it has been checked that this approximation provides results consistent with the evaluation of the lattice energy levels, albeit with slightly larger $\chi^2$ values.

Regarding pseudoscalar meson masses and decay constants from the lattice, we fit the ratios, $h_0=m_K/m_\pi$, $h_1=m_\pi/f_\pi$, $h_2=m_K/f_K$ and $h_3=m_K/f_\pi$, which are, in principle, more stable against possible discretization effects.
Thus, we also minimize
\begin{equation}
  \chi^2_f=\sum_{ij}\frac{\left(h_{i,j}^{p
  \\}-h_{i,j}^l\right)^2}{\Delta h^{l\;2}_{i,j}}\,,
\end{equation}
where $i$ denotes the different ratios, $j=1,\cdots,n$ are the measurements, and $n$ is the length of lattice data. The superscripts $l$ and $p$ indicate values from lattice simulations and predicted by one-loop ChPT, respectively.

\begin{figure*}
\begin{center}
  \includegraphics[width=0.48\textwidth]{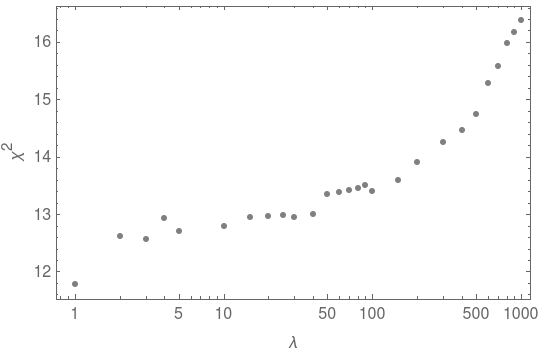}\includegraphics[width=0.48\textwidth]{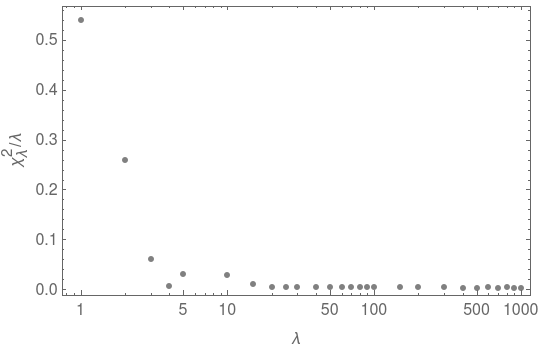}\\
 \end{center}
\caption{The minimized function $\chi^2$ in Eq.~\eqref{eq:ftest} (left) and variation of $\chi_\lambda^2/\lambda$ with $\lambda$ (right)}.
\label{fig:chi}
\end{figure*}

\begin{figure}
\centering
\includegraphics[width=0.50\textwidth]{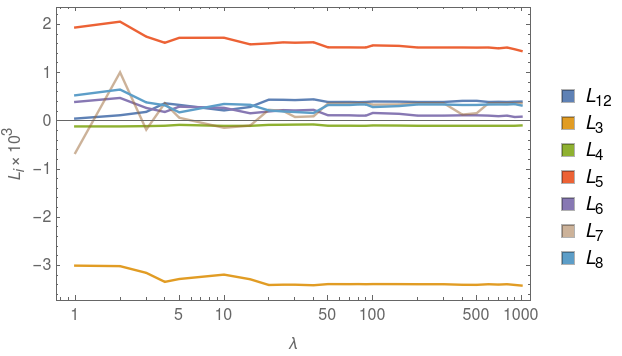}
\caption{Low energy constants of the IAM for HadSpec(236)+MILC data as a function of $\lambda$.}
\label{fig:plotlecs}
\end{figure}

Finally, as already discussed in section~\ref{sec:iamcc}, the coupled-channel version of the IAM generates unphysical LHC contributions arising from the on-shell coupled-channel approximation considered.
These contributions produce small violations of unitarity, which translate into undesirable phase shift peaks at low energies and in the resonance region, starting below $s=4\,m_K^2-4\,m_\pi^2$ (this energy corresponds to $880$ MeV for the HadSpec lighter pion mass).
These small peaks are enhanced when there are lattice data around that energy.
To eliminate these unphysical artifacts, a term that minimizes $S$-matrix unitarity violations at a degree controlled by a parameter $\lambda$ is added to the $\chi^2$,
\begin{equation}
\chi_{\lambda}^2=\lambda\,\sum_{ij}\int  \vert (S\,S^\dagger)_{ij}-\delta_{ij}\vert^2\, ds\,.\label{eq:chilambda}
\end{equation}
In summary, the total $\chi^2$-like minimization function reads as
\begin{equation}
\chi^2=\chi^2_E+\chi^2_f+\chi_{\lambda}^2\,.\label{eq:ftest}
\end{equation}
In Fig.~\ref{fig:chi} we show the value of $\chi^2$ and $\chi^2_\lambda$ in Eq.~\eqref{eq:chilambda} as a function of $\lambda$ for the minimization of the Hadron Spectrum Collaboration $\rho$-meson phase-shift data at $m_\pi=236$ MeV~\cite{Wilson:2015dqa} together with decay constants from MILC~\cite{Bazavov:2010hj}. The LEC values obtained are given in Fig.~\ref{fig:plotlecs}.
Clearly, for $\lambda\sim 40$ the LECs become stable while $\chi^2_\lambda/\lambda$ gets significantly reduced. 
One could also choose a higher value of $\lambda$, however, at the cost of increasing $\chi^2$. Thus, we set the value of $\lambda$ to $40$ .


There is an additional caveat that one should take into account; ChPT is built as an expansion in meson masses and, as such, the chiral series is only expected to converge for light pions. In order to study the convergence radius of ChPT we perform first individual fits of lattice data sets and discard pion mass results for which the fit does not pass the Pearson's $\chi^2$ test at a $90$\% upper confidence limit. This restricts the lattice data sets to pion masses below around $430$ MeV. Results presented in the following sections beyond that pion mass are merely qualitative.

As a final remark, we want to point out that the uncertainties for our final global fit are evaluated using the bootstrap method and hence, the errors should be understood in terms of probability, i.e., our central values are given by the median of the distribution and the uncertainties are expressed in terms of the 68\% and 95\% confidence intervals.
  
\section{ChPT: Decay constant analysis}\label{sec:chpt}

In this section, we attempt to perform a global fit of  pseudoscalar meson masses and decay constants $\lbrace m_\pi, m_K,f_\pi,f_K\rbrace$ from~\cite{Bruno:2016plf,Blum:2014tka,Bazavov:2010hj,Bazavov:2009bb,Aubin:2008ie,Noaki:2009sk,Aoki:2008sm}. These data are simulated on the chiral trajectories $m_s=m_s^0$~\cite{Blum:2014tka,Bazavov:2009bb,Aubin:2008ie,Aoki:2008sm}, $m_s=0.6\,m_s^0$~\cite{Bazavov:2010hj}, $m_s\simeq \left\{1.6\,m_s^0,2\,m_s^0\right\}$~\cite{Noaki:2009sk} and $\text{Tr}{\cal M}=C$~\cite{Bruno:2016plf}. The free parameters are the LECs $L_i$, with $i=\{4,\dots,8\}$, which appear in Eqs.~\eqref{eq:pimass}-\eqref{eq:feta},
as well as the variables, $C B_0$ and $k B_0$, which fix the chiral trajectories, $\text{Tr}{\cal M}=C$  and $m_s=k$, respectively, according to Eqs.~\eqref{eq:trm} and \eqref{eq:msp}.

A few aspects need to be considered before. First, the role of the renormalization scheme used in  the lattice simulations to fix quark masses.
Here, we do not adjust quark masses values but pseudoscalar meson masses, which, in principle, should be independent of the renormalization scheme.
Still, we checked if the pseudoscalar meson masses in the lattice data sets with different renormalization schemes are compatible.
For example, we notice that UKQCD Collaboration uses the MS scheme at $3$ GeV~\cite{Blum:2014tka}, while the MILC Collaboration uses the same scheme at $2$ GeV~\cite{Bazavov:2010hj,Bazavov:2009bb,Aubin:2008ie}. When we compare both sets of data, we do not observe any substantial inconsistency, but instead, their values do agree quite well.

Second, other important issue is the size of the pion masses used in the simulations.
We observe that in general the JL/TWQCD~\cite{Noaki:2009sk} and PACS-CS Collaborations~\cite{Aoki:2008sm} have larger pion and kaon masses.
For instance, the JL/TWQCD pion and kaon masses are larger than $300$ and 600 MeV, respectively.
These values might be too large for the perturbative ChPT expansion and indeed we are not able to fit these data sets in combination with MILC and UKQCD data.
Thus, in this fit we only include data from~\cite{Bruno:2016plf,Blum:2014tka,Bazavov:2010hj,Bazavov:2009bb,Aubin:2008ie}. The JL/TWQCD and PACS-CS data are studied in separated analysis in the next section.

Third, we should discuss possible finite volume and lattice spacing effects. In~\cite{Bruno:2016plf,Blum:2014tka,Bazavov:2010hj,Bazavov:2009bb,Aubin:2008ie}, the dependence of the decay constant determinations with the lattice spacing was studied carefully and the results were extrapolated to the continuum. These extrapolated data are the input of the fit we show here.
Another difficulty that we find to study data from~\cite{Aoki:2008sm} (PACS-CS) is the following. In~\cite{Aoki:2008sm}, the chiral trajectory is set in such a way that the physical point of the strange quark is determined and later fixed onto the chiral trajectory of the simulation. Thus, the $m_K$ dependence on $m_\pi$ in principle should agree with that from MILC~\cite{Bazavov:2010hj,Bazavov:2009bb,Aubin:2008ie} and UKQCD~\cite{Blum:2014tka}, since these simulations are also performed at the physical strange-quark mass.
However, we found substantial discrepancies in the behavior of the chiral trajectory in~\cite{Aoki:2008sm} with those from MILC and UKQCD. These inconsistencies may be due to finite volume and discretization effects, which can be partly absorbed by the free parameters. The result from analyzing PACS-CS data, pseudoscalar meson masses and decay constants~\cite{Aoki:2008sm} together with $\rho$-meson phase-shift data~\cite{Feng:2014gba}, is shown in the next section.
 
Lastly, it is also pertinent to discuss the relevance of the scale setting. Different lattice collaborations use different methods to set the scale. 
While all of them should agree at the physical point, i.e., for physical quark masses at zero lattice spacing, different schemes might approach this point with different slopes. 
It implies that, for unphysical pion masses, lattice observables might be scale dependent quantities. 
Thus, from a rigorous point of view, one should only compare among lattice results using the same scale setting procedure, or at least, include this dependence as an additional uncertainty.  Unfortunately, on one side, there is not enough lattice data from the same collaboration to study the dependence of the $\rho$ meson mass with the scale setting. On the other side, an extension of the IAM considering this effect is not available yet. 
In our particle case, the ensembles in~\cite{Bruno:2016plf} (see Table II of~\cite{Bruno:2016plf} for the pseudoscalar meson mass and decay constants of the CLS collaboration) consider two different scale setting methods, called here \textit{scale settings A} and \textit{B}. In the first one, \textit{scale setting A}, the lattice spacing is determined by fixing the chiral extrapolations of $f_\pi$ and $f_K$ to the physical point.
The second one, \textit{scale setting B}, uses the Wilson flow ($t_0$) to set the scale by assuming that, for all  different ensembles, the data over the $\text{Tr}{\cal M}=C$ trajectory intersects the $m_{ud}=m_s$ symmetric line  at $\phi_4=1.15$, with $\phi_4=8\,t_0\left(m_K^2+\frac{1}{2}m^2_\pi\right)$. 
This method requires small corrections in the quark masses from the ones used in the simulations~\cite{Bruno:2016plf}, which translates into small shifts for the pseudoscalar masses and decay constants.
Nevertheless, the CLS $\rho$-meson phase-shift data in~\cite{Andersen:2018mau} for the \textit{scale setting B} were not shifted accordingly, and hence, these corrections could lead to a conflict among the CLS decay constant and phase shift data.
Then,  we take here the no-shifted values, first rows of Table II in~\cite{Bruno:2016plf}. For each ensemble $\beta$, these two scale settings lead to  different lattice spacing values $a_\beta$. Namely, $\{a_{3.4}, a_{3.46}, a_{3.55}, a_{3.7}\}=\{0.079,0.071,0.061,0.0481\}$ fm for \textit{scale setting A} and $\{a_{3.4},a_{3.46}, a_{3.55}$, $a_{3.7}\}=\{0.086,0.076,0.064,0.0498\}$ fm for \textit{ B}~\cite{Bruno:2016plf}.
Nevertheless, we find that \textit{scale setting B} produces systematically smaller values of $f_\pi$ than \textit{A} for the same pion masses.
For instance, we see a difference of around $4$ MeV in $f_\pi$ for pion masses of around $200-300$ MeV between the two scale settings.
This difference is not small, since changes of $80$ MeV in $m_\pi$ imply variations on $f_\pi$ of around $4$ MeV in these data.
 Because of these discrepancies, we are only able to find an optimal $\chi^2$ when data with \textit{scale setting A} are included. Notice that this is the method that fixes the scale using the $f_\pi$ and $f_K$ physical quantities.\footnote{However, we show in section~\ref{sec:trm} that phase-shift lattice data in this scale setting cannot be reproduced globally. The reason is that the data for the ensembles N200 \& N401 produce lower $\rho$-meson masses than the predictions in the IAM. This problem is tackled in section~\ref{sec:gl}.} In section~\ref{sec:trm} we analyze the decay constant data in combination with $\rho$-meson phase-shift data for both scale settings and  discuss the main differences. 

\begin{table}[htb]
\begin{center}
{\renewcommand{\arraystretch}{2}
\setlength\tabcolsep{1cm}
 \begin{tabular}{cc}
 \toprule
  Fit I ($\chi^2/\mathrm{d.o.f}=1$)&LEC$\times 10^3$\\
  \hline
   $L_4 $ &    $ -0.060(6)$ \\
  $ L_5$  &     $ 0.91(2)$ \\ 
  $ L_6$  &     $ 0.15(2)$\\
   $L_8$  &     $ 0.03(3)$ \\
   \hline
 \end{tabular}}
 \end{center}
\caption{Values of the LECs obtained in Fit I.}
\label{tab:fit1l}
\end{table}
\begin{table}[htb]
\begin{center}
{\renewcommand{\arraystretch}{2}
\setlength\tabcolsep{0.5cm}
 \begin{tabular}{cc}
 \toprule
  Fit I &$C(k)\,B_0\times 10^{-3}(\mathrm{MeV}^2)$\\
  \hline
  $[a\,B_0]_{ \beta=3.4}$&$316(6)$\\
  $[b\,B_0]_{\beta=3.55}$&$295(6)$\\ 
  $[c\,B_0]_{\beta=3.7}$&$298(6)$\\
  $\quad[k\,B_0]_{m_s^0}$  &     $257(6)$ \\
  \hline
 \end{tabular}}
 \end{center}
\caption{Values of parameters in the different chiral trajectories analyzed.}
\label{tab:fit1c}
\end{table}

In conclusion, it is only possible to do a combined fit of data from~\cite{Bruno:2016plf} (\textit{scale setting A}) and~\cite{Blum:2014tka,Bazavov:2010hj,Bazavov:2009bb,Aubin:2008ie}.\footnote{This fit passes the Pearson's test.}
In Tables~\ref{tab:fit1l}~and~\ref{tab:fit1c}, the values of the fitting parameters obtained from this analysis are presented. This result is called Fit I.
We notice that the LECs in this fit are not very sensitive to small variations of the $CB_0$ and $kB_0$ parameters, being thus quite stable.
Furthermore, they are in line with the compilation of the FLAG Review~\cite{Aoki:2019cca}, which only includes results for $m_s=k$ data.
However, note that we are obtaining much smaller LEC errors compared to the FLAG average. Notice also that since these data include variations of the strange-quark mass, one is able to fix well the strange-quark mass dependence of the pseudoscalar decay constants for the pion masses studied.

\begin{figure*}
\begin{center}
{\renewcommand{\arraystretch}{2}
\setlength\tabcolsep{0.3cm}
 \begin{tabular}{cc}
 \includegraphics[scale=0.29]{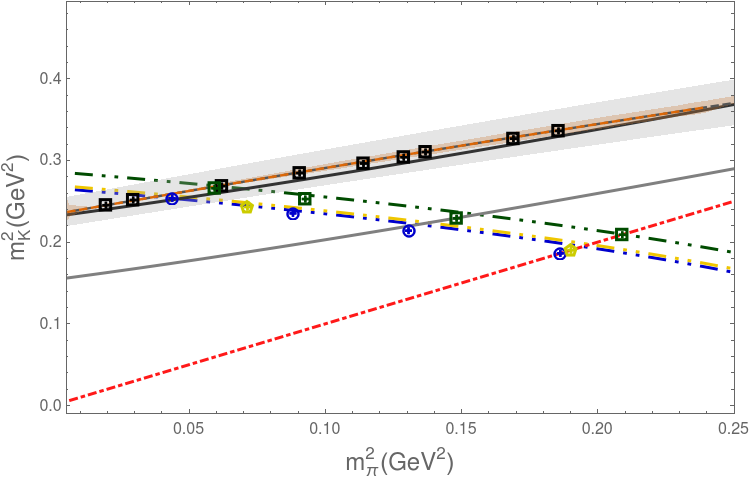}&\includegraphics[scale=0.28]{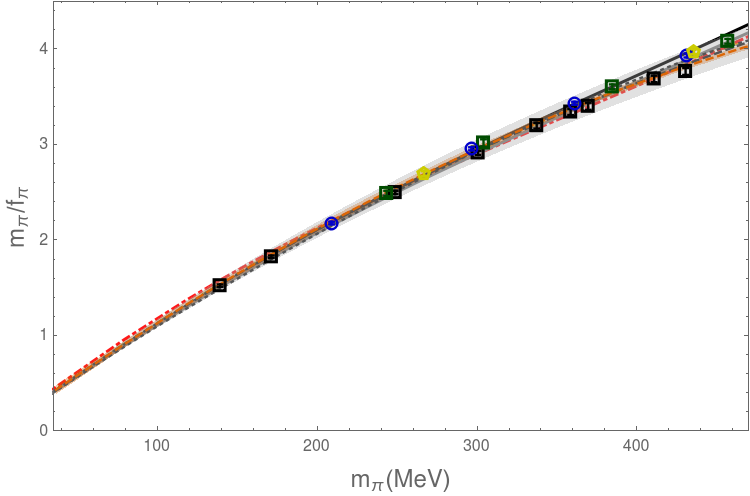}\\
\includegraphics[scale=0.28]{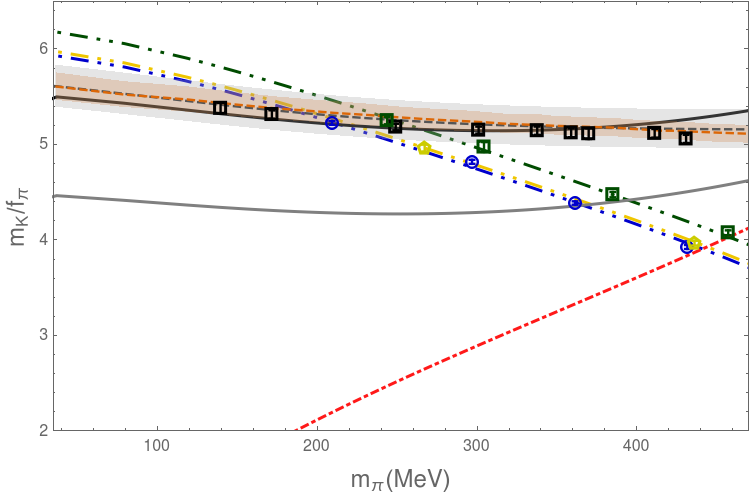}&\includegraphics[scale=0.28]{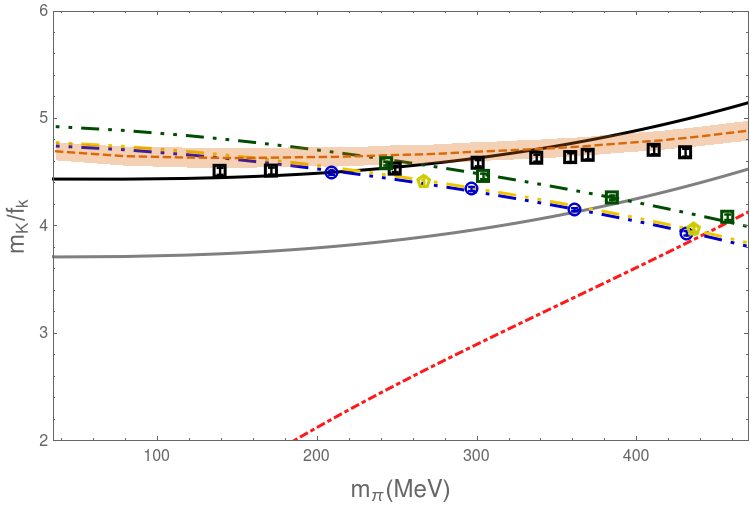}\\
\vspace{-0.4cm}&\\\hspace{0.5cm}\includegraphics[scale=0.32]{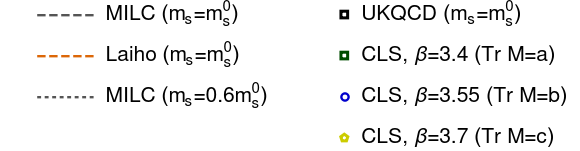}&\hspace{0.4cm}\includegraphics[scale=0.325]{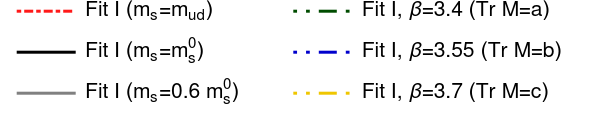}\\
 \end{tabular}}
\end{center}
\vspace{-0.5cm}
\caption{Chiral trajectories (top-left) and the ratios $m_\pi/f_\pi$ (top-right), $m_K/f_\pi$ (bottom-left), and $m_K/f_K$ (bottom right) obtained in Fit I over  the chiral trajectories $m_s=k$ and $\mathrm{Tr\,M=C}$. The light-brown and orange bands correspond to the errors of the MILC and Laiho data, in black and orange dashed lines, respectively. }
\label{fig:mpikfpik}
\end{figure*}

The various chiral trajectories studied are shown in Fig.~\ref{fig:mpikfpik} (top-left panel), where one can see that the kaon mass squared data for the $\text{Tr}{\cal M}=C$ trajectory~\cite{Bruno:2016plf} differ considerably from the $m_s=k$ ones~\cite{Blum:2014tka,Bazavov:2010hj,Bazavov:2009bb,Aubin:2008ie}. In addition, two of the ensembles simulated in~\cite{Bruno:2016plf}, the ones with $\beta=3.55$ and $3.7$, lead to very similar curves and hence to similar values of $CB_0$ in Table~\ref{tab:fit1c}.
Furthermore, the UKQCD~\cite{Blum:2014tka}, MILC~\cite{Bazavov:2010hj,Bazavov:2009bb} and Laiho~\cite{Aubin:2008ie} lattice data are in very good agreement. Indeed, ChPT is able to reproduce well the data on these two different trajectories. 

The ratios $m_\pi/f_\pi$, $m_K/f_\pi$ and $m_K/f_K$ are also depicted in Fig.~\ref{fig:mpikfpik}.
For the ratio $m_\pi/f_\pi$, it is worth noting that all data, independently of the chiral trajectory, lie almost on the same curve.
This suggests that the ratio $m_\pi/f_\pi$ is indeed quite independent on $m_s$. We discuss this further in Sect.~\ref{sec:gl}.
In fact, all lattice data for this ratio fall into the gray error band plotted, which is just an extrapolation of the percentage error of this ratio at the physical point determined by MILC~\cite{Bazavov:2009bb}.
For this collaboration only $m_\pi$, $m_K$ and $f_\pi$ data are provided, which are shown with dashed black lines. The $m_s=0.6\,m_{s}^0$ trajectory from~\cite{Bazavov:2010hj} is denoted by a solid gray line.
The data of Laiho~\cite{Aubin:2008ie} is represented by dashed-orange lines.
UKQCD data are denoted by black squares, while CLS data~\cite{Bruno:2016plf} are given by dark-green squares ($\beta=3.4$), blue circles ($\beta=3.55$) and yellow pentagons ($\beta=3.7$).
Note that the UKQCD Collaboration and Laiho data sets provide very similar values of $f_K$.
In addition, we include in Fig.~\ref{fig:mpikfpik} the chiral prediction for the SU(3) $m_s=m_{ud}$  trajectory.
Both, $m_s=0.6\,m_{s}^0$ and $m_s=\hat m$ trajectories, lead to a substantial reduction of the ratios $m_K/f_\pi$ and $m_K/f_K$.
For pion masses larger than $430$ MeV, the ChPT prediction begins to differ from the data, which suggests the breakdown of the chiral series.

\section{IAM: Rho phase shifts analyses}\label{sec:iama}
\subsection{Chiral trajectories $m_s=k$}\label{sec:ms}

In this section we analyze the $\rho$-meson phase-shift data from~\cite{Dudek:2012xn,Wilson:2015dqa,Bulava:2016mks,Feng:2014gba,Alexandrou:2017mpi} and pseudoscalar meson masses and decay constants from~\cite{Blum:2014tka,Bazavov:2010hj,Bazavov:2009bb,Aubin:2008ie,Noaki:2009sk}.
All these data are taken from simulations over chiral trajectories where the strange-quark mass is kept fixed to the physical value, $m_s=m_s^0$, except for the JL/TWQCD, where $k\simeq \{1.6, 2\}\, m_s^0$~\cite{Noaki:2009sk}. In fact, the pion and kaon masses used in the simulations of~\cite{Noaki:2009sk} are larger than in the other simulations. This simulation is studied independently and discussed at the end on in this section. 

First of all, we perform individual fits to the pseudoscalar masses and decay constant ratios from UKQCD~\cite{Blum:2014tka}, MILC~\cite{Bazavov:2010hj,Bazavov:2009bb} and Laiho~\cite{Aubin:2008ie} together with the $\rho$-meson phase shift data from the HadSpec Collaboration~\cite{Dudek:2012xn,Wilson:2015dqa} corresponding to $m_\pi=\{236,391\}$ MeV. The LECs obtained in these fits are shown in the second, third and fourth columns of Table~\ref{tab:lecsms}, respectively. Although some small differences among the individual fits are observed for $L_5$ and $L_6$, they provide in general compatible LEC values within uncertainties. Thus, we conduct a simultaneous analysis of the UKQCD, MILC and Laiho decay constants and HadSpec phase shifts, which is denoted as MUL+HS in the fifth column of Table~\ref{tab:lecsms}. As expected, the fit provides a good description of all data with consistent LECs.

\begin{figure*}
 \begin{center}
 {\renewcommand{\arraystretch}{4}
\setlength\tabcolsep{0.5cm}
 \begin{tabular}{cc}
 \hspace{-1cm}  \includegraphics[width=0.453\textwidth]{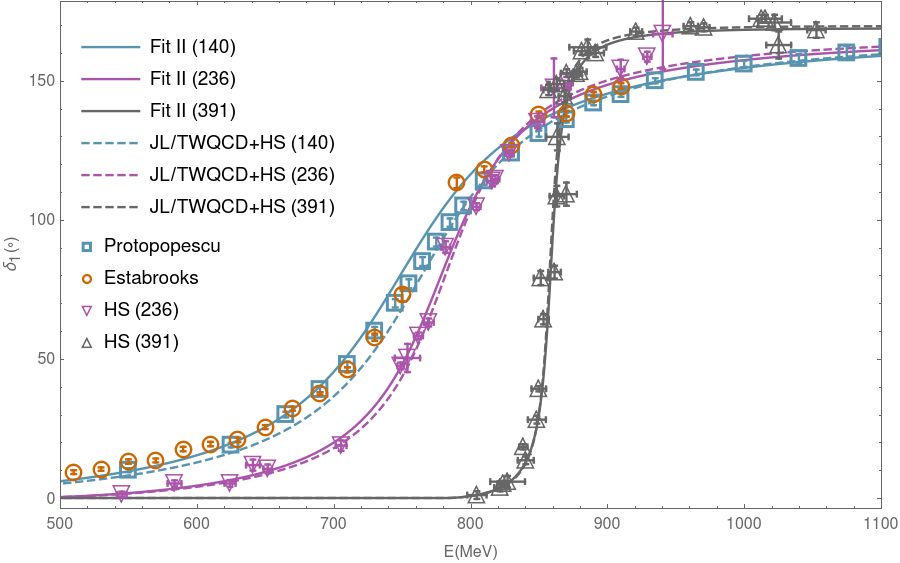}& \hspace{-0.8cm} \includegraphics[width=0.45\textwidth]{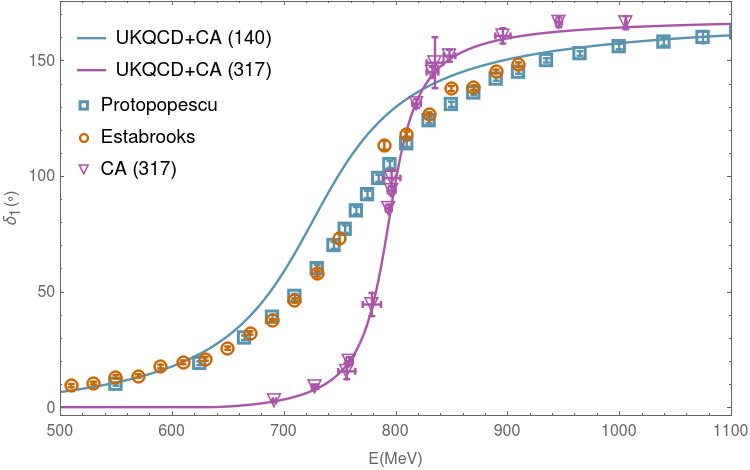}\\
\hspace{-1cm} \includegraphics[width=0.45\textwidth]{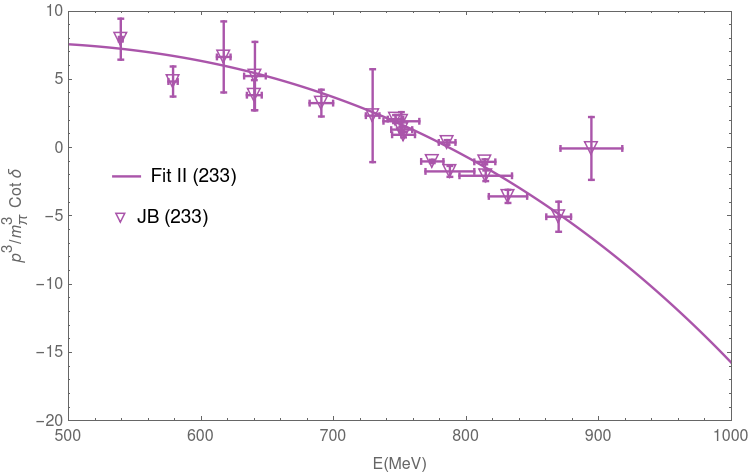}&\hspace{-0.8cm}\includegraphics[width=0.45\textwidth]{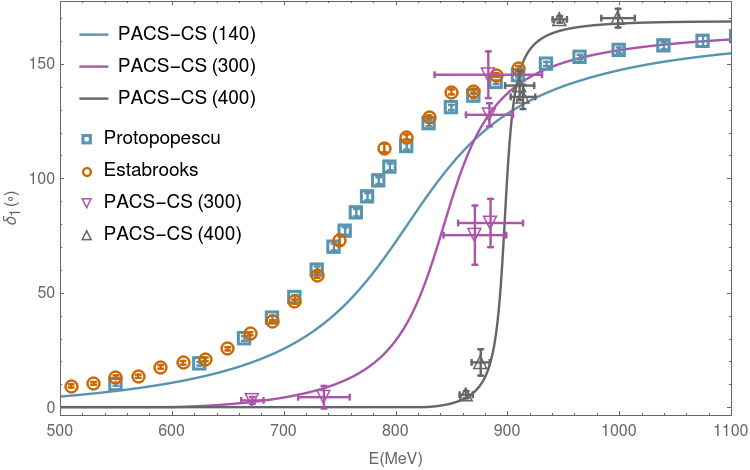}\\
  \end{tabular}}
  \end{center}
  \caption{Lattice phase shift data analyzed and fit results obtained as explained in the text. For each fit we also plot the extrapolations to the physical point in comparison with the available experimental data. The pion mass value (in MeV) for each simulation is given in parenthesis. }
 \label{fig:ph}
 \end{figure*}

Finally, we include the phase-shift results from~\cite{Bulava:2016mks} (JB) at $m_\pi=233$ MeV. This is denoted as Fit II in the sixth column of Table~\ref{tab:lecsms}. Notice that this fit  encompasses a large bunch of data on $m_s=k$ ($k=\{1,0.6\}\,m_s^0$). The LECs obtained in these fits are very similar to the previous ones suggesting
 consistency among the different data sets. Results for $\rho$ phase shifts 
together with the fitted lattice data are plotted in Fig.~\ref{fig:ph}. 
As shown in Fig.~\ref{fig:ph} (left, top), the extrapolation of Fit II results to the physical point (light-blue solid line) is very close to experimental data, depicted as light-blue squares~\cite{Protopopescu:1973sh} and orange circles~\cite{Estabrooks:1974vu}.

\begin{figure*}
\begin{center}
{\renewcommand{\arraystretch}{1.5}
\setlength\tabcolsep{0.5cm}
 \begin{tabular}{ll}
 \includegraphics[scale=0.3]{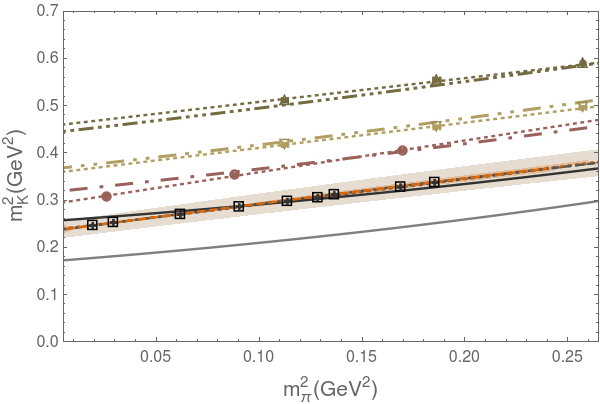}&\includegraphics[scale=0.3]{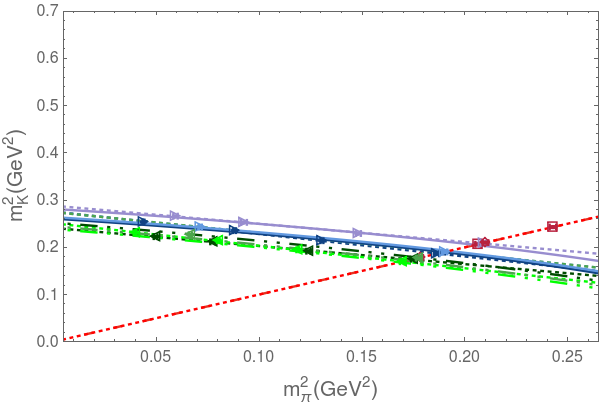}\\
 \includegraphics[scale=0.3]{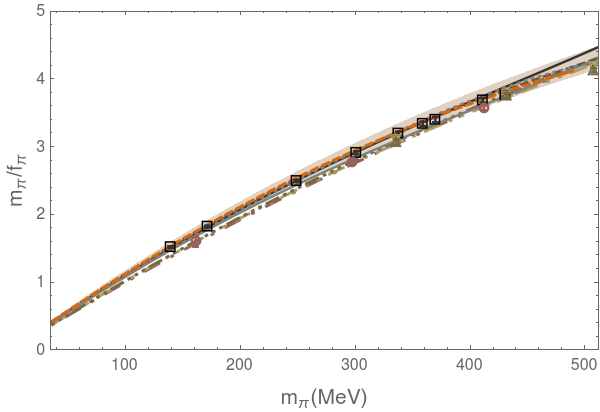}&\includegraphics[scale=0.3]{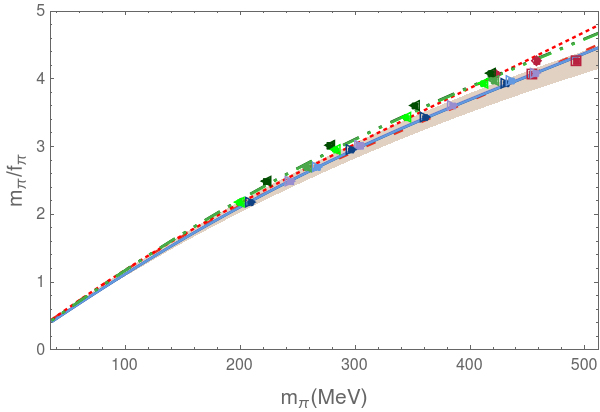}\\
 \includegraphics[scale=0.3]{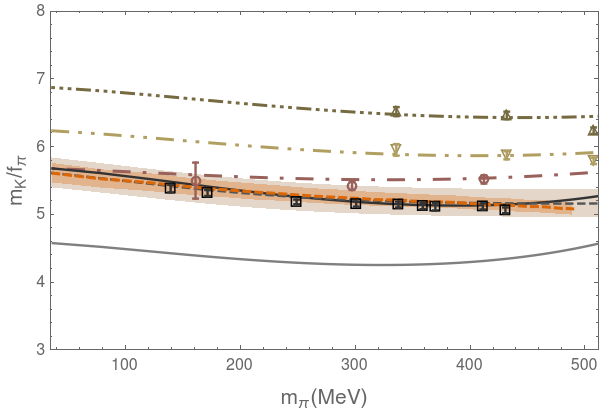}&\includegraphics[scale=0.3]{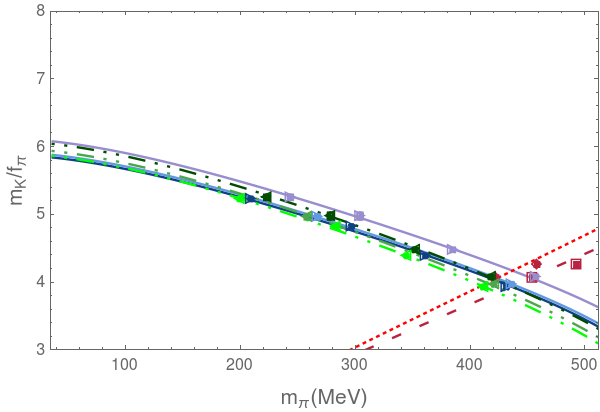}\\
 \includegraphics[scale=0.3]{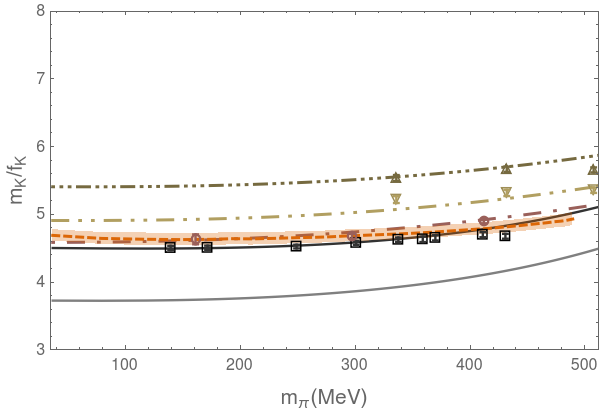}&\includegraphics[scale=0.3]{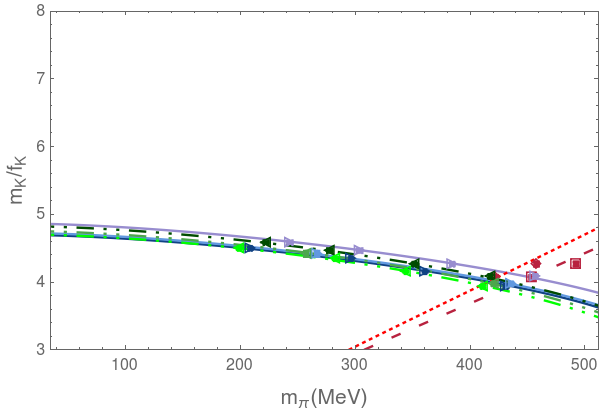}\\
 \includegraphics[scale=0.22]{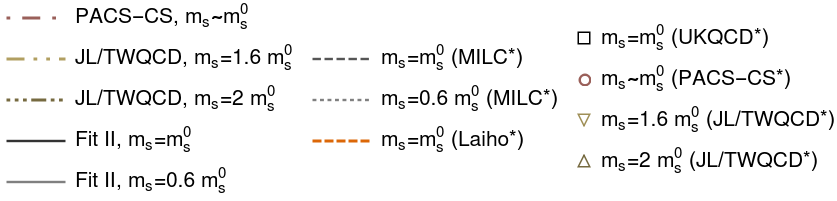}&\includegraphics[scale=0.22]{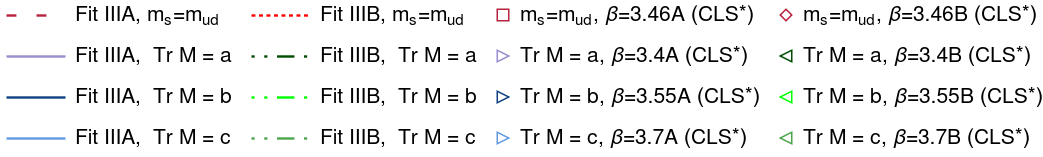}\\
 \end{tabular}}
\end{center}
\caption{Chiral trajectories (top) and ratios $m_\pi/f_\pi$, $m_K/f_\pi$, and $m_K/f_K$ obtained in fits II, PACS-CS, and JL/TWQCD, (left), and IIIA(B) (right),  from Tables~\ref{tab:lecsms},~\ref{tab:lecsmsother},~\ref{tab:lecstrma}, and~\ref{tab:lecstrmb}. Values with $(*)$ denote lattice data.}
\label{fig:mpikfpik2}
\end{figure*}

Regarding decay constant ratios and pseudoscalar meson masses, results from Fit II are very similar to those obtained in Sect.~\ref{sec:chpt} over $m_s=k$ trajectories, and are shown in Fig.~\ref{fig:mpikfpik2}.

Unfortunately, we could not obtain additional consistency with the lattice data from~\cite{Feng:2014gba,Alexandrou:2017mpi,Noaki:2009sk}. Thus, in the following, we analyze the remaining lattice results separately. The simulation of~\cite{Alexandrou:2017mpi} (CA) for $\rho$-meson phase-shift data does not include decay constant determinations, thus, we analyze this data with the UKQCD meson and decay constant values. If other decay constant data are used instead, as for example, those from MILC, the results are very similar.
The resulting LECs, given in the second column of Table~\ref{tab:lecsmsother}, are, in general, compatible with the values from Fit II, but we find slightly different values for $L_4$ and $L_5$, and larger discrepancies for $L_8$. These differences have a large impact on the phase-shift values. As shown in the right-top panel in Fig.~\ref{fig:ph}, the extrapolation to the physical point provides results incompatible with experimental data.

Concerning the JL/TWQCD collaboration decay constant data~\cite{Noaki:2009sk}, we only find good partial fits if we  include the three and two lightest pion mass data points for the trajectories $m_s=1.6\,m_s^0$ and $m_s=2\,m_s^0$, respectively. This can be due to the breakdown of the ChPT expansion for such large $m_s$ values. Since in these simulations decay constant determinations are provided but not $\rho$ phase shifts, we analyze them together with the Hadron Spectrum Collaboration (HS) $\rho$-meson phase shift results at $m_\pi=\{236,390\}$ MeV. The only purpose of this fit is to show the qualitative behavior of the pseudoscalar meson mass and decay constant ratios over trajectories with larger $m_s$ values than the physical one.
The corresponding LECs obtained in the fit are given in the third column in Table~\ref{tab:lecsmsother}.
A comparison with the result from Fit II in Table~\ref{tab:lecsms} shows up sizable discrepancies between both fits, which might be due to inconsistencies of the JL/TWQCD data with data included in Fit II apart from the breaking of the chiral series. 
These phase shift results are also plotted in the top-left panel of Fig.~\ref{fig:ph} in dashed lines.
Nevertheless, the extrapolation to the physical point of this fit turns out to be also very close to the experimental data.

For the PACS-CS collaboration, both $\rho$-meson phase shift~\cite{Feng:2014gba} and decay constant~\cite{Aoki:2008sm} data are available and analyzed together.
The LECs are given in the fourth column of Table~\ref{tab:lecsmsother}, and also the $L_i$'s, $i=4,6-8$, differ considerably from the Fit II values. As explained in Sect.~\ref{sec:chpt} these data have larger kaon masses for the same trajectory $m_s=m_s^0$ than data in Fit II. This can be due to sizable finite volume effects in these simulations. As a consequence, these data are in disagreement with the data included in Fit II.
In this case, the extrapolation to the physical point, depicted in the bottom-right panel of Fig.~\ref{fig:ph}, fails substantially to describe the experimental data.

Let us note that these $\rho$-meson phase-shift data were analyzed before in~\cite{Hu:2017wli} using the UChPT model in~\cite{Oller:1998hw}. Even though this model neglects the LHC contribution, which now is taken into account, we obtain here similar results  to the ones of~\cite{Hu:2017wli}. 

The chiral trajectories and decay constant ratios for  these fits are shown in Fig.~\ref{fig:mpikfpik2}.
We find that the pseudoscalar meson mass data on $m_s=k$ trajectories fit very well into a linear formula $m^2_K=a\,m^2_\pi+b$ with slope $a=0.5$, depicted in dotted lines.
This behavior is qualitatively similar to the leading order ChPT prediction.
For the ratios of decay constants we find similarities with the results of Fit I over the trajectory $m_s=m_s^0$. The ratios $m_K/f_\pi$ and $m_K/f_K$ in other $m_s=k$ trajectories as a function of the pion mass are parallel to the ones over $m_s=m_s^0$ and take higher values. For the $m_\pi/f_\pi$ ratio, only the JL/TWQCD and PACS-CS data are just a bit out of the error band.

Finally, the  $\rho(770)$ pole position on the second Riemann sheet obtained in the different fits are given in Table~\ref{tab:poms}. While the values obtained in Fit II and in the JL/TWQCD\,\&\,HS fits are compatible with the most precise theoretical prediction~\cite{GarciaMartin:2011jx}, the results obtained for UKQCD\,\&\,CA and PACS-CS provide smaller and larger values, respectively.  In order to write down results that can be compared to the BW values provided in lattice articles, we perform a refit of the IAM solution to the BW formula in Eq.~\eqref{eq:bw}. As we show in Fig.~\ref{fig:bw} in the Appendix~\ref{app:BWIAM}, the data is also well described by a Breit-Wigner (BW) parameterization. The BW mass, coupling and width, normalized to the pion mass, are shown in Table~\ref{tab:poms}, where we also provide the result for the extrapolation to the physical point.
\begin{table*}[htb]
\begin{center}
{\renewcommand{\arraystretch}{1.7}
\setlength\tabcolsep{0.4cm}
  \begin{tabular}{cccccc}
  \toprule
  LEC$\times 10^3$ & MILC+HS & UKQCD+HS &Laiho+HS& MUL+HS & Fit II\\
  \hline
  $L_{12}$&$0.3(4) $&$-0.1(4) $ &$0.1(2) $ &$0.3(1) $ &$0.2(1)$\\
  $L_3$&$-3.4(4) $ &$-3.1(4)$ &$-3.4(2) $ &$-3.4(1) $ &$ -3.4(1)$\\
  $L_4$&$0.03(2) $ &$0.04(1) $ &$0.03(2) $ &$0.05(2) $ &$0.04(1) $\\
  $L_5$&$1.2(2)$ &$0.90(3) $ &$0.90(5) $ &$0.93(3) $ &$0.94(2) $ \\
  $L_6$&$0.5(1)$&$0.24(3)$&$0.2(1) $ &$ 0.25(4)$ &$0.24(2) $ \\
  $L_7$&$0.6(3) $ &$0.5(2) $ &$0.5(1) $ &$0.40(6) $ &$0.44(6) $ \\
  $L_8$&$-0.5(1) $ &$-0.3(1) $ &$-0.2(1) $ &$-0.3(1) $ &$-0.27(4) $ \\
   \hline
  \end{tabular}}
 \end{center}
 \caption{Values of the LECs obtained from the different fits to the lattice trajectories $m_s=k$ described in the main text. Uncertainties are obtained from the minimization with the MINUIT program.}
 \label{tab:lecsms}

\end{table*}

\begin{table}[htb]
\begin{center}
{\renewcommand{\arraystretch}{1.6}
\setlength\tabcolsep{0.15cm}
  \begin{tabular}{cccc}
  \toprule
  LEC$\times 10^3$ & UKQCD+CA & JL/TWQCD+HS &PACS-CS\\
  \hline
  $L_{12}$&$0.4(3) $&$0.2(3) $ &$0.4(1) $ \\
  $L_3$&$-3.7(3) $ &$-3.7(3)$ &$-3.6(1) $ \\
  $L_4$&$-0.01(1) $ &$0.26(2) $ &$0.32(2) $ \\
  $L_5$&$1.3(1)$ &$1.08(4) $ &$1.0(1) $ \\
  $L_6$&$0.27(2)$&$0.7(1)$&$1.09(2) $ \\
  $L_7$&$0.3(2) $ &$0.8(3) $ &$1.0(1) $ \\
  $L_8$&$-0.06(4) $ &$-0.7(1) $ &$-1.52(5) $ \\
   \hline
  \end{tabular}}
 \end{center}
 \caption{Values of the LECs obtained from different $m_s=k$ fits as described in the main text. Uncertainties are obtained from the minimization with the MINUIT program.}
 \label{tab:lecsmsother}

\end{table}

\begin{table*}[htb]
\begin{center}
{\renewcommand{\arraystretch}{1.6}
\setlength\tabcolsep{0.1cm}
  \begin{tabular}{ccccccc}
  \toprule
 &$m_\pi$(MeV) &$\widetilde{E}_0$&$\widetilde{m}_\rho^{BW}$&$g^{BW}$&$\widetilde{\Gamma}^{BW}$&$\chi^2/\mathrm{d.o.f}$\\
  
  \hline 
  Fit II&$140$ &$5.42(4) - 0.51(8)\,i$ &$ 5.5(1)$ & $  5.9(1)$ & $ 1.0(1)$&$0.3$\\
  &$235$ &$3.304(6) - 0.195(3)\,i$ &$3.340(6)$& $  5.92(2)$& $ 0.399(3)$&\\
  &$390$ &$2.193(5) - 0.0170(5)\,i$ &$  2.195(4) $  & $5.81(2)$& $ 0.034(1)$&\\
  \hline
  UKQCD\&CA&$140$ &$5.26(3) - 0.46(1)\,i$&$    5.33(3)$& $5.75(3)$&$ 0.93(1)$&$0.5$\\
  &$317$ &$2.51(1) - 0.062(2)\,i$ &$  2.52(1)$ &  $ 5.79(4)$ & $0.126(4)$\\
  \hline
  TW/JLQCD\&HS&$140$ & $5.55(2) - 0.49(1)\,i$     &$  5.62(2)$ &$  5.66(3)$& $1.02(1)$&$1.4$\\
  &$235$ & $3.32(1) - 0.183(2)\,i$&$  3.35(1)$ &$  5.68(2)$   & $0.370(4)$&\\
  &$390$ &$2.19(3) - 0.0156(6)\,i$ & $  2.19(4) $ & $ 5.6(1)$  &$ 0.03(1)$&\\
  \hline
    PACS-CS&$140$ &$5.87(6) - 0.58(2)\,i$ & $    5.97(6)$  & $5.93(7)$ & $1.17(3)$&$1.2$\\
    &$300$ & $2.83(3) - 0.12(1)\,i$  &$ 2.86(3)$ &$ 5.98(7)$ &$ 0.25(1)$&\\
    &$400$ &$2.18(2) - 0.016(3)\,i$  &$  2.18(2)$ &$ 5.91(7) $ &$ 0.032(6)$&\\
   \hline
  \end{tabular}}
 \end{center}
 \caption{Pole positions obtained in the IAM, $E_0$, and BW parameters $m_\rho^{BW}$, $g^{BW}$ and $\Gamma^{BW}$ obtained from the refit of the IAM solution to the BW formula in Eq.~(\ref{eq:bw}) normalized to the pion mass, i.e., $\widetilde{E}_0$ stands for $E_0/m_\pi$. }
 \label{tab:poms}
\end{table*}

\subsection{Chiral trajectories $\text{Tr}{\cal M}=C$}\label{sec:trm}

\begin{figure*}
 \begin{center}
  \includegraphics[scale=0.37]{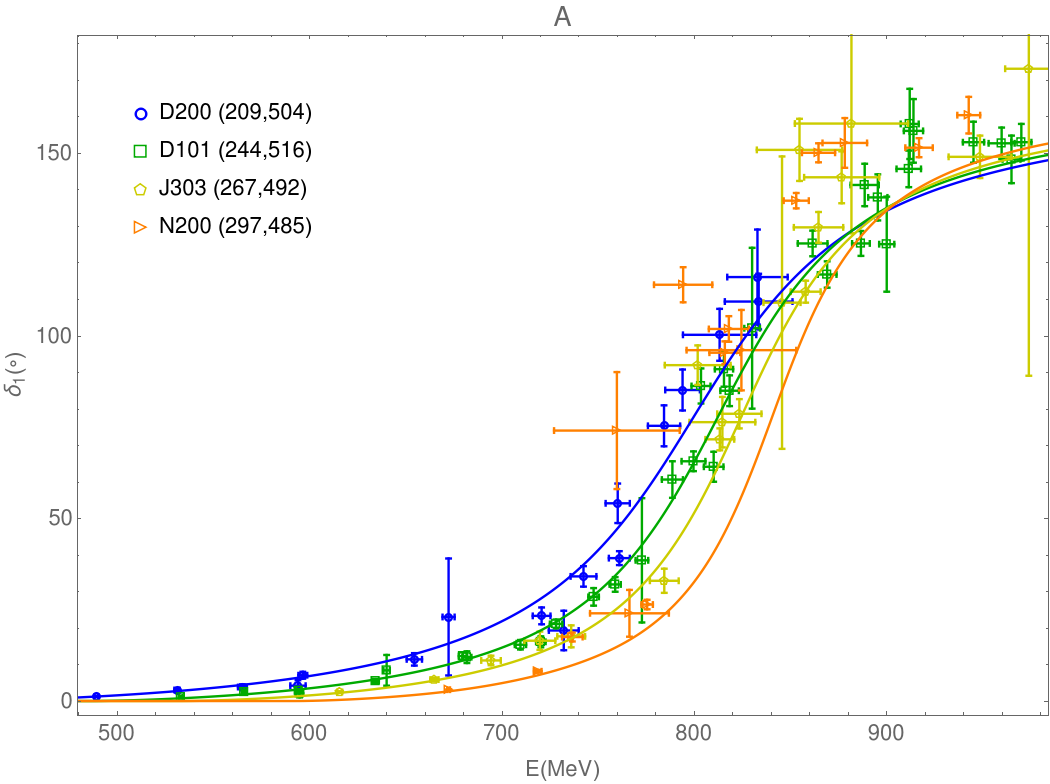}
 \end{center}
\caption{Phase shift lattice data in comparison with the result of Fit III A (global fit of $\text{Tr}{\cal M}=c$ lattice data with scaling method A).}
\label{fig:agl}
\end{figure*}

\begin{figure*}
 \begin{center}
  \includegraphics[scale=0.37]{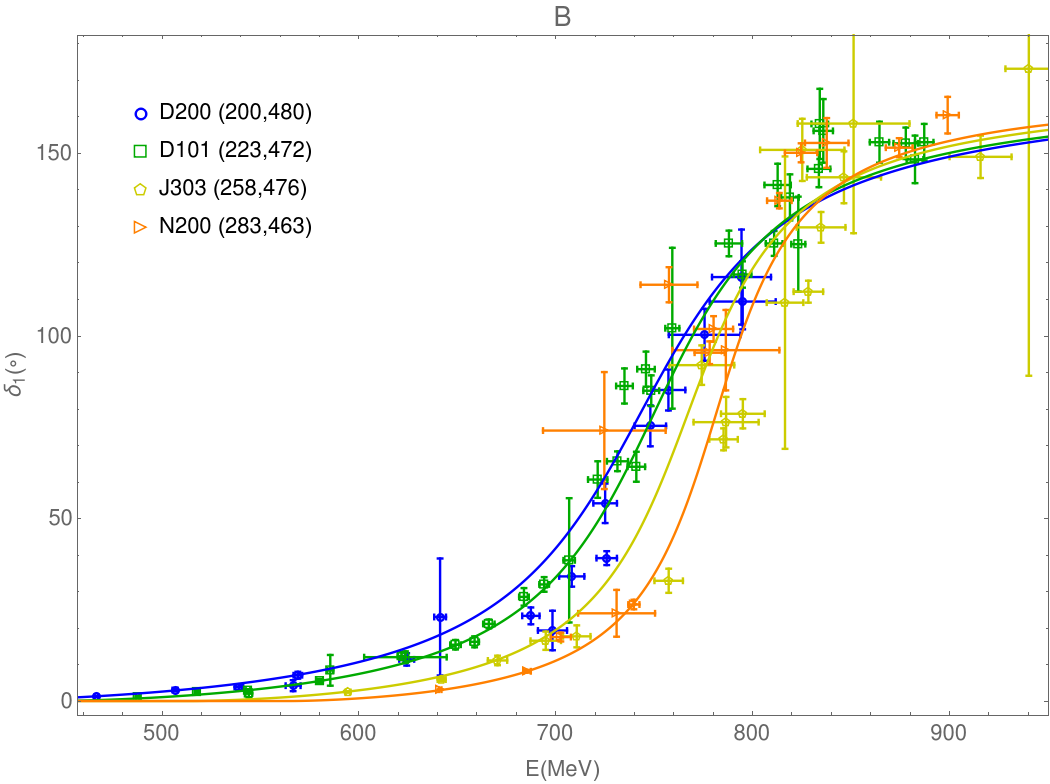}
 \end{center}
\caption{Phase shift lattice data in comparison with the result of Fit III B (global fit of $\text{Tr}{\cal M}=c$ lattice data with scaling method B).}
\label{fig:bgl}
\end{figure*}

In this section we show the outcome of the analysis of $\rho$-meson phase-shift~\cite{Andersen:2018mau} and decay constant~\cite{Bruno:2016plf} data of the CLS Collaboration over trajectories where $\text{Tr}{\cal M}=2m_{ud}+m_s=C$, see Tables II of ~\cite{Bruno:2016plf} and 6-11 of ~\cite{Andersen:2018mau}. Thus, in these trajectories the kaon becomes lighter  as the pion mass increases. 
Two different scale setting methods were considered in~\cite{Bruno:2016plf}. These two methods lead to differences of around $10-20$ MeV in $m_\pi$, $20-45$ MeV in $m_K$, and $4-8$ MeV for $f_\pi$ and $f_K$. These differences entail several difficulties. As discussed in section~\ref{sec:chpt}, we could only find an optimal solution to the minimization problem of Fit I, that also includes $m_s=k$ data, when {\it scale setting A} was taken for the pseudoscalar meson masses and decay constants over the $\text{Tr}{\cal M}=C$ trajectories. When the {\it scale setting B} was considered instead, the global $\chi^2$ minimum was found to be around twice larger than with the {\it scale setting A}.
On the contrary, we observe that, when using {\it scale setting A}, the dependence of $\rho$-phase shift data with the pion mass of~\cite{Andersen:2018mau} cannot be described well within the IAM for all ensembles.
While the ensembles D101, J303 and D200 are well described, the ensemble N200 (or N401) cannot be reproduced. This is because the IAM predicts higher values of the $\rho$ meson mass for the pion mass used in this ensemble, see Fig.~\ref{fig:agl} (Fit IIIA).
Interestingly, by using {\it scale setting B}, we find a solution describing all $\text{Tr}{\cal M}=C$ lattice data, i.e., pseudoscalar meson mass and decay constant ratios and $\rho$-meson phase shift (excluding $m_s=k$ data). These phase-shift results  are plotted in Fig.~\ref{fig:bgl} (Fit IIIB).

Nevertheless, it is possible to perform fits of decay constant and $\rho$-meson phase shift data for ensembles with the same  gauge coupling $\beta$ \cite{Andersen:2018mau}. Namely, C101, D101 ($\beta=3.4$), N401 ($\beta=3.46$), N200, D200 ($\beta=3.55$) and J303 ($\beta=3.7$).
 Several of these ensembles use the same pion mass but different volume or lattice spacing. On one side, the ensembles C101 and D101 are simulated with the largest lattice spacing but D101 uses a volume $2.4$ times bigger than C101. On the other side, the ensembles N200 and N401 were simulated in the smallest volume but N200 has a lattice spacing $1.13$ times smaller than N401. Finally, J303 has the biggest volume and smallest lattice spacing.\footnote{The volumes of the C101 and D101 ensembles are $L^3\times T=48^3 \times 96$ and $64^3 \times 128$ respectively, both with a lattice spacing $a=0.086$ fm ({\it scale setting B)}. The lattice spacings for N401 and N200 are $a=0.076$ fm and $0.064$ fm, respectively, and both have the same volume $48^3\times 128$. J303 has $a=0.05$ fm and is simulated in a volume $L^3\times T= 64^3\times 192$. The volume and lattice spacing used for D200 are $64^3\times 128$ and $0.064$ fm, respectively. }
In this way, possible differences between individual fits in these pairs might highlight finite volume and lattice spacing effects. 
The resulting LECs are shown in Tables~\ref{tab:lecstrma} and~\ref{tab:lecstrmb} for the {\it A } and {\it B }, respectively. The ensembles C101 and D101 are fitted separately in order to study the finite volume effect. Overall, the values of the LECs $L_3$ and $L_5$ are approximately stable, but we find large differences for the others. 

We also attempt to perform combined fits including most ensembles for different $\beta$ in order to check whether these effects can be absorbed in the LECs. Since the D101 and N200 ensembles supersede the C101 and N401 ones, accordingly,  we only include the ensembles D101, N200, D200 and J303. These fits are denoted as Fit III A and III B for the {\it A} and {\it B scale settings}, respectively, and they also include the corresponding pseudoscalar meson mass and decay constant ratio data. The LECs obtained are given in the last columns of Tables~\ref{tab:lecstrma} and~\ref{tab:lecstrmb}.

In general, the LECs of Fit IIIA agree better with those obtained for the $m_s=k$ trajectories.
The chiral trajectories and decay constant ratios of these fits are depicted in Fig.~\ref{fig:mpikfpik2} (right),
where we also show the result of the $m_s=k$ fits for comparison (left panel).
Results for the fits IIIA and B are plotted in like blue-solid and green-double-dot-dash lines, respectively. The kaon mass dependence on the pion mass for the trajectories $\text{Tr}{\cal M}=C$ also fit well into straight lines, $m^2_K=a\,m^2_\pi+b$, but now with a slope close to $a=-0.5$ instead of $0.5$, as we found for the $m_s=k$ ones. In this way, the IAM is able to reproduce very well the $\mathrm{Tr}\,{\cal M }=C$ trajectories, which appear as three close decreasing curves intersecting the symmetric line, $m_s=m_{ud}$. At pion masses of $300$ MeV, the kaon mass is around $60$ MeV lower than for the $m_s=k$ trajectory.

The ratios of decay constants are also well reproduced. For the {\it scale setting A}, the ratio $m_\pi/f_\pi$ agrees well with the $m_s=k$ data, emphasizing that this ratio is almost independent of $m_s$. In the case of {\it scale setting B}, it falls a bit out of the $m_s=m_s^0$ error band, depicted in a light-brown color.
Note that this behavior is different from the $m_s=0.6\,m_s^0$ trajectory (MILC, dotted-gray), which lies inside the error band and does not show any substantial difference with the $m_s=m_s^0$ curve.
This suggests that there could be small dependencies with the strange-quark mass. We comment more on this issue in the next section.

\begin{figure*}
 \begin{center}
  \includegraphics[scale=0.4]{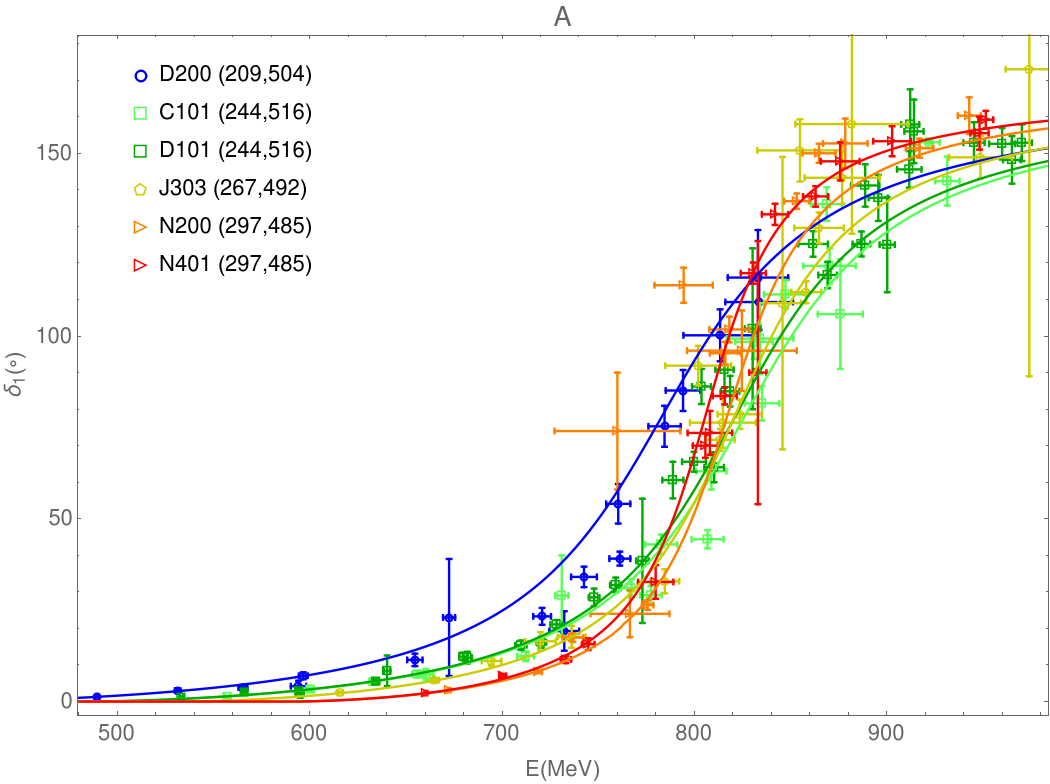}
 \end{center}
\caption{Phase shift lattice data~\cite{Andersen:2018mau} and individual fits (depending of $\beta$) obtained with the {\it scale setting A}. The values in brackets stand for the pion and kaon masses (in MeV), respectively.}
\label{fig:pha}
\end{figure*}

\begin{figure*}
 \begin{center}
  \includegraphics[scale=0.4]{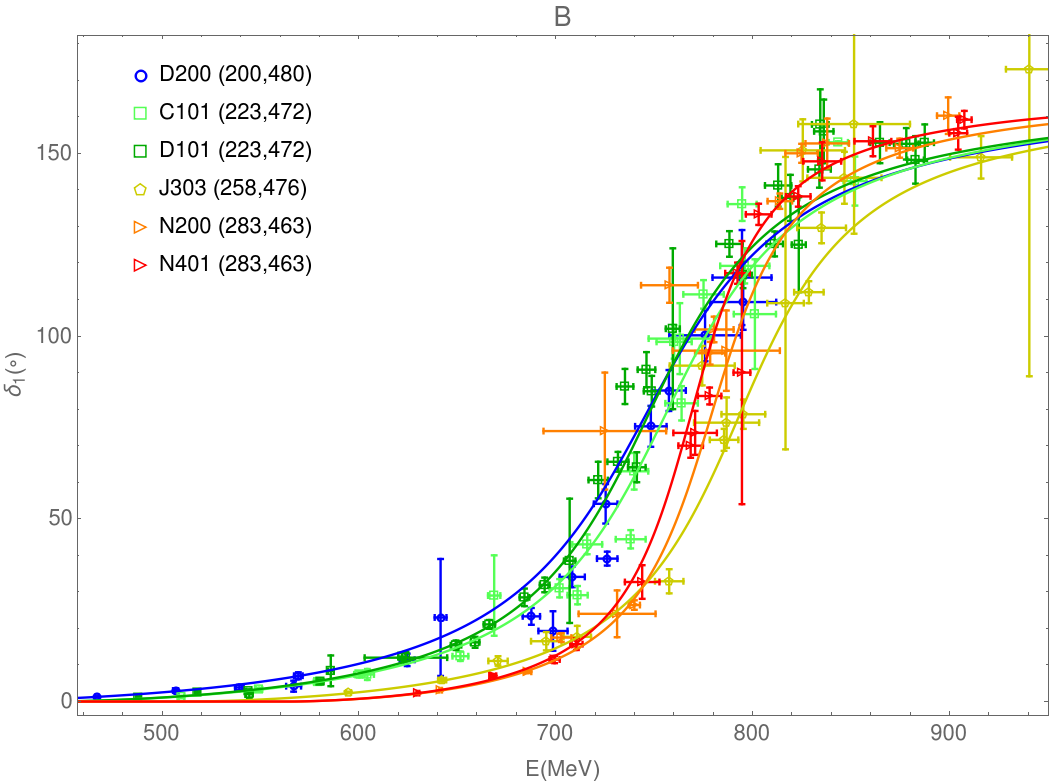}
 \end{center}
 \caption{Phase shift lattice data~\cite{Andersen:2018mau} and individual fits (depending of $\beta$) obtained with the {\it scale setting B}.
   The values in brackets stand for the pion and kaon masses (in MeV), respectively.}
\label{fig:phb}
\end{figure*}

Results for $\rho$-phase shifts are provided in Figs.~\ref{fig:agl} and~\ref{fig:bgl}.
As commented before, except for the ensemble N200 in {\it scale setting A}, all the other phase shifts can be described qualitatively well in these fits. 
In Figs.~\ref{fig:pha} and~\ref{fig:phb} we show the $\rho$-meson phase shifts obtained for the different gauge coupling fits. 
The IAM allows one to describe the $\rho$-meson phase-shift data in $\text{Tr}{\cal M}=C$ trajectories for every ensemble.
Nevertheless, note that one can not observe a trend of the overall data indicating that the $\rho$-meson mass\footnote{Understood as the energy for which the phase shift is $90^\circ$.} increases monotonically with the pion mass. Namely, for {\it scale setting A}, the N200 and N401 ensembles give rise to a lighter $\rho$-meson mass than the ensembles D101 and C101 even when they are simulated with heavier pion masses. In addition, the $\rho$-meson mass takes about the same value for the J303 and D101 ensembles, although the pion mass used in J303 is around $20$ MeV larger. Similar results have been observed in a recent two-loop SU(2) IAM analysis of the same CLS data~\cite{Niehus:2020gmf}.

At low energies, phase shifts decrease as the pion mass grows, as expected from the $p$-wave centrifugal barrier and the chiral expansion.
For {\it scale setting B} one observes that the trend of the $\rho$-meson mass dependence on the pion mass is flatter. Noticeably, the $\rho$-meson becomes lighter for pion masses around $300$ MeV in both scale settings.
In both cases, systematic effects due to a finite volume and lattice spacing are reflected in around $8$ MeV difference in the $\rho$-meson mass between the C101 and D101 and $14$ MeV between the N200 and N401 ensembles, respectively.

\begin{figure*}
 \begin{center}
  \includegraphics[scale=0.4]{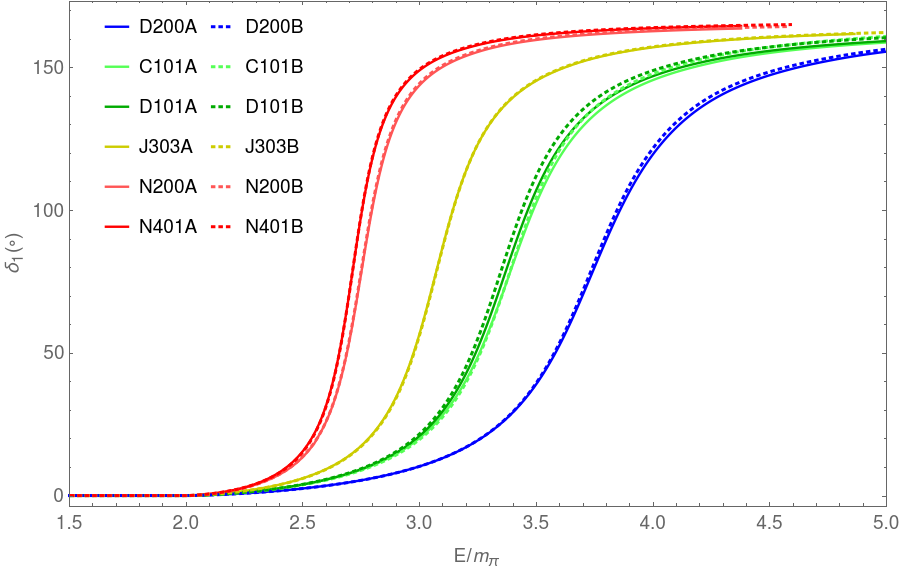}
 \end{center}
\caption{Comparison between the individual fit solutions using {\it A} and {\it B scale settings}.}
\label{fig:compascaling}
\end{figure*}

The corresponding pole positions and couplings for both scale setting are given in Table~\ref{tab:polosfit3}. In addition, given the large discrepancies observed between the scale settings we perform a new fit with their average for each gauge coupling $\beta$,  denoted as Fit C in Table~\ref{tab:polosfit3}. For comparison, the result of the global fit including both data on $m_s=k$ and Tr${\cal M}=C$ trajectories, discussed in the next section (Fit IV) is also shown.
The values are normalized to the pion mass, so that the dependence of the ratio $m_\rho/m_\pi$ with the pion mass and scale setting used is visible.
Overall, we see that the results for different lattice spacings are quite similar. For the ensemble J303 the dependence on the scale setting considered is negligible, while for other ensembles it produces shifts of less than 1\% for the normalized $\rho(770)$ mass and less than 5\% for the couplings.
Regarding finite volume and lattice effects, the systematic differences between C101 and D101 are of around 1\% in the normalized $\rho$ mass, and 1.5\% between the N200 and N401, while these are of less than 2\% in the couplings in both cases.
Finally, the comparison between the individual fit solutions obtained using {\it A} and {\it B scale settings } is given in Fig.~\ref{fig:compascaling},
where it can be seen that the differences produced in phase shifts as a function of $E/m_\pi$ are in general reasonably small, and  negligible for the J303 ensemble.
The largest difference is coming from the size of the lattice spacing used in the simulation, i.e., the difference observed between the N200 and N401 ensembles. 

Finally, we can compare the result of Fit C in Table~\ref{tab:polosfit3} with the result of Fit IV which includes also $m_s=k$ data. There are small differences between these two fits of less than 3\% in the normalized $\rho$-meson mass and less than 6\% in the couplings. We discuss this further below.
 
 \begin{table*}[htb]
\begin{center}
{\renewcommand{\arraystretch}{1.7}
\setlength\tabcolsep{0.2cm}
  \begin{tabular}{ccccccc}
  \toprule
  LEC$\times 10^3$ &\multicolumn{2}{c}{ $\beta=3.4$}&$\beta=3.46$&$\beta=3.55$&$\beta=3.7$&Fit IIIA\\
  Ensemble:&C101&D101&N401&N200, D200&J303&{\small{D200, N200}}\\
 &&&&&&J303, D101\\
 $\chi^2/\mathrm{d.o.f}$&$2.0$&$2.9$&$1.1$&$1.5$&$0.6$&$2.2$\\
  \hline
  $L_{12}$&$0.20(1)$ &$0.28(5) $&$0.18(3)$&$-0.1(1) $ &$-0.1(2) $ &  $0.25(5)$\\
  $L_3$&$-2.85(3)$&$-3.00(5) $ &$-3.16(3)$&$-2.74(2)$ &$-2.6(4) $ & $ -2.98(5)$\\
  $L_4$&$0.005(2)$&$-0.038(4) $ &$-0.090(2)$&$-0.04(1) $ &$-0.11(3) $ &$-0.08(1)$\\
  $L_5$&$1.10(1)$&$0.94(3)$ &$1.07(1)$&$1.27(2) $ &$0.9(1) $ & $1.00(4)$\\
  $L_6$&$0.72(4)$&$0.40(3)$&$0.40(1)$&$0.77(2)$&$0.31(2) $ &$0.40(4)$\\
  $L_7$&$0.44(2)$&$0.3(1) $ &$0.43(6)$&$0.98(3) $ &$0.8(1) $ & $0.22(1)$\\
  $L_8$&$-0.2(1)$&$-0.17(6) $ &$0.13(3)$&$0.23(1) $ &$-0.08(5) $ &$-0.05(5)$\\
   \hline
  \end{tabular}}
 \end{center}
 \caption{Values of the LECs obtained in the fits $\text{Tr}{\cal M}=C$ with the {\it scale setting A}.}
 \label{tab:lecstrma}
\end{table*}
\begin{table*}[htb]
\begin{center}
{\renewcommand{\arraystretch}{1.7}
\setlength\tabcolsep{0.2cm}
  \begin{tabular}{ccccccc}
  \toprule
  LEC$\times 10^3$ &\multicolumn{2}{c}{ $\beta=3.4$}&$\beta=3.46$&$\beta=3.55$&$\beta=3.7$&Fit IIIB\\
  Ensemble:&C101&D101&N401&N200, D200&J303&{\small{D200, N200}}\\
 &&&&&&J303, D101\\
 $\chi^2/\mathrm{d.o.f}$&$1.8$&$3.2$&$1.0$&$1.5$&$0.5$&$2.4$\\
  \hline
  \hline
  $L_{12}$&$-0.46(1)$&$-0.3(1) $&$0.0(2)$&$-0.1(1) $ &$-0.1(1) $ &$-0.27(2)$\\
  $L_3$&$-2.28(1)$&$-2.5(1) $ &$-3.0(2)$&$-2.8(1)$ &$-2.7(1) $ &$ -2.60(3)$\\
  $L_4$&$-0.37(2)$&$-0.40(1) $ &$-0.28(1)$&$-0.26(2) $ &$-0.22(1) $ &$-0.32(1)$\\
  $L_5$&$0.86(6)$&$0.92(2)$ &$1.09(4)$&$1.29(5) $ &$0.92(4) $ &$1.12(4)$ \\
  $L_6$&$0.29(3)$&$0.23(3)$&$0.37(3)$&$0.74(4)$&$0.36(1) $ &$0.55(2)$ \\
  $L_7$&$1.3(2)$&$1.0(1) $ &$0.7(3)$&$0.7(4) $ &$0.9(1) $ &$1.26(5)$ \\
  $L_8$&$-0.23(6)$&$-0.18(1) $ &$0.2(1)$&$0.3(1) $ &$-0.13(1) $ &$0.13(5)$ \\
   \hline
  \end{tabular}}
 \end{center}
 \caption{Values of the LECs obtained in the fits $\text{Tr}{\cal M}=C$ with the {\it scale setting B}.}
 \label{tab:lecstrmb}
\end{table*}

\begin{table}[htb!]
\begin{center}
{\renewcommand{\arraystretch}{1.6}
\setlength\tabcolsep{0.12cm}
  \begin{tabular}{ccccccc}
  \toprule
 & &$m_\pi$(MeV)&$\widetilde{E}_0$&$\widetilde{m}_\rho^{BW}$&$g^{BW}$&$\widetilde{\Gamma}^{BW}$\\
   \hline
  D200 &A&209 & $3.75 - 0.30\,i$ &  $3.80$ &$ 6.20$ &  $ 0.60$\\
       &B&200   & $  3.74 - 0.29\,i$& $  3.79$  & $ 6.10$&  $0.57$\\
       &C&204 & $  3.73 - 0.29\,i$& $ 3.79$  & $6.16$& $0.59$\\
       \hline
        &IV&204  &  $3.72 -0.27\,i$ & $ 3.77$ & $ 6.00$ & $ 0.55$\\
        \hline
 C101 &A&244  & $ 3.39 - 0.25\,i$&$ 3.44$ & $ 6.43$   &$ 0.51$\\
       &B&223&  $   3.39 - 0.23\,i$ &$3.44$ &$  6.15$  & $0.46$\\
       &C&233  & $  3.37 - 0.23\,i$ &$3.42 $&$  6.22 $& $ 0.47$\\
       \hline
  D101 &A&244   & $ 3.37 - 0.24\,i$ & $ 3.42$ &$ 6.38 $ &$ 0.49$\\
       &B&223   & $ 3.35 - 0.23\,i$ &$ 3.40 $&$6.24$ &$ 0.46$\\
       &C&233  & $   3.36 - 0.23 \,i$& $3.41$& $6.21$ &$ 0.46$\\
       \hline
       &IV&233   & $ 3.31 -0.20 \,i$& $ 3.35 $&$  6.01$  &$0.41$\\
       \hline
  J303 &A&267 & $ 3.07 - 0.18 \,i$ &$ 3.11$ &$   6.30$ & $0.37$\\
       &B&258 &$  3.07 - 0.18 \,i $ & $ 3.11$ & $  6.27 $&$0.36$\\
       &C&263  & $  3.07 - 0.18\,i$&$  3.10$ & $  6.31$ &$0.37$\\
       \hline
       &IV&263  & $  3.00 -0.15\,i$& $ 3.03 $& $  5.99 $ &$ 0.31$\\
       \hline
  N200 &A&297  & $ 2.75 - 0.11 \,i$& $ 2.78$ & $  6.18$ &$ 0.24$\\
       &B&283 &$ 2.75 - 0.11 \,i$& $ 2.77$  & $ 6.13$ &$0.23$\\
       &C&290 &  $2.76 - 0.12\,i$& $  2.78  $& $ 6.17$ &$0.24$\\
       \hline
       &IV&290&  $ 2.77 -0.11 \,i $& $ 2.79$ & $  5.97$ & $0.22$\\
       \hline
  N401 &A&297    &$ 2.72 - 0.10 \,i$ &$ 2.74$  & $6.07$ &$ 0.21$\\
       &B& 283  & $2.72 - 0.10 \,i$ &$ 2.74$&$   6.03$ &$ 0.21$\\
       &C&290  &  $2.74 - 0.11 \,i$ & $ 2.76$  & $6.10$ &$0.22$\\
       \hline
       &IV&290 &$ 2.77 -0.11 \,i$  &$  2.79$  & $5.97$   &$ 0.22$\\
   \hline
  \end{tabular}}
 \end{center}
 \caption{$\rho$-meson pole positions and couplings obtained for the individual $\text{Tr}{\cal M}=C$ fits given in Tables~\ref{tab:lecstrma} and~\ref{tab:lecstrmb} and for each scale setting. The IAM pole position is denoted by $E_0$, while the BW parameters $m_\rho^{BW}$, $g^{BW}$ and $\Gamma^{BW}$ are obtained by refitting the IAM solution to the Breit-Wigner formula. The fit C stands for the average of both scale settings, while IV denotes the global fit discussed in section~\ref{sec:gl}, included for comparison, which is obtained performing a resampling of the lattice spacing and lattice data. The quantities with tilde are normalized to the pion mass, i. e. , $\widetilde{E}_0$ stands for $E_0/m_\pi$.}\label{tab:polosfit3}
\end{table}

\section{Global fit over $\text{Tr}{\cal M}\boldsymbol{=C}$ and $\boldsymbol{m_s=k}$ trajectories: Fit IV}\label{sec:gl}

In this section we perform a simultaneous analysis of lattice data over both  $m_s=k$ and $\text{Tr}{\cal M}=c$ trajectories.
This final study will be denoted as Fit IV and it analyzes  lattice
$\rho$-meson phase shift data in $N_f=2+1$ of~\cite{Dudek:2012xn,Wilson:2015dqa,Bulava:2016mks,Andersen:2018mau} in combination with pseudoscalar meson masses and decay constants from~\cite{Bruno:2016plf,Blum:2014tka,Bazavov:2010hj,Bazavov:2009bb,Aubin:2008ie}. Thus, this analysis takes into account all data included in the fits II and III of Sects.~\ref{sec:ms} and~\ref{sec:trm}.

As discussed in Sects.~\ref{sec:chpt} and~\ref{sec:trm}, we were not able to find a solution with the IAM describing data on $\text{Tr}{\cal M}=C$ trajectories using either of the scale settings in~\cite{Andersen:2018mau,Bruno:2016plf} in combination with data over $m_s=k$. Hence, in order to attempt a global fit some remarks are necessary. First of all, the ensemble C101 of~\cite{Andersen:2018mau} will be discarded since the simulation for the ensemble D101 is performed in a larger volume.
Note that even when the N401 ensemble has larger lattice spacing than the N200, the former has more data points and its  uncertainties are smaller, therefore, we include both ensembles in the present analysis. 
Secondly, it is important to highlight that, according to Tables~\ref{tab:poms} and~\ref{tab:polosfit3}, the CLS result for the ratio $m_\rho/m_\pi$ of the D101 ensemble is very close to the one from the HadSpec (HS) collaboration at $m_\pi=236$ MeV, Fit II; the difference is only of around 2\%. This fact points out that the pion masses used in these simulations should also be very similar.
Nevertheless, only the average between the {\it scale setting A} and {\it B} has a similar pion mass ($m_\pi=233$ MeV).
This facts motivates us to consider that the average between both scale settings provides a reasonable estimate to be used in order to perform a global fit of data. Hence, we perform a  bootstrap of the lattice spacing for the $\text{Tr}{\cal M}=C$ ensembles assuming that for every $\beta$, it is normally distributed around the average of {\it scale setting A} and {\it B} and the standard deviation being half the difference between them.
Not only the lattice spacings, $a_\beta$'s, but also decay constant ratios and energy levels (normalized respect to the pion mass) for each ensemble are generated from a normal distribution accordingly to their lattice data errors. Regarding the lattice energy levels, the resampling is performed assuming a multivariate normal distribution with the original covariance matrix.\footnote{Phase-shift data are then obtained from a first order Taylor expansion around the lattice data energies.}

Remarkably, following this strategy we could reproduce decay constant and phase-shift data simultaneously on both trajectories.
The resampling is performed 300 times and the error is evaluated from that sample. This number of fits turns out to be enough,
since the average, median and fit solution (taking the average of the lattice spacing) are indeed very close to each other.
Namely, they produce differences in the $\rho$-meson mass of less than $1-2$ MeV. Since we are interested on interpreting the results in terms of probability and confidence intervals, our central results are represented by the median or first quartile and the uncertainties will be described by the 68\% and 95\% confidence intervals (CI), which are represented as darker and lighter error bands, respectively.

Finally, let us remind that here we are only considering the systematic error associated to the scale setting for the CLS data~\cite{Andersen:2018mau,Bulava:2016mks}, which has much larger effects (as discussed in Sect.~\ref{sec:chpt}) than the one observed in the HS data~\cite{Dudek:2012xn,Wilson:2015dqa,Bolton:2015psa}. The latter was investigated in~\cite{Hu:2017wli}, where the two different lattice spacings from~\cite{Bolton:2015psa} were considered, leading to a difference in the $\rho$ mass of less than $0.3\%$, which is neglected here (see Table II of \cite{Hu:2017wli}).

\begin{figure*}
\includegraphics[scale=0.55]{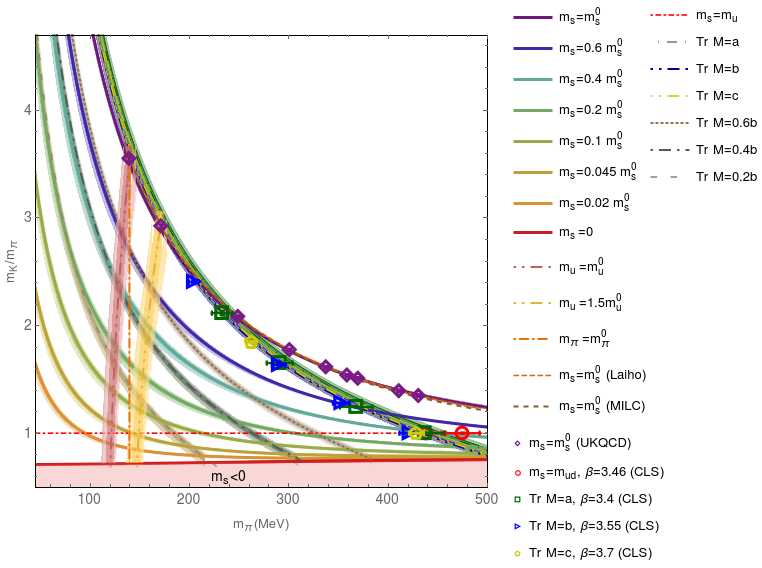}\\
\caption{Chiral trajectories ($m_K/m_\pi$ ratio) considered in Fit IV in comparison with lattice data.}
\label{fig:mkmpi}
\end{figure*}

\begin{figure*}
\includegraphics[scale=0.55]{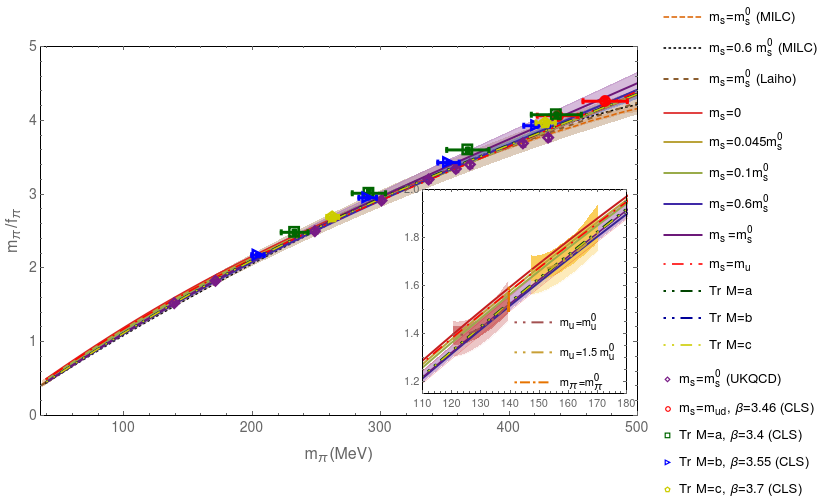}\\
\caption{The ratio $m_\pi/f_\pi$ obtained in Fit IV in comparison with the lattice data.}
\label{fig:mpifpi}
\end{figure*}

\begin{figure*}
\begin{tabular}{ccc}
 \includegraphics[scale=0.37]{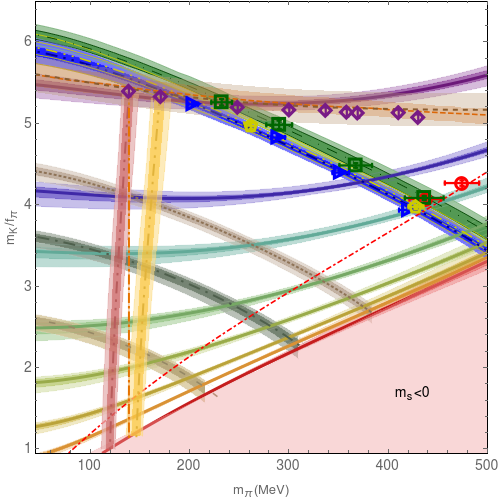} &\quad\includegraphics[scale=0.37]{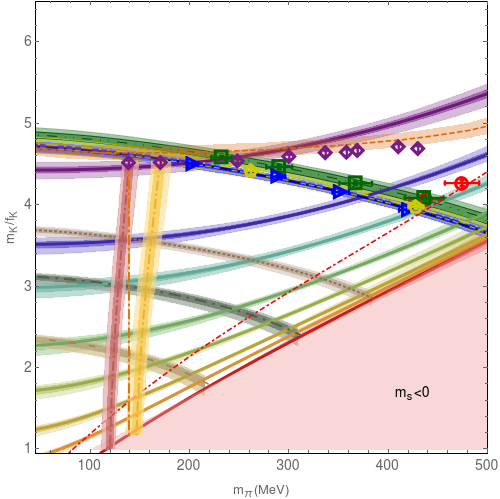}&\includegraphics[scale=0.4]{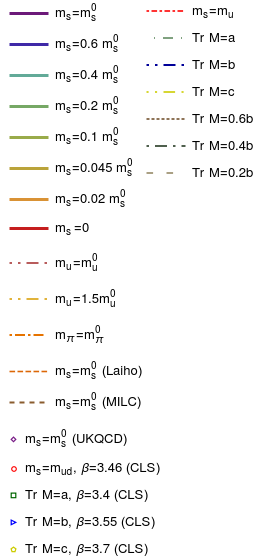}\\
 \end{tabular}
\caption{Decay constant ratios, $m_K/f_\pi$ and $m_K/f_K$, obtained in Fit IV in comparison with the lattice data.}
\label{fig:mkfk}
\end{figure*}

\subsection{Results for meson masses and decay constants}

In Figs.~\ref{fig:mkmpi},~\ref{fig:mpifpi} and~\ref{fig:mkfk}, the chiral trajectories and pseudoscalar meson mass and decay constant ratios studied are plotted.
The lattice data fitted correspond to the extrapolation to the continuum limit with finite volume effects corrected.
In more detail, for the $m_s=m_s^0$ trajectory we include the UKQCD~\cite{Blum:2014tka} (purple diamonds), MILC~\cite{Bazavov:2010hj,Bazavov:2009bb} (black dashed curves with light-brown error bands\footnote{The data was sent to us by C. Bernard and the error band is extrapolated from the physical point as suggested by him.}) and Laiho~\cite{Aubin:2008ie} (orange dashed curves and error bands) lattice data. For other $m_s=k$ trajectories there is not much data  except for the ratio $m_\pi/f_\pi$ extracted by MILC~\cite{Bazavov:2010hj} for $m_s=0.6\,m_s^0$ (gray dotted line). The $\text{Tr}{\cal M}=C$ data from the CLS Collaboration are given for the different lattice gauge couplings $\beta=3.4$ (green squares), $3.46$ (red circles), $3.55$ (blue triangles) and $3.7$ (yellow pentagons). The error in the pion mass (x-axis) corresponds to half the difference between the pion mass using the two {\it A} and {\it B scale settings}. 
Although in principle chiral trajectories for the several gauge couplings $\beta$ are different, in practice, we obtain very similar curves when the error in the lattice spacing is considered, which only start to separate more clearly when these cross the symmetric line.
This is, we get $c_{(\beta=3.55)}\simeq c_{(\beta=3.7)}$ and only a small difference for $\beta=3.4$.
In addition, we include in Figs.~\ref{fig:mkmpi},~\ref{fig:mpifpi} and~\ref{fig:mkfk} the IAM prediction for the trajectories $m_s=\lbrace0,0.02,0.045, 0.1,0.2,0.4,0.6\rbrace\,m_s^0$, which are almost parallel to the $m_s=m_s^0$ one. Furthermore, in order to highlight the relevance of the strange-quark mass, we also include the prediction for the trajectories $m_{u}=\lbrace 1,1.5\rbrace \,m_u^0$\footnote{In this case, it is understood that $m_d=m_u$.} and $m_\pi=m_\pi^0$. These three trajectories start at a small value of $m_s$ ($m_s B_0=2 \,\mathrm{MeV}^2$), then, they cross the symmetric line and end up at the $m_s=m^0_s$ curve. All ratios $m_K/m_\pi$, $m_\pi/f_\pi$, $m_K/f_\pi$ and $m_K/f_K$ are reproduced well inside the $95$ \% CI till $m_\pi\simeq 400$ MeV, when the ChPT predictions start to deviate. Therefore, the predictions for pion masses between $m_\pi=400-500$ MeV are merely qualitative.

The extrapolation to the physical point for the mass and decay constant ratios is given in Table~\ref{tab:physdecayra}. The central value represents the median, while the first upper and down indices show the limits of the 68\% CI. The upper (down) limits of the 95\% CI are obtained by summing the absolute values of the first and second upper (down) indices. These extrapolated ratios are compatible with the experimental values, which are inside our $68\%$CI.
\begin{table}
\begin{center}
{\renewcommand{\arraystretch}{2}
\setlength\tabcolsep{0.10cm}
\begin{tabular}{cccc}
\toprule
 $m_K^0/m_\pi^0$&$m_\pi^0/f_\pi^0$&$m_K^0/f_\pi^0$&$m_K^0/f_K^0$\\\hline
 $3.55_{-0.02 (0.05)}^{+0.02(0.04)}$&$1.51 _ {-0.012 (0.03)}^{+0.013 (0.03)}$&$5.33_{-0.05(0.13)}^{+0.05(0.11)}$&$4.45 _ {-0.03 (0.10)}^{+0.04 (0.09)}$\\
 \hline
 \multicolumn{4}{c}{Experiment [PDG]}\\\hline
 $3.5371(1)$&$1.513(2)$&$5.350(7)$&$4.48(2)$\\
 \hline
 \end{tabular}}
\end{center}
\caption{In the first row, the values of the ratios of pseudoscalar masses and decay constants extrapolated to the physical point for Fit IV. The uncertainties quoted should  be interpreted in terms of probability. The central value represents the median, the first upper and down indices gives the $68$\% CI, while the sum of the absolute values of the two upper (down) indices provides the upper (down) limits of the $95$\% CI. The experimental values are shown in the second row~\cite{Tanabashi:2018oca}.}
\label{tab:physdecayra}
\end{table}

The values of the LECs and remaining fit parameters are given in Table~\ref{tab:lecsgl}, where errors also represent $68\%$ and $95\%$ CI. The quark condensate $\Sigma_0$ can be estimated for a given strange-quark mass from Table~\ref{tab:lecsgl}. For instance, taking $m_s^0=95$ MeV we obtain $\Sigma_0^{1/3}=247$ MeV,
in close agreement with the MILC result~\cite{Bazavov:2009fk}, $245(5)(4)(4)$ MeV. 
Finally, the correlation matrix of the parameters is given in Eq.~\eqref{eq:cor} in the Appendix~\ref{app:cov}. 

\begin{table}
\begin{center}
{\renewcommand{\arraystretch}{2}
\setlength\tabcolsep{0.15cm}
 \begin{tabular}{cc}
 \toprule
 ($\chi^2/\mathrm{d.o.f}=1.2$)&LECs$\times 10^{3}$\\\hline
  $L_{12}$&$0.36^{+0.02(0.06)}_{-0.02(0.02)}$\\
  $L_3$&$-3.44^{+0.04(0.07)}_{-0.04(0.06)}$\\
  $L_4$&$-0.08^{+0.03(0.05)}_{-0.04(0.03)}$\\
  $L_5$&$0.98^{+0.07(0.06)}_{-0.05(0.04)}$\\
  $L_6$&$0.24^{+0.08(0.16)}_{-0.06(0.05)}$\\
  $L_7$&$0.008^{+0.09(0.12)}_{-0.14(0.15)}$\\
  $L_8$&$0.098^{+0.10(0.11)}_{-0.11(0.16)}$\\
  \hline
 & $c\times 10^{-3}$, $k\times 10^{-3}$\\
  \hline
  Tr\,M($\beta=3.4$)&$268^{+14(8)}_{-18(20)}$\\
  Tr\,M($\beta=3.55$)&$254^{+11(7)}_{-18(18)}$\\
  Tr\,M($\beta=3.7$)&$257^{+12(7)}_{-17(19)}$\\
  $m_s B_0$&$224^{+14(10)}_{-18(20)}$\\
  \hline
 \end{tabular}}\end{center}
 \caption{Values of the parameters obtained in Fit IV. The errors can be interpreted in terms of probability. The central value represents the median, the first upper and down indices gives the $68$\% CI, while the sum of the absolute values of the two upper (down) indices provides the upper (down) limits of the $95$\% CI.  }
 \label{tab:lecsgl}
 \end{table}

\subsubsection*{Strange-quark dependence of the pion mass and decay constant}

\begin{figure}
 \begin{center}
  \includegraphics[scale=0.3]{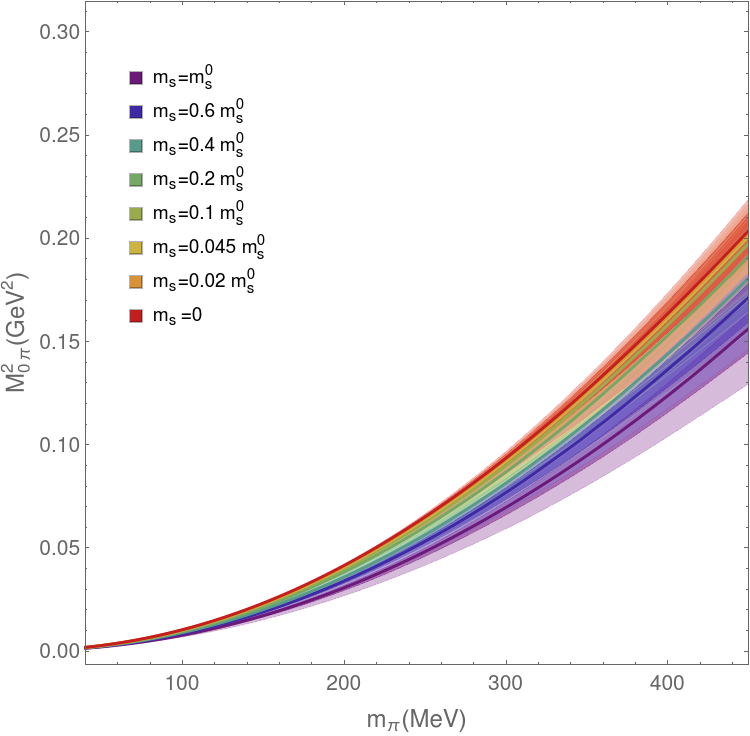}
 \end{center}
\caption{The squared leading order mass, $M^2_{0\pi}$, of Eqs.~(\ref{eq:ml}) and~(\ref{eq:pimass}),  as a function of $m_\pi$ for different values of the strange-quark mass, $m_s$. }
\label{fig:mpiza}
\end{figure}

\begin{figure}
  \hspace{-0.5cm}\includegraphics[scale=0.3]{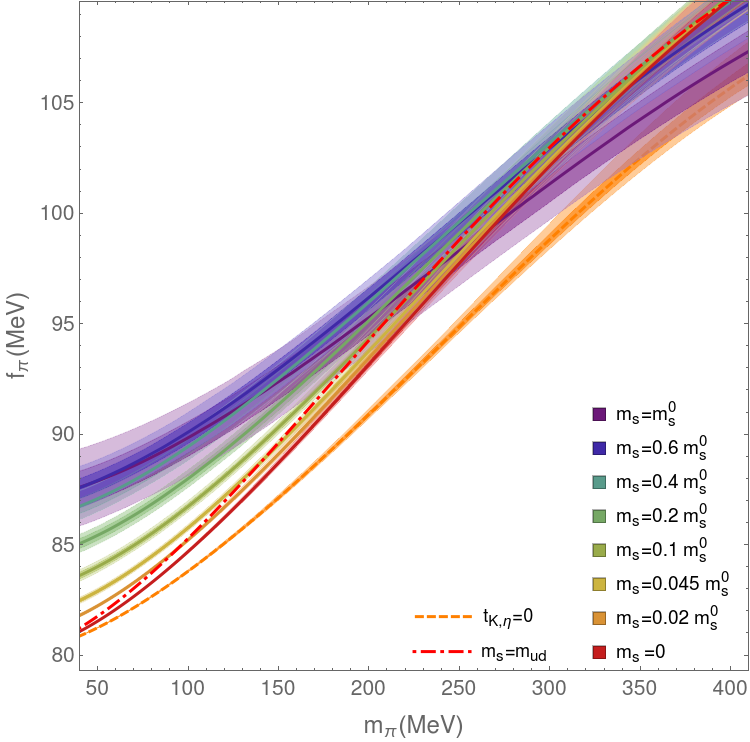}
 \caption{The pion decay constant, $f_\pi$, see Eq.~(\ref{eq:fpis}), as a function of $m_\pi$ for different strange-quark masses, $m_s$. The continuous and dot-dashed red lines represent the solution for the pion decay constant from Fit IV at different values of $m_s$ and in the symmetric line, while the orange-dashed curve is the solution when $t_{K,\eta}=0$.}
\label{fig:fpims}
\end{figure}

From Fig.~\ref{fig:mpifpi} one sees that the ratio $m_\pi/f_\pi$ does not depend much on $m_s$. However, this does not necessarily mean that $f_\pi$ is independent of the strange-quark mass. In fact, both $m_\pi$ and $f_\pi$ depend on $m_s$. This dependence is shown explicitly in Figs.~\ref{fig:mpiza} and~\ref{fig:fpims}. In Fig.~\ref{fig:mpiza}, the squared leading order mass, $M^2_{0\pi}$, is depicted as a function of the pion mass for different strange-quark masses. Indeed, one can see that $m_{ud}$ kept constant does not imply that the pion mass is constant as well. In fact, our analysis at one-loop level predicts that the pion mass grows with $m_s$ for a constant value of $m_{ud}$; while for $m_s=0$ one obtains $m_K \simeq 1/ \sqrt 2 m_\pi$ (see the red line in Fig.~\ref{fig:mkmpi}) and $M_{0\pi} \simeq m_\pi$ (notice the almost quadratic behavior of the red curve in Fig.~\ref{fig:mpiza}), consistently with the leading order ChPT prediction, effects coming from the kaon and eta particles in $f_\pi$ become more relevant as $m_s$ increases. Although this effect is invisible for very light pion masses and small for physical pion masses (for a constant value of $m_{ud}$, the difference between the physical pion mass and the $m_\pi$ value at $m_s=0$ in Fig.~\ref{fig:mpiza} is around $14\%$.), it becomes larger for heavy pions. On the contrary, in Fig.~\ref{fig:fpims}, where the dependence of $f_\pi$ on $m_\pi$ for different strange-quark masses is shown, one sees that this dependence is more noticeable for light pion masses and smaller for heavier pion masses, when the uncertainties for $f_\pi$ increase. At the physical point, one indeed finds a difference of $5$~MeV in $f_\pi$ between its value at $m_s=0$ and $m_s=m_s^0$. This is intrinsically connected with the contribution of the terms which involve kaons and etas in Eqs.~\eqref{eq:pimass} and~\eqref{eq:fpis},
\begin{align}
m_\pi^2=&M_\pi^2+\Delta_{K,\eta} m_\pi^2\nonumber\\
f_\pi=&F_\pi+\Delta_{K,\eta}{f_\pi},\label{eq:fpibis}
\end{align}
with
\begin{eqnarray}
&M_\pi^{2}= M_{0\,\pi}^2\left[1+\mu_\pi+\frac{8 M_{0\,\pi}^2}{f_0^2}\left(2L_6^r+2L_8^r-L_4^r-L_5^r\right)\right]\,,\nonumber\\\label{eq:mpip}\\
&F_\pi= f_0\left[1-2\mu_\pi+\frac{4 M_{0\,\pi}^2}{f_0^2}\left(L_4^r+L_5^r\right)\right],\label{eq:fpip}
\end{eqnarray}
which are 
\begin{align}
\Delta_{K,\eta}m^2_\pi=&M_{0\,\pi}^2\left[-\frac{\mu_\eta}{3}+\frac{16 M_{0\,K}^2}{f_0^2}\left(2L_6^r-L_4^r\right)\right],\\ 
\Delta_{K,\eta}f_\pi=&f_0\left[ -\mu_K+\frac{8 M_{0\,K}^2}{f_0^2}L_4^r\right].
\end{align}
These terms, which involve kaon and eta meson loops (tadpoles) and kaon mass contact terms, called $t_{K,\eta}$ from now on, contribute slightly for $m_s=0$, around $1$~MeV in the $f_\pi$ value, but they account for about $6-7$ MeV at $m_s=m_s^0$. Later, we show that this relatively small variation in $f_\pi$ is translated into a visible difference in the $\rho$ mass.

Beyond that, one might wonder what would happen in a world where kaons and etas are not present. In order to answer that question, we could explicitly set to zero the  $t_{K,\eta}$ contribution in Eqs.~\eqref{eq:pimass} and \eqref{eq:fpis}, and define $M_\pi$ and $F_\pi$ in Eqs.~\eqref{eq:mpip} and ~\eqref{eq:fpip} as the pion mass and decay constant in this world.
In such a way, one obtains a dependence of the new pion decay constant, $F_\pi$, with the pion mass, as the orange-dashed line in Fig.~\ref{fig:fpims}. 
Effectively, this limit is equivalent to set the coupling of pions to kaons and etas to zero, which can be achieved when, first, $f_{K,\eta}$ in Eqs.~\eqref{eq:fk} and~\eqref{eq:feta} are sent to infinity,\footnote{Note that it can only achieved if one breaks the SU(3) symmetry in the partially conserved axial current (PCAC), i.e., one has to differentiate between the pion and the kaon and eta decay constants in the chiral limit. See explanation in Sect.~\ref{sec:nf2}.} and, second, when $m_{K,\eta}$ are set to zero. This is discussed in Sect.~\ref{sec:nf2}.
 Furthermore, note  that this is different from the so-called SU(2) formalism (corresponding to take the limit $m_s\to\infty$), where the effect of pions interacting with kaons and etas is absorbed in the SU(2) LECs, the constant $B_0$ and $f_0$, which are fixed, together with $f_\pi$, to get observables in the physical world with more than two flavors.

When $t_{K,\eta}=0$, $f_\pi$ reduces its value around $6-7$ MeV at the physical point, while it approaches the value of the chiral trajectories $m_s=0$ and $m_s=m_{ud}$ as the pion mass decreases, reaching $f_0$ in the chiral limit. These interacting terms involving kaons and etas have a similar effect to the one caused by a change in $m_s$ for light pion masses. In fact, varying $m_s$ from zero rises the contribution of these terms, increasing the value of $f_\pi$.

We can also compare the ratios obtained here with the ones of the $N_f=2$ simulation of~\cite{Guo:2016zos}. For the pion masses used in that simulation, $m_\pi=225$ and $315$ MeV, the ratios $\{m_K/m_\pi,m_\pi/f_\pi,m_K/f_K\}$ are $\{2.25,2.31,4.508\}$ and $\{1.67,2.98,4.507\}$,\footnote{Where we have divided $f_K$ from~\cite{Guo:2016zos} by a factor $\sqrt{2}$ according to the normalization of $f_K$ used here.} for the light and heavy pion mass, respectively. Looking at Figs.~\ref{fig:mkmpi},~\ref{fig:mpifpi} and~\ref{fig:mkfk}, we see that the deviations from the mean values in Fit IV at the $m_s=m_s^0$ trajectory are of less than $\{1\%, 1\%, 2\%\}$ for the light pion mass, and around  $\{2\%,3\%,4\%\}$ for the heavy pion mass. These relative differences are small and the ratios in Table III of~\cite{Guo:2016zos}  are compatible with our error bands. This indicates that the setup of the simulation of~\cite{Guo:2016zos} is in line with the result of this analysis for the $m_s=m_s^0$ chiral trajectory. Thus, possible deviations in the $\rho$-meson parameters with the ones obtained here might be caused by a different reason. Notice also that the method used to determine these ratios in~\cite{Guo:2016zos} is to take $m_K/f_K$ to the physical point in a strange-quark quenched approximation. As a consequence, the values of $f_\pi$ obtained are consistent with the ones from Fit IV at $m_s=m_s^0$ and its extrapolation to the physical value. However, since the real world has more than two flavors, that approach misses the $m_s$ dependence of $f_\pi$ as discussed before.  

\subsection{Results for the $\rho$-meson phase shift and pole parameters}

\begin{figure*}
\begin{tabular}{cc}
 \includegraphics[scale=0.45]{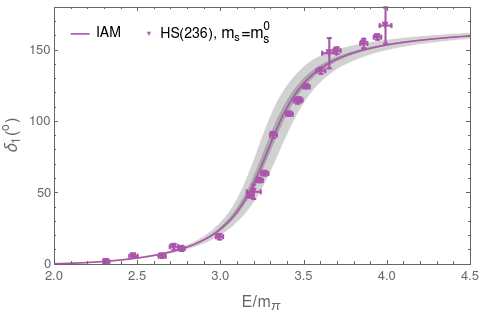} &\includegraphics[scale=0.45]{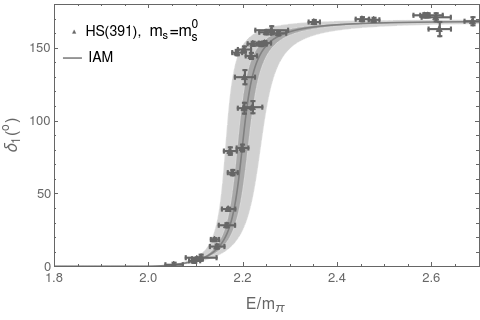}\\
 \end{tabular}
\caption{Result of Fit IV in comparison with the HS data at $m_\pi=236$ and $391$ MeV~\cite{Wilson:2015dqa,Dudek:2012xn}}
\label{fig:msphase}
\end{figure*}

\begin{figure*}
\begin{tabular}{cc}
 \includegraphics[scale=0.45]{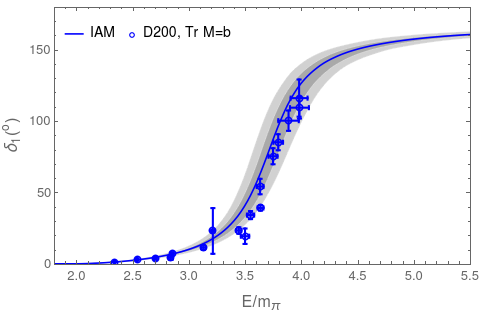} &\includegraphics[scale=0.45]{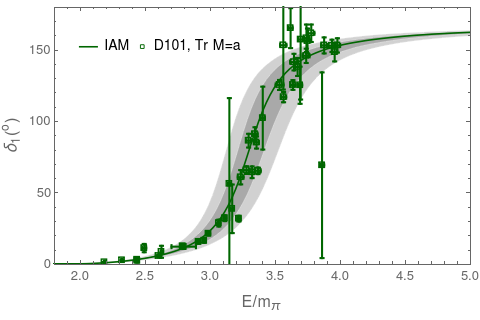}\\
  \includegraphics[scale=0.45]{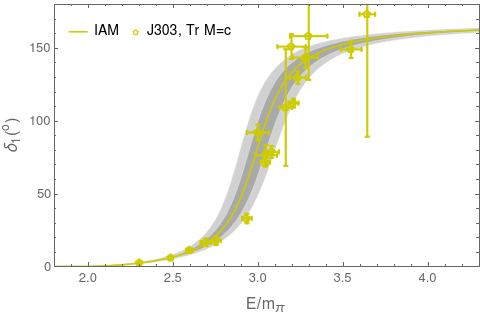}&\includegraphics[scale=0.45]{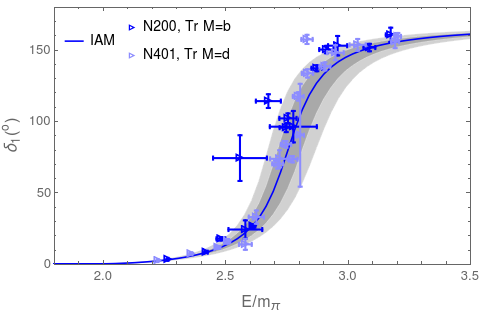}\\
 \end{tabular}
\caption{Result of Fit IV in comparison with the $\text{Tr}{\cal M}=C$ data of the CLS ensembles, D200, D101, J303, N200 and N401~\cite{Andersen:2018mau}.}
\label{fig:trmphase}
\end{figure*}

\begin{figure}
  \includegraphics[scale=0.5]{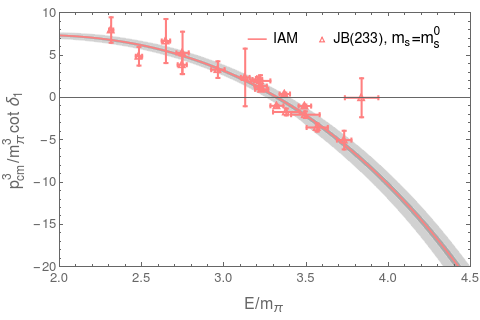}
\caption{Result of Fit IV in comparison with the JB data at $m_\pi=233$ MeV~\cite{Bulava:2016mks}.}
\label{fig:bul233}
\end{figure}

Phase shift lattice data and solutions from Fit IV at the corresponding pion masses are depicted in Figs.~\ref{fig:msphase},~\ref{fig:trmphase}, and~\ref{fig:bul233}, where one can see that lattice data are very well described also inside the $95$ \% CI. The only exceptions are few data points from the CLS data for the N200 and N401 ensembles (right-bottom panel in Fig.~\ref{fig:trmphase}), which lie outside the error band and are also far from the bulk of data. Beyond that, most of data for these ensembles are inside of the 95\% CI error bands. This is, the N200 and N401 ensembles are compatible within uncertainties, as concluded also in~\cite{Andersen:2018mau}.
Note that the error bands are larger for the $\text{Tr}{\cal M}=C$ data since they include the variation of the lattice spacing $a_\beta$ between {\it A} and {\it B scale settings}.

\begin{figure}
\includegraphics[scale=0.45]{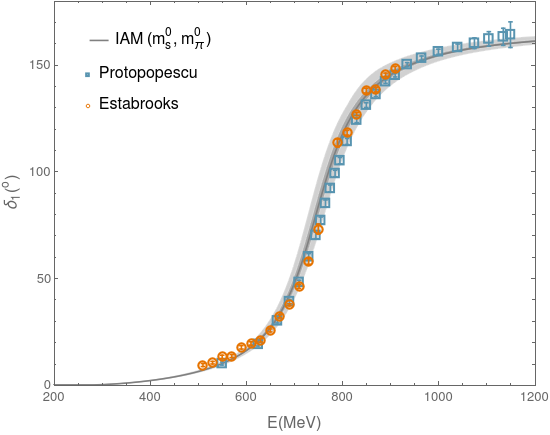}\\
\caption{Extrapolation to the physical point of the Fit IV solution in comparison with the experimental data.}
\label{fig:exp}
\end{figure}

\begin{figure}
 \includegraphics[scale=0.4]{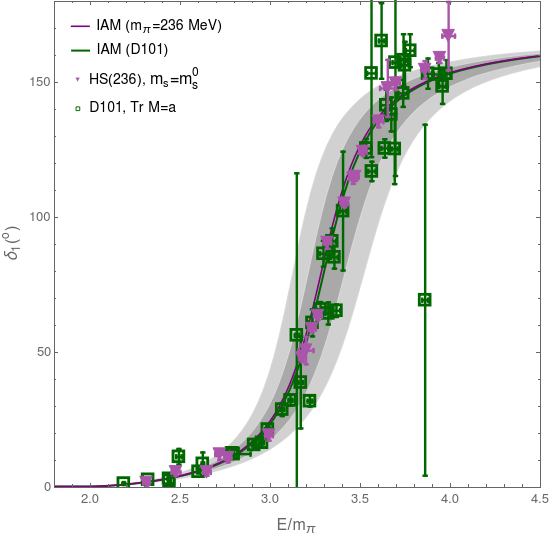}
\caption{Phase shift lattice data corresponding to the D101 ensemble in comparison with the HadSpec data for $m_\pi=236$ MeV and the IAM solution for D101.}
\label{fig:hsbul}
\end{figure}

The extrapolation to the physical point in comparison with experimental data is plotted in Fig.~\ref{fig:exp}, where one can see that it indeed provides an excellent description of the experiment.
In Fig.~\ref{fig:hsbul}, the CLS D101 ensemble and HS data for $m_\pi=236$ MeV are plotted together.
We can see that indeed both results are compatible.

\begin{figure}
\includegraphics[scale=0.4]{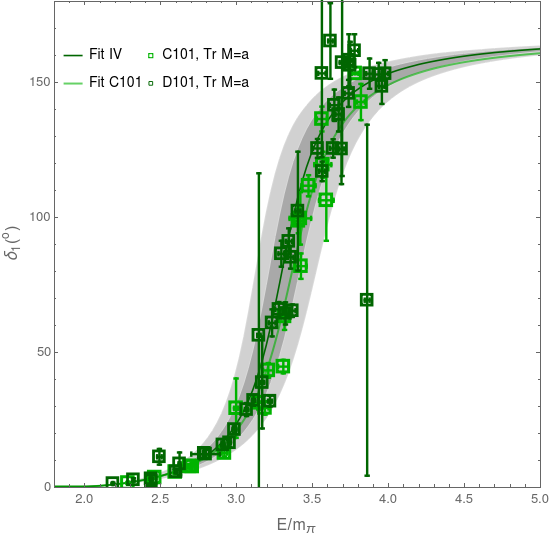}\\
\caption{Comparison between the IAM solutions in Fit IV for D101, with error bands, and the individual fit to the C101 data using the averaged lattice spacing.}
\label{fig:compacd}
\end{figure}

The phase-shift solution for the pion mass used in the D101 ensemble (dark green) is shown  in comparison with the result from an individual fit of the C101 data (light green) in Fig.~\ref{fig:compacd}.\footnote{The average between the lattice spacings in {\it scale settings A} and {\it B} is taken in this individual fit of the C101 data in order to compare with the solution from Fit IV.} 
Note that even when the C101 ensemble was not included in the global Fit IV, both solutions, and the bulk of data itself, lie well inside the 95\% CI.
The difference between both fits for the energy at which the phase shift crosses $90^\circ$ ($E\,(\delta=90^o)/m_\pi$) is of around $3\%$. This indicates that the deviations due to the volume size are not large.


\begin{figure*}
 \begin{center}
  \includegraphics[scale=0.37]{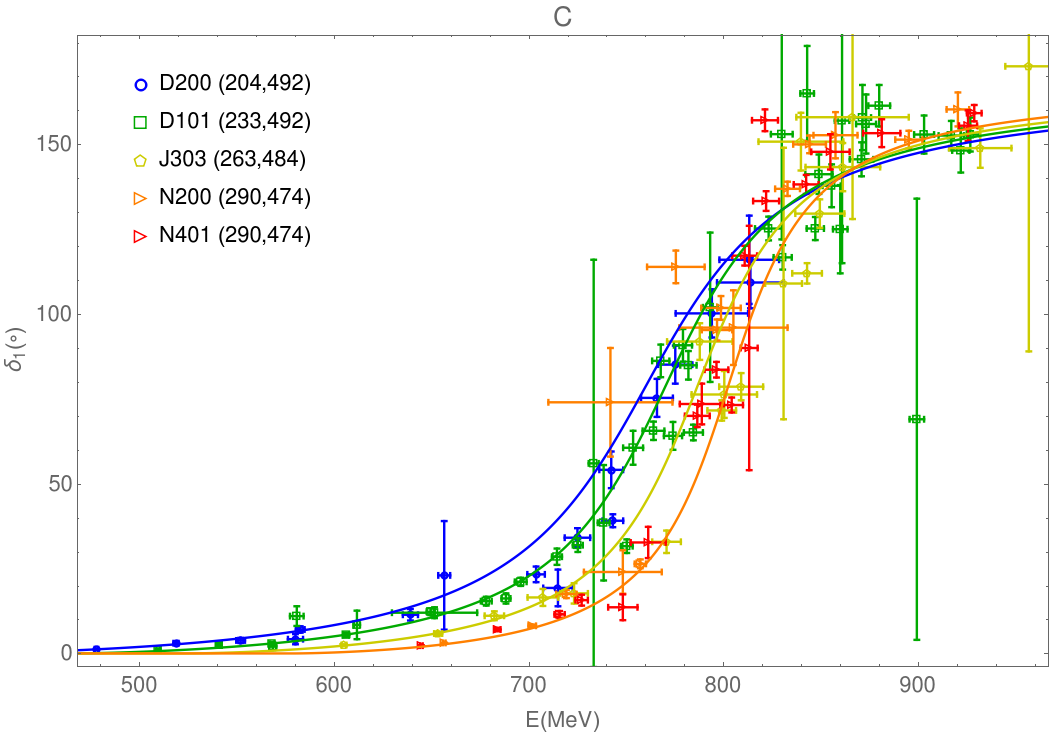}
 \end{center}
\caption{Phase shift lattice data and global Fit IV solution (only the solution for the average lattice spacing is shown).}
\label{fig:glphaseall}
\end{figure*}

\begin{figure*}
\begin{tabular}{cc}
\includegraphics[width=0.473\textwidth]{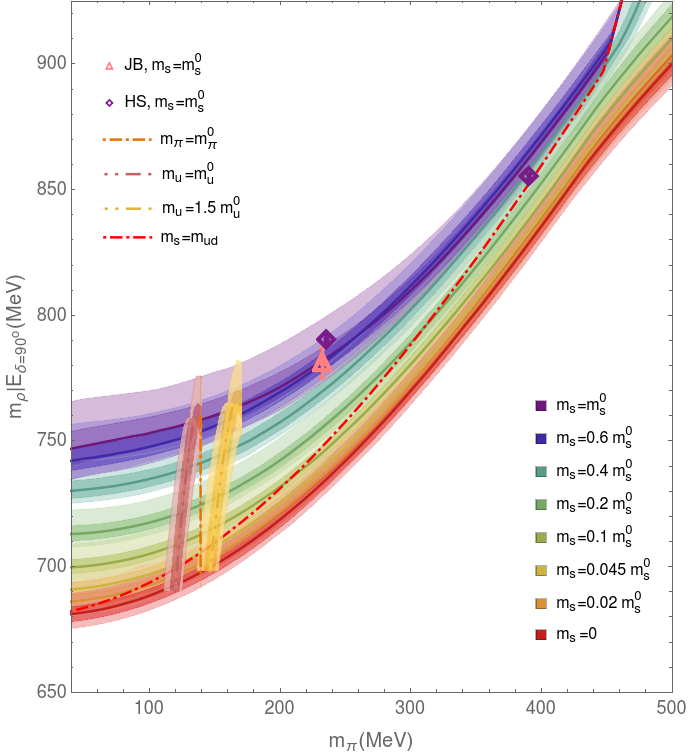} &\includegraphics[width=0.45\textwidth]{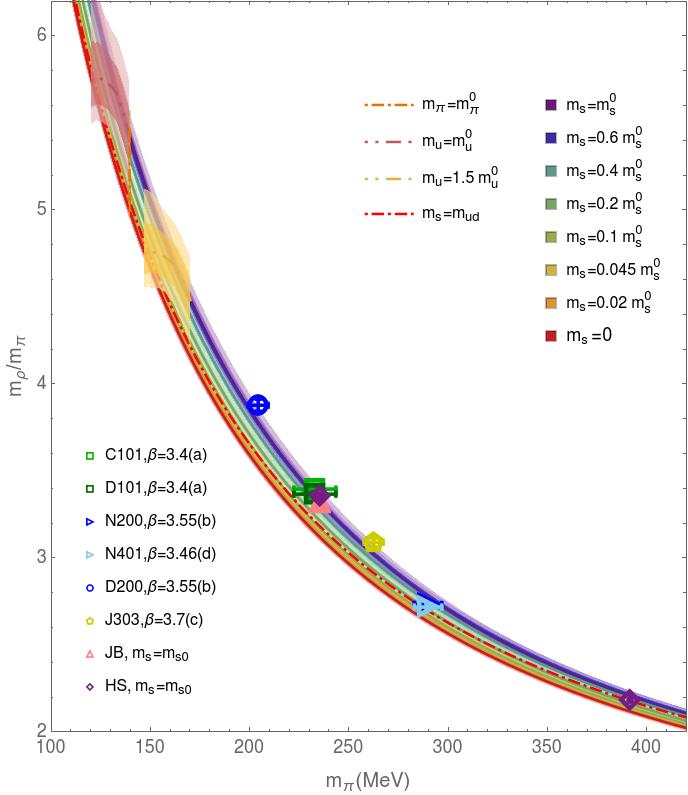}\\
 \end{tabular}
\caption{The $\rho$-meson mass (left) and normalized $\rho$ mass (respect to the pion mass) (right) as a function of the pion mass for the $m_s=k$, $m_u=c$, $m_\pi=m_\pi^0$, and $m_s=m_u$ trajectories.}
\label{fig:mrhomsgl}
\end{figure*}

\begin{figure*}
\begin{tabular}{cc}
\includegraphics[width=0.463\textwidth]{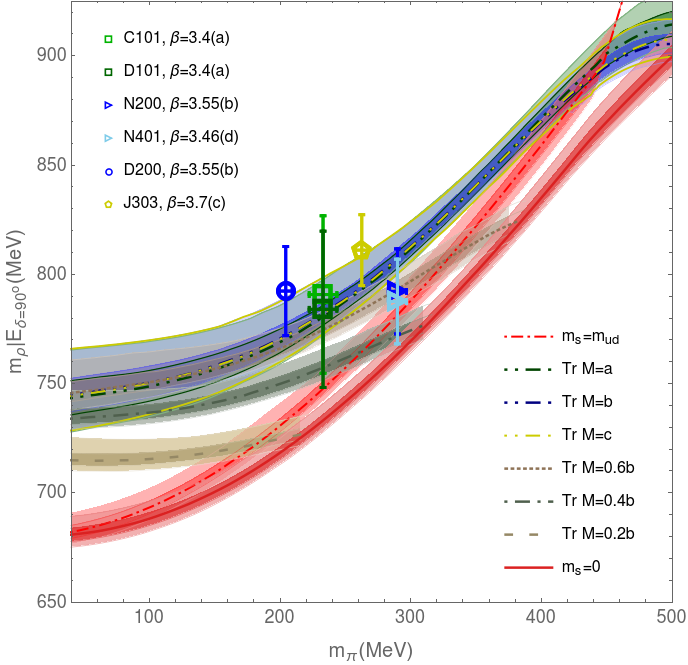} &\includegraphics[width=0.45\textwidth]{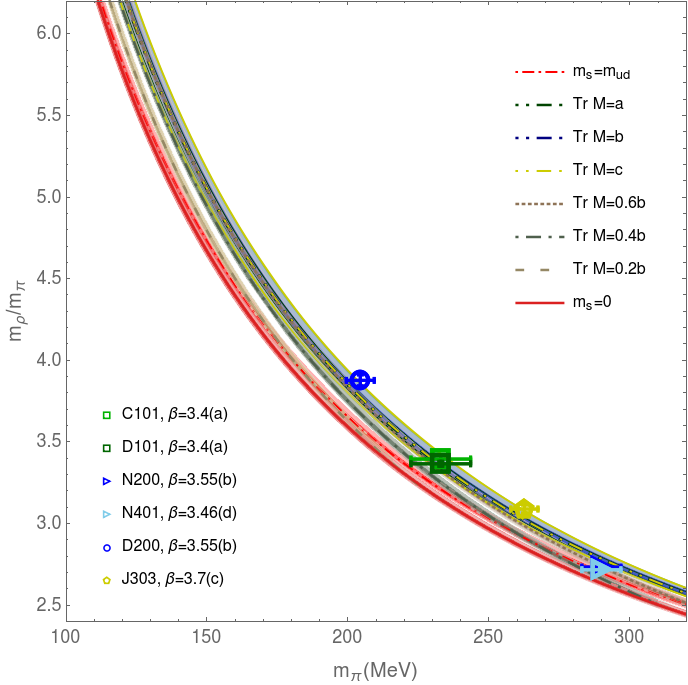}\\
 \end{tabular}
\caption{The $\rho$-meson mass (left) and normalized $\rho$ mass (respect to the pion mass) (right) as a function of the pion mass for the $\text{Tr}{\cal M}=C$, $m_s=0$ and $m_s=m_u$ trajectories.}
\label{fig:mrhotrmgl}
\end{figure*}

\begin{figure*}
\begin{tabular}{cc}
\includegraphics[width=0.47\textwidth]{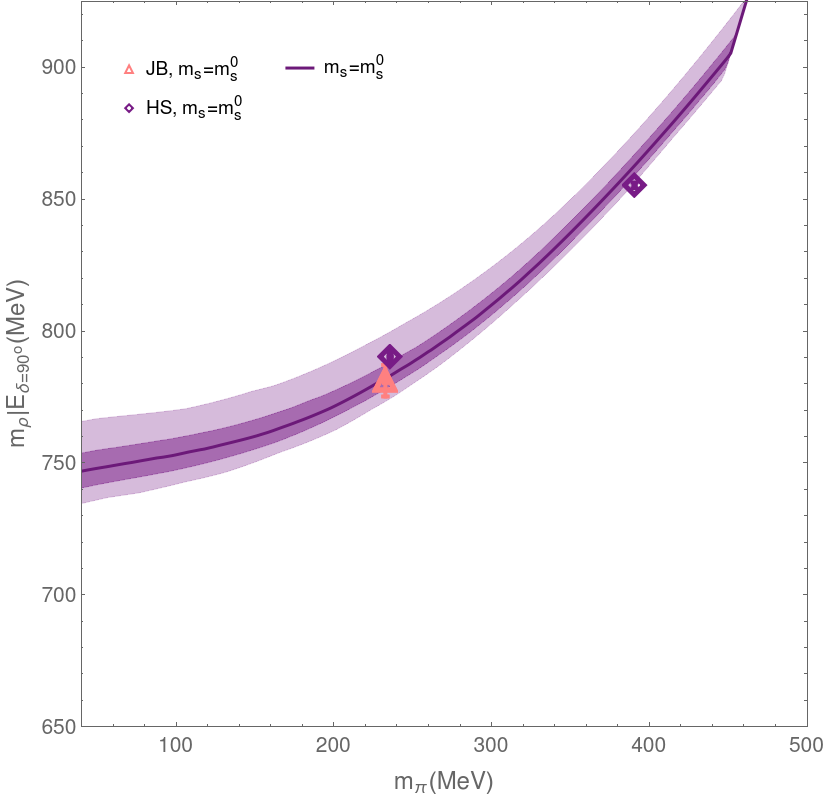} &\includegraphics[width=0.47\textwidth]{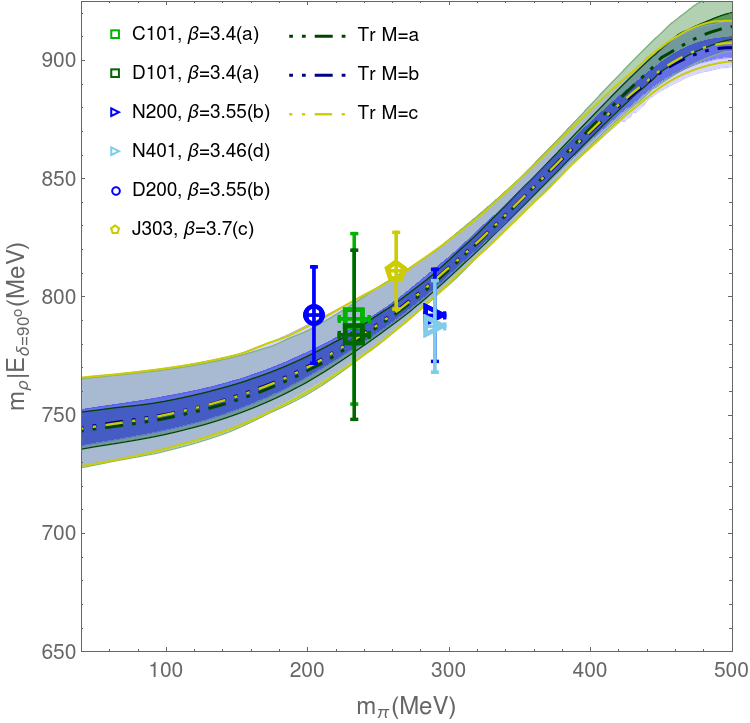}\\
 \end{tabular}
\caption{The $\rho$-meson mass as a function of the pion mass over the $m_s=m_s^0$ (left) and $\mathrm{Tr}{\cal M}=\mathrm{Tr}{\cal M}^0$ (right) trajectories in comparison with the lattice data.}
\label{fig:mrhomstr}
\end{figure*}

 To show the trend of the $\text{Tr}{\cal M}=C$ data, the mean solution of Fit IV together with the lattice data are represented in Fig.~\ref{fig:glphaseall}.\footnote{This correspond to using the averaged lattice spacing values as the data plotted in the figure.} As can be seen, these data are now well described. Phase shift data corresponding to higher pion masses fall more to the right and the $\rho$-meson mass increases monotonically with the pion mass.
 The dependence of the $\rho$-meson mass\footnote{Defined as the value of the energy for which $\delta=90^o$.} with the pion mass is depicted in Figs.~\ref{fig:mrhomsgl} and~\ref{fig:mrhotrmgl} for the $m_s=k$ and $\text{Tr}{\cal M}=C$ trajectories, respectively, where we also show the values of the $\rho$-meson mass given in the corresponding lattice papers.\footnote{For the $\text{Tr}{\cal M}=C$ trajectories, the error due to the use of {\it scale settings A} and {\it B} is also depicted.} The $m_\rho/m_\pi$ ratios are also represented in the right panels.
For clearness, we present again the lattice data and the resulting curves separately in Fig.~\ref{fig:mrhomstr}. 
In both trajectories $m_\rho$ increases with $m_\pi$. Furthermore, for the trajectories $m_s=m_s^0$ and $\text{Tr}{\cal M}=\text{Tr}{\cal M}^0$, we find almost identical results till pion masses of around $400$ MeV, when these start to separate.
The reason for this behavior is well understood. On one side, the $\rho(770)$ meson becomes a bound state at pion masses around $m_\pi=450$ MeV in the $m_s=m_s^0$ trajectory (above this value, the $\pi\pi$ threshold is plotted in Figs.~\ref{fig:mrhomsgl} and~\ref{fig:mrhomstr} (left) instead). On the other side, it starts to decay into $K\bar K$ in the $\text{Tr}{\cal M}=\text{Tr}{\cal M}^0$ trajectory when the $\rho$-meson pole crosses this threshold and the kaon gets lighter than the pion. Indeed, it becomes a pole in the IV Riemann sheet as defined in Eqs.~\eqref{eq:IIRS2}~and~\eqref{eq:sigmaRS}.
 Conversely, other $\text{Tr}{\cal M}=C$ trajectories tend to be flatter than the $m_s=k$ ones. This is actually in line with the  trend of lattice data. 

\begin{figure*}
\begin{tabular}{cc}
\hspace{-0.5cm} \includegraphics[scale=0.45]{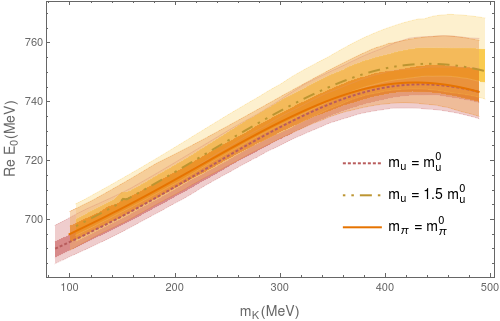} &\includegraphics[scale=0.45]{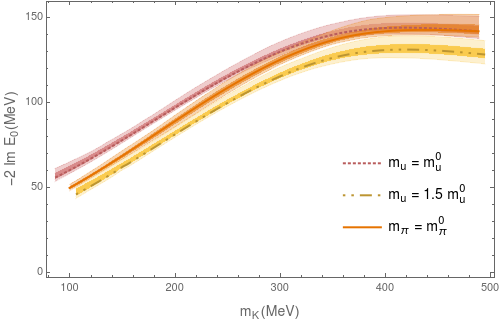}\\
 \end{tabular}
\caption{The real and imaginary parts of the $\rho$-meson pole position as a function of $m_K$ over the $m_u=c$ and $m_\pi=m_\pi^0$ trajectories.}
\label{fig:mrhomupi}
\end{figure*}

\begin{figure*}
\begin{tabular}{cc}
\hspace{-0.5cm} \includegraphics[scale=0.365]{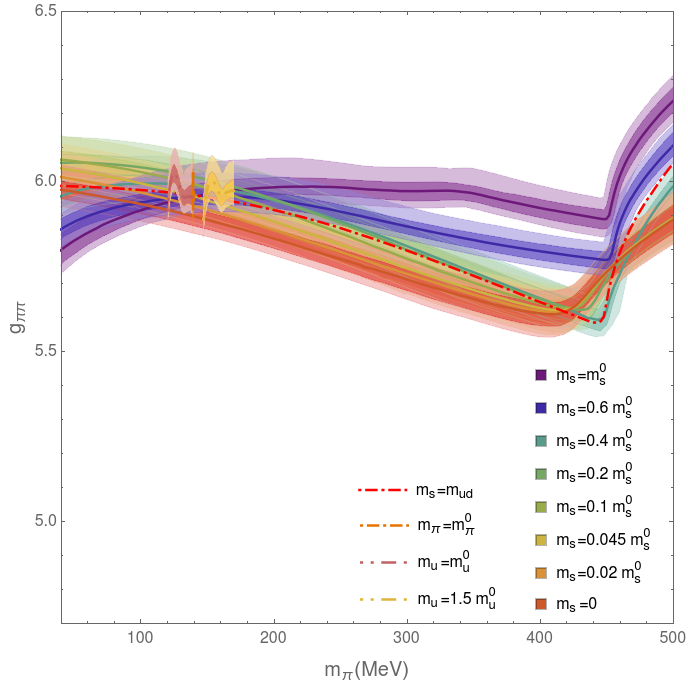} &\includegraphics[scale=0.365]{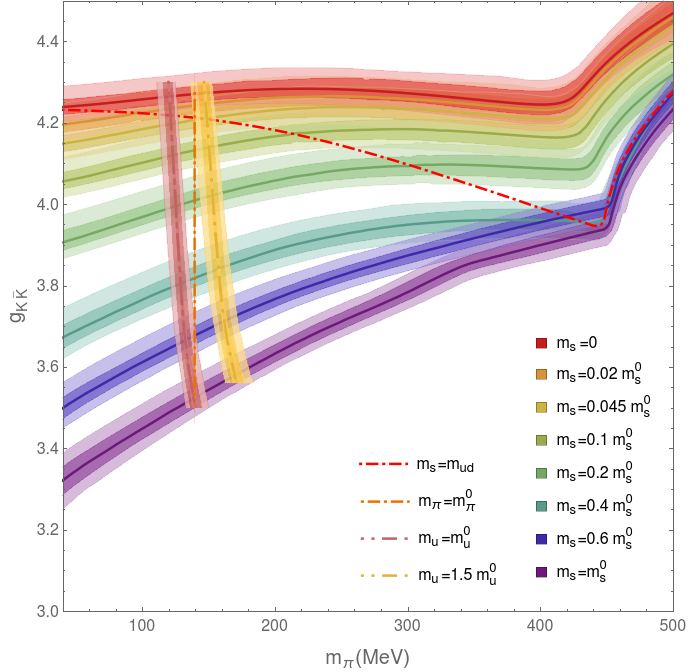}\\
 \end{tabular}
\caption{The couplings of the $\rho$-meson pole to the $\pi\pi$ and $K\bar{K}$ channels, $g_{\pi\pi}$ and $g_{K\bar{K}}$, for the $m_s=k$, $m_u=c$, $m_\pi=m_\pi^0$ and $m_s=m_u$ trajectories. }
\label{fig:mrhomsggl}
\end{figure*}

\begin{figure*}
\begin{tabular}{cc}
\hspace{-0.5cm} \includegraphics[scale=0.365]{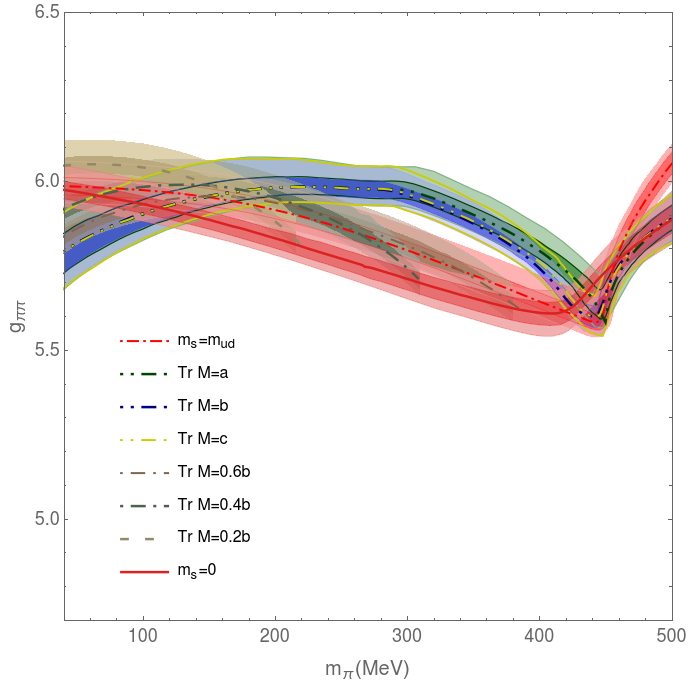} &\includegraphics[scale=0.365]{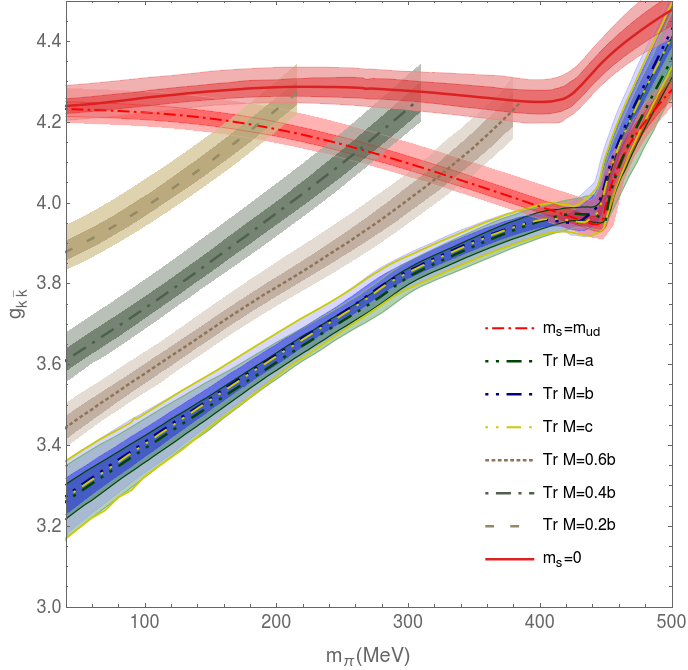}\\
 \end{tabular}
\caption{The couplings of the $\rho$-meson pole to the $\pi\pi$ and $K\bar{K}$ channels, $g_{\pi\pi}$ and $g_{K\bar{K}}$, for the $\text{Tr}{\cal M}=C$ trajectories in comparison with the $m_s=0$ and $m_s=m_u$ ones.}
\label{fig:mrhotrmggl}
\end{figure*}

It is relevant to note that close to the physical point we do not observe any relevant change in the $\rho$-meson mass. This suggests that the $\rho$-meson properties are quite stable against small variations of the strangeness around the physical point, however, its coupling to the $K\bar{K}$ channel is still large, around $60\%$ of its coupling to $\pi\pi$. \footnote{This can be seen in Figs.~\ref{fig:mrhomsggl} and~\ref{fig:mrhotrmggl}, as discussed later in this section. }.  Nevertheless, for $m_s$ values below $0.5\,m_s^0$, the $\rho$-meson mass starts decreasing considerably reaching a value inside the interval $[675,\,695]$ MeV for $m_s=0$. This behavior is even more clear in the $m_{u,\pi}=c$ trajectories where the mass of the $u$ quark (or pion) is kept fixed and only $m_s$ varies. Since the $\rho$ meson starts to decay into $K\bar{K}$ for lighter strange quarks, the effect in the real part of the pole, see Fig.~\ref{fig:mrhomupi}, becomes significant.
 
\begin{figure*}
\includegraphics[scale=0.37]{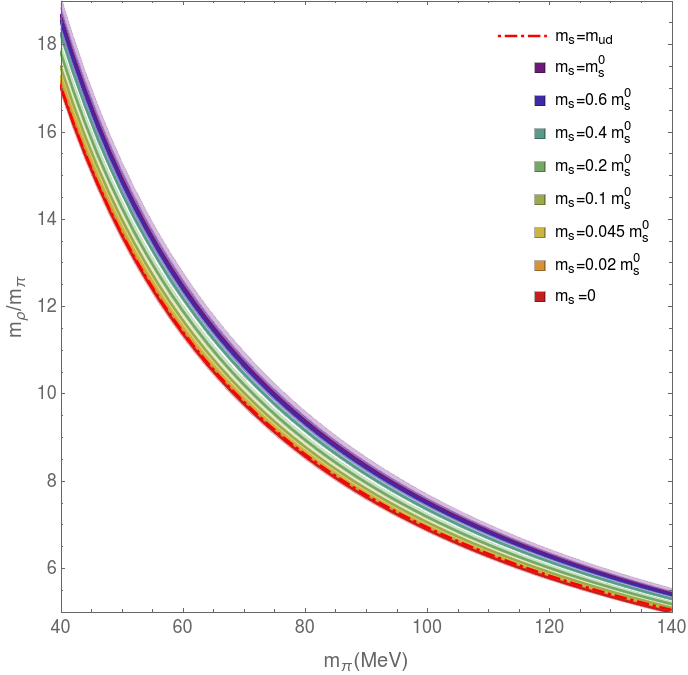}
\caption{$m_\rho/m_\pi$ ratio as a function of $m_\pi$ for the $m_s=k$ and $m_s=m_{ud}$ trajectories continued towards lighter pion masses.}
\label{fig:mrhoms140}
\end{figure*}

\begin{figure*}
\hspace{-0.4cm}\includegraphics[scale=0.4]{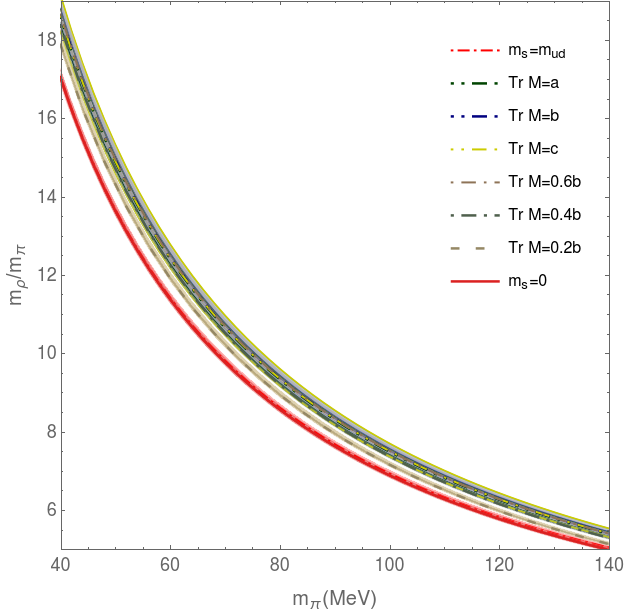}
\caption{$m_\rho/m_\pi$ ratio as a function of $m_\pi$ for the $\text{Tr}{\cal M}=C$, $m_s=m_{ud}$ and $m_s=0$ trajectories continued towards lighter pion masses}
\label{fig:mrhotrcm140}
\end{figure*}

This behavior of the $\rho$-meson mass is also visible in the corresponding $m_\rho/m_\pi$ plots (right panel of Figs.~\ref{fig:mrhomsgl} and~\ref{fig:mrhotrmgl}), where the errors in the y-axis are reduced. These plots also show that the error due to the lattice spacing (or scaling setting used) is smaller than the reduction of the $\rho$-meson mass around the $m_s=0$ limit.
The behavior of the $\rho$-meson mass and width respect to the kaon mass is depicted in Fig.~\ref{fig:mrhomupi} for the $m_u=c$ and $m_\pi=m_\pi^0$ trajectories.
When $m_K$ decreases, both mass and width decrease, as commented before. In Figs.~\ref{fig:mrhoms140} and~\ref{fig:mrhotrcm140}, the continuation of the $m_\rho/m_\pi$ ratios for smaller pion masses are also depicted. This difference with respect to the physical point is more abrupt as the quark masses get smaller.
The symmetric line is also plotted in dot-dashed lines.


\begin{figure*}
\hspace{-0.5cm}\includegraphics[scale=0.4]{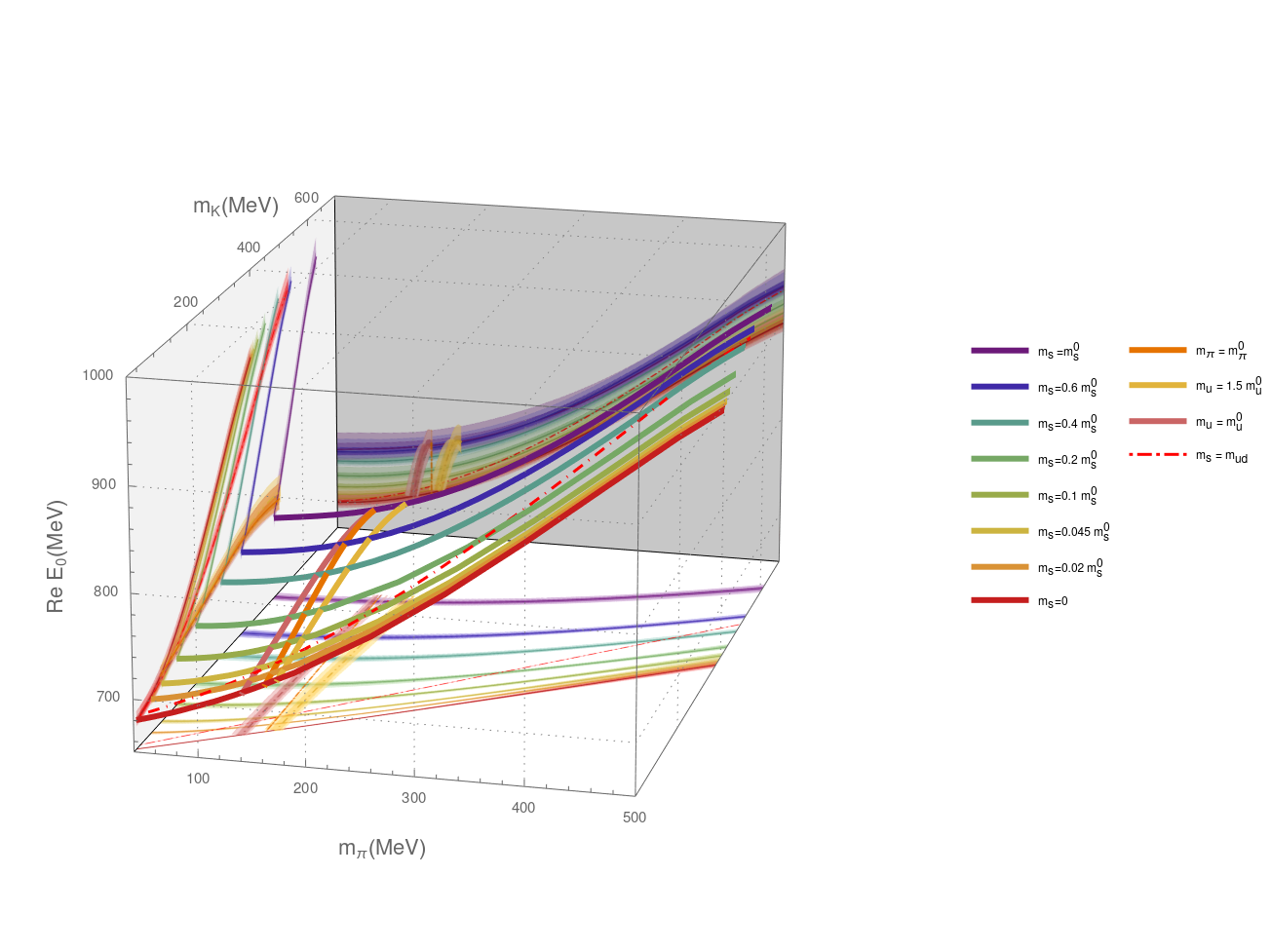}\\
\caption{The real part of the $\rho$-meson pole position, $\mathrm{Re}E_0$, as a function of $m_\pi$ and $m_K$ for the $m_s=k$ trajectories, $m_u=c$, $m_\pi=m_\pi^0$, and $m_s=m_{ud}$.}
\label{fig:3dresms}
\end{figure*}

\begin{figure*}
\hspace{-0.5cm}\includegraphics[scale=0.4]{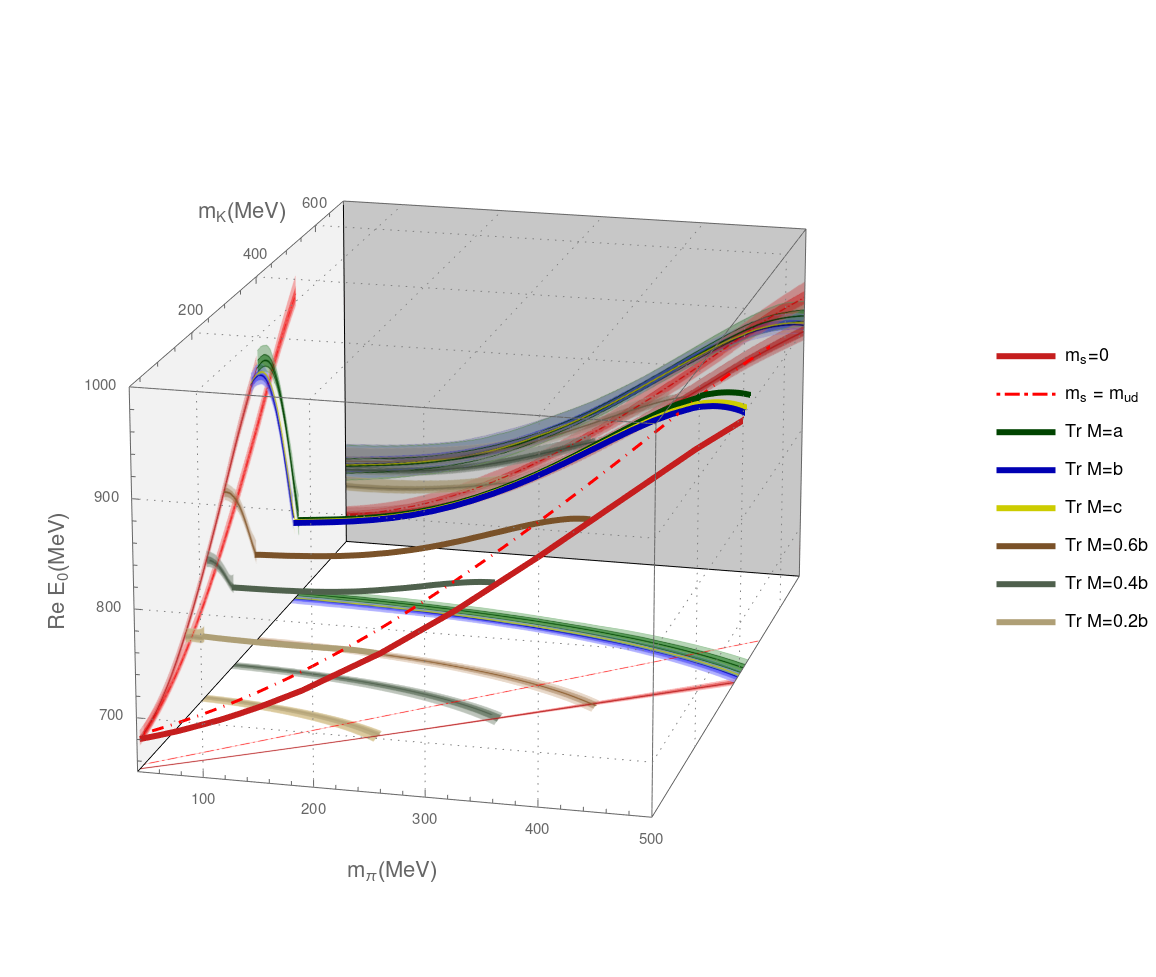}\\
\caption{The real part of the $\rho$-meson pole position, $\mathrm{Re}E_0$, as a function of $m_\pi$ and $m_K$.}
\label{fig:3drestrm}
\end{figure*}

\begin{figure*}
\hspace{-0.5cm}\includegraphics[scale=0.35]{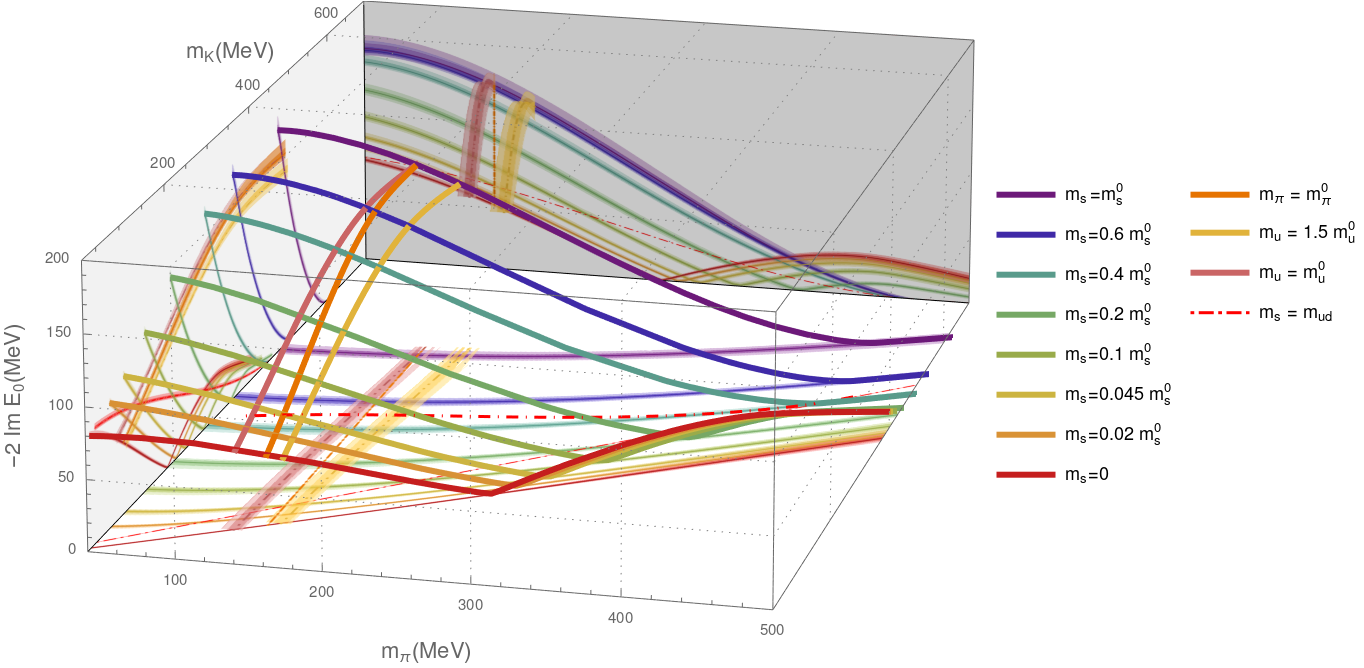}\\
\caption{The width of the $\rho$-meson pole position, $-2\mathrm{Im}E_0$, as a function of $m_\pi$ and $m_K$ for the $m_s=k$, $m_u=c$, $m_\pi=m_\pi^0$ and $m_s=m_{ud}$ trajectories.}
\label{fig:3dimsms}
\end{figure*}

\begin{figure*}
\hspace{-0.5cm}\includegraphics[scale=0.35]{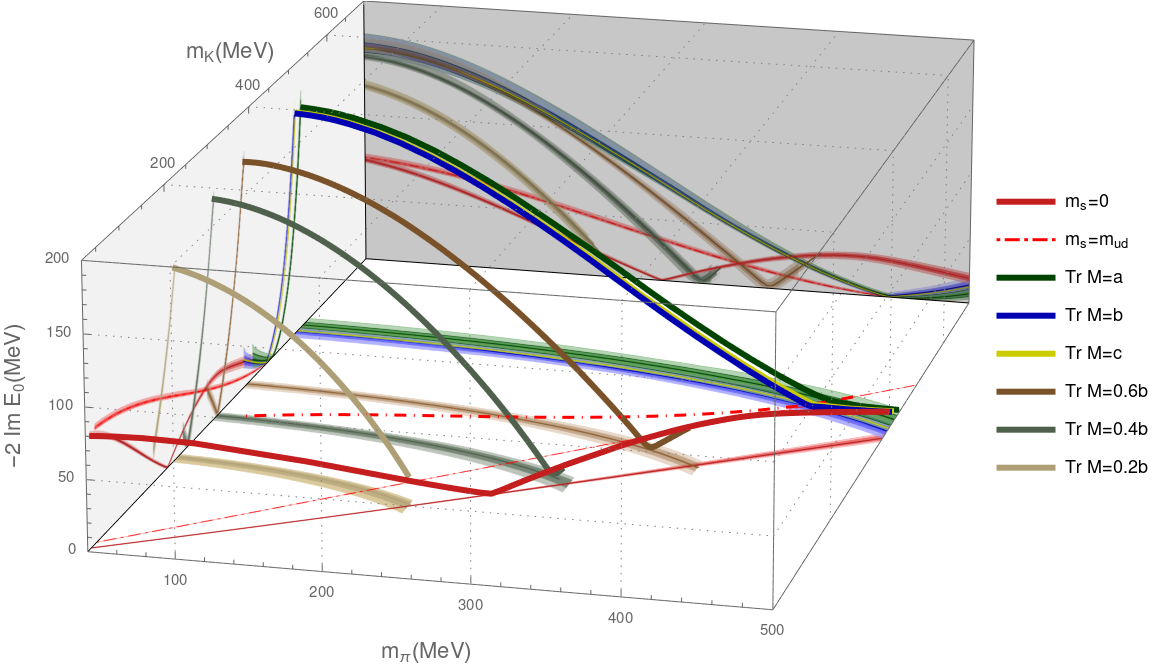}\\
\caption{The width of the $\rho$-meson pole position, $-2\mathrm{Im}E_0$, as a function of $m_\pi$ and $m_K$ for $Tr{\mathcal M}=K$ in comparison with the $m_s=0$ and $m_s=m_{ud}$ trajectories.}
\label{fig:3dimstrm}
\end{figure*}

In Figs.~\ref{fig:3dresms},~\ref{fig:3drestrm},~\ref{fig:3dimsms} and~\ref{fig:3dimstrm} we provide the real and imaginary parts of $\rho$-meson pole position in a 3D plot respect  both, the pion and kaon mass.
To render some references, we also give in Table~\ref{tab:poles} the pole positions at $m_s=\{0,0.6,1\} m_s^0$ for pion masses near the chiral limit, physical point and when the $\rho$ gets bound ($m_\pi\sim 450$ MeV in the last two trajectories).
\begin{table}
\begin{center}
{\renewcommand{\arraystretch}{2}
\setlength\tabcolsep{0.15cm}
 \begin{tabular}{cccc}
 \toprule
 $m_s/m_s^0$&$m_\pi$ &$\mathrm{Re}E_0$&$\Gamma$\\
 \hline
 $0$&$\sim 0$&$678^{+3(5)}_{-3(3)}$ &$77^{+1(2)}_{-1(2)}$\\
    &$140$~MeV&$695^{+3(4)}_{-3(3)}$ &$48^{+1(3)}_{-1(2)}$\\
    &$450$~MeV&$867^{+3(6)}_{-2(3)}$ &$27^{+2(1)}_{-2(1)}$\\\hline
 $0.6$&$\sim 0$&$732^{+5(9)}_{-3(5)}$ &$162^{+3(5)}_{-2(2)}$\\
    &$140$~MeV&$744^{+4(6)}_{-4(2)}$ &$140^{+3(4)}_{-2(2)}$\\
    &$450$~MeV&$905^{+4(6)}_{-4(2)}$ &$0$\\
    \hline
 $1$&$\sim 0$&$735^{+7(10)}_{-7(5)}$ &$164^{+5(6)}_{-4(2)}$\\
    &$140$~MeV&$747^{+5(11)}_{-5(4)}$ &$141^{+4(7)}_{-3(2)}$\\
    &$450$~MeV& $908^{+3(7)}_{-4(3)}$&$0$\\\hline
 \end{tabular}}
 \end{center}
 \caption{$\rho(770)$ meson mass, $\mathrm{Re}\,E_0$, and width, $\Gamma=-2 \mathrm{Im} E_0$, extracted from the pole position for several strange-quark masses at the chiral limit, physical pion mass and $m_\pi\simeq 450$, i.e., when it becomes a bound states in the $m_s=0,1m_s^0$ trajectories. The central value represents the median, the first upper and down indices gives the $68$\% CI, while the sum of the absolute values of the two upper (down) indices provides the upper (down) limits of the $95$\% CI.}
 \label{tab:poles}
\end{table}

Let us start analyzing the $\rho$-meson behavior in the $m_s=m_s^0$ trajectory. See Figs.~\ref{fig:3dresms},~\ref{fig:3drestrm}. In this case, we obtain a $\rho$-meson pole position $E_0=(735-i\,82)$~MeV near the chiral limit, while at the physical pion mass we get $E_0=(747-i\,70)$~MeV, see Table~\ref{tab:poles}, consistently with previous analyses~\cite{RuizdeElvira:2017aet}. 
  Nevertheless, as $m_\pi$ increases, the $\rho$-meson mass moves slower than the $\pi\pi$ threshold, so that, eventually, the $\rho(770)$ meson becomes a $\pi\pi$ bound state with a mass around $908$~MeV for a pion mass of around 450 MeV. Note that, in the case of the $m_s=m_s^0$ trajectory, the $\rho$-meson pole is always below the $K\bar{K}$ threshold for the pion masses analyzed here. We do not start to appreciate significant changes in the behavior of the $\rho$-meson mass till the strange-quark mass is reduced in half its physical value.
For instance, for $m_s=0.6\,m_s^0$, we still get a pole at $732-i\,81$~MeV in the chiral limit, which transforms into a bound state with a mass of $905$~MeV also for pion masses of around $450$~MeV. 
For lighter strange-quark mass trajectories relevant changes are observed. Both, $\rho$-meson mass and width, decrease consistently, so that we obtain $E_0=678-i\,38$~MeV for $m_s=0$ in the chiral limit.
Furthermore, for $m_s \le 0.4\, m_s^0$, both the $\pi\pi$ and $K\bar{K}$ thresholds get closer to each other as $m_\pi$ increases, in such a way  the kaon becomes lighter at a given point.
In this regime, the $\rho$ meson becomes a pole in the fourth Riemann sheet\footnote{The Riemann sheet that is reached when only the $K\bar K$ cut is crossed continuously from the first Riemann sheet.} when its mass gets below the $\pi\pi$ threshold. In this case, the $\rho$-meson decays only into $K\bar K$ and its width starts increasing again until it gets a maximum, after which the $\rho$ eventually becomes a $K\bar K$ bound state as $m_\pi$ increases.

This behavior is even more noticeable for the $\text{Tr}{\cal M}=C$ trajectories, depicted in Figs.~\ref{fig:3dimsms} and~\ref{fig:3dimstrm}. In this case, the strange-quark mass decreases as $m_\pi$ grows reaching the symmetric $m_s=m_{ud}$ line for pion masses of around 450 MeV. Once the symmetric line is crossed, the  $K\bar K$ channels opens below the two-pion threshold and the $\rho(770)$ meson becomes again a pole on the fourth Riemann sheet. Nevertheless, the kaon mass in this trajectory decreases till it ends up in the $m_s=0$ line (red-solid curve). Hence, the  $\rho$-meson mass (width) starts decreasing (increasing) at a given point (when the kaon gets lighter than the pion after crossing the symmetric line) till ending at the zero strangeness line. This behavior suggests that strangeness plays an important role in the $\rho(770)$ meson near the SU(3) flavor limit. This can also be inferred from the increase of its coupling to $K\bar{K}$, as discussed below.\footnote{At the symmetric line, the coupling to $K\bar{K}$ grows $20\%$ of its value at $m_s=m_s^0$ for physical pions.}

The pion mass dependence of the $\rho$-meson couplings to the $\pi\pi$ and $K\bar K$ channels, $g_{\pi\pi}$ and  $g_{K\bar{K}}$ as defined in Eq.~\eqref{eq:iamco}, are shown in Figs.~\ref{fig:mrhomsggl} and~\ref{fig:mrhotrmggl} for the $m_s=k$ and $\text{Tr}{\cal M}=C$ trajectories, respectively.
On one hand, $g_{\pi\pi}$ varies smoothly with $m_\pi$ before the the $\rho$-meson transition into a bound state, decreasing as it approaches the $K\bar{K}$ threshold and increasing with $m_s$. Once the $\rho$ meson becomes bound, its coupling rises sharply till $m_\pi\simeq480$ MeV. Overall, it takes values $g_{\pi\pi}\simeq 5.5-6.3$ for the range of pion masses studied. On the other hand, and contrary to the $g_{\pi\pi}$ behavior, $g_{K\bar{K}}$ decreases significantly with the mass of the strange quark. In addition, while the pion-mass dependence of $g_{K\bar{K}}$ flattens for lighter strange quarks, it becomes larger as $m_s$ reaches the physical value. All in all, it takes values within the range $g_{K\bar{K}}\simeq 3.2$ to $4.6$.

\begin{figure*}
\begin{tabular}{cc}
\hspace{-0.5cm} \includegraphics[scale=0.35]{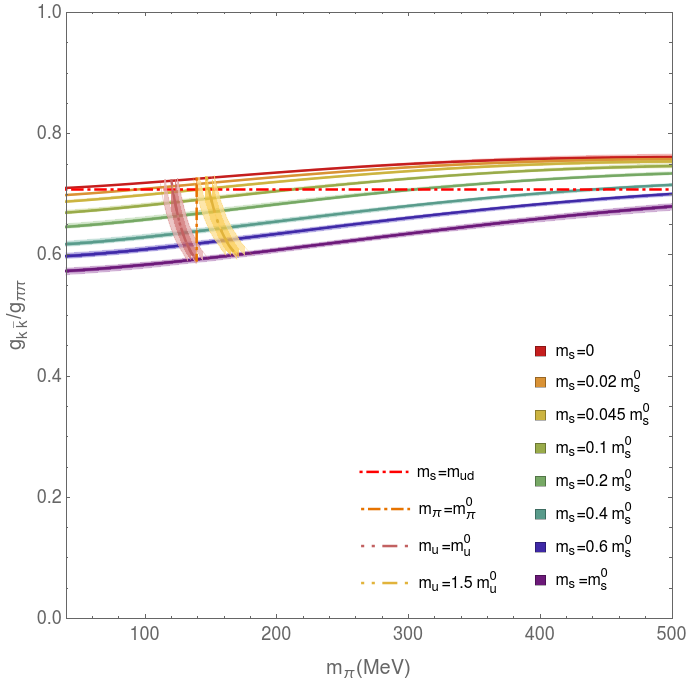} &\includegraphics[scale=0.35]{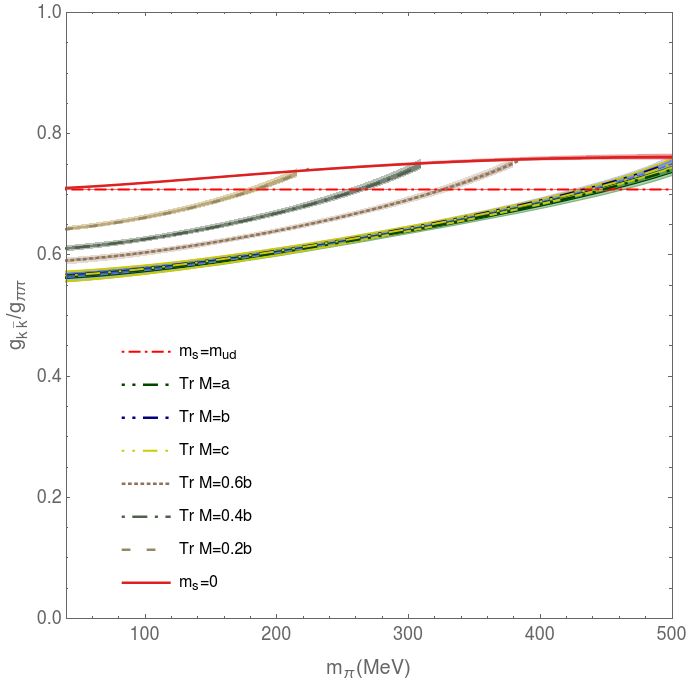}\\
 \end{tabular}
\caption{Ratio of the $g_{K\bar{K}}$ and $g_{\pi\pi}$ couplings for different trajectories.}
\label{fig:cosym}
\end{figure*}

More information can be extracted when the ratio of both couplings is depicted, we refer to Fig.~\ref{fig:cosym}. Remarkably, for those regions where $m_\pi \le m_K$ one observes the ratio $g_{K\bar{K}}/g_{\pi\pi}\le 1/\sqrt 2$. On the contrary, $g_{K\bar{K}}/g_{\pi\pi}> 1/\sqrt 2$ when $m_\pi> m_K$. In the symmetric line we obtain exactly $g_{K\bar{K}}/g_{\pi\pi}=1/\sqrt 2$.
  This is not a coincidence. In the SU(3) limit, the decomposition of a $I=1$, $I_3=0$ state of the antisymmetric octet representation into two-Goldstone--Boson states with well defined isospin reads
\begin{align}
  \vert Y=0, I=1, I_3=&0\rangle_{8_A}=\frac{1}{\sqrt 6}\vert K\bar K\rangle_1 -\frac{1}{\sqrt 6}\vert \bar K K\rangle_1\nonumber\\
  &\qquad+\frac{2}{\sqrt 3}\vert \pi\pi\rangle_1,
\end{align}
where, $Y$ stands for the hypercharge and $I$ is the isospin. Thus, taking into account the kaon degeneracy due to strangeness, the $\rho$-meson coupling to pions should be a factor $\sqrt 2$ times larger than for kaons.
Notably, the IAM analysis presented here reproduces exactly the SU(3) limit prediction.

\begin{figure*}
\begin{tabular}{cc}
\hspace{-0.5cm} \includegraphics[scale=0.35]{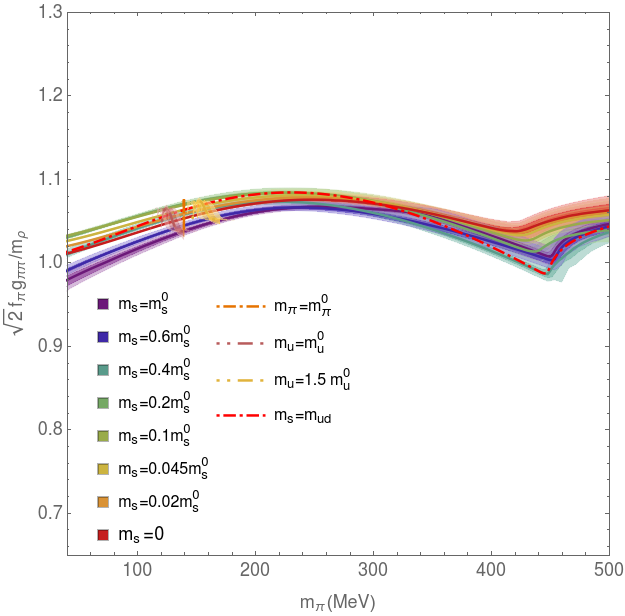} &\includegraphics[scale=0.35]{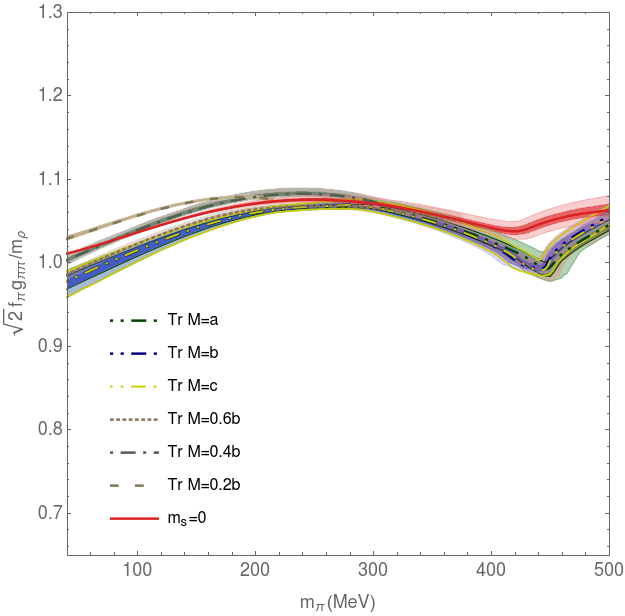}\\
 \end{tabular}
\caption{The quantity $\sqrt{2}f_\pi g_{\pi\pi}/m_\rho$ as a function of $m_\pi$ for $m_s=k,m_u=c,m_\pi=m^0_\pi$, and $m_s=m_u$ trajectories (left), and over $\mathrm{Tr}{\cal M}=C$ (right). Deviations from one reflect violations of the KSFR relation.}
\label{fig:ksfr}
\end{figure*}

In Fig.~\ref{fig:ksfr}, the ratio $\sqrt{2}g_{\pi\pi}f_\pi/m_\rho$ is depicted.\footnote{In this figure, $m_\rho$ means the real part of the pole position, $\mathrm{Re}E_0$, so that we are able to plot the ratio for a larger range of pion masses and after the transition. Since pole positions are slightly lower than the energy corresponding to $\delta=90^0$, this ratio gives values lower in around $1.7\%$, than if $E\,(\delta=90^\circ)$ is taken.} This ratio lies within the interval $[0.95,1.1]$, i.e., close to $1$, for the quark masses studied in this work, as predicted by the KSFR relation~\cite{Riazuddin:1966sw}.
 Thus, we find that KSFR is qualitatively valid, being more accurate near the chiral limit, specially for the $m_s=\{0,m_{ud}\}$ curves, and well applicable also around the physical point, with deviations from KSFR of less than 4\%. The largest deviations (still smaller than 8\%) are found for pion masses between $200$ and $300$~MeV.


\begin{figure*}
\begin{center}
{\renewcommand{\arraystretch}{2}
\setlength\tabcolsep{0.3cm}
 \begin{tabular}{ccc}
  \multicolumn{3}{c}{\hspace{1.8cm}\includegraphics[scale=0.4]{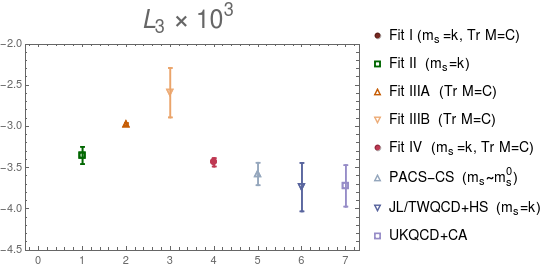}}\\
  \includegraphics[scale=0.4]{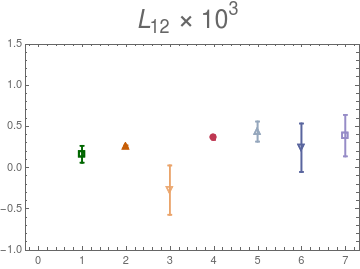}&\includegraphics[scale=0.4]{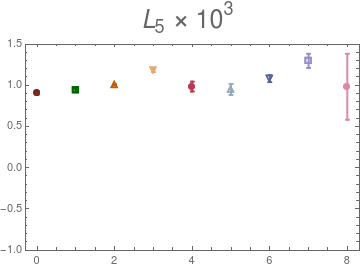}& \includegraphics[scale=0.27]{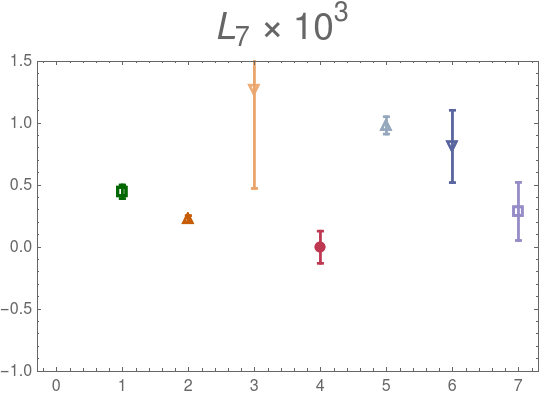}\\
   \includegraphics[scale=0.4]{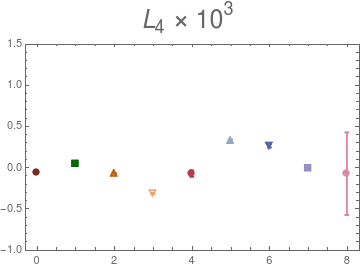}&\includegraphics[scale=0.4]{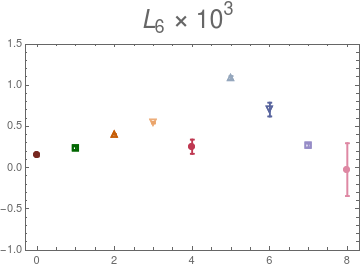}& \includegraphics[scale=0.4]{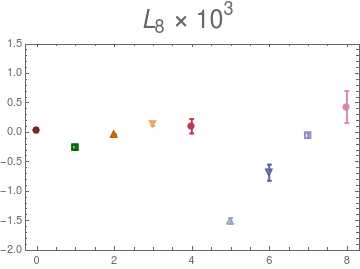}\\
 \end{tabular}}
\end{center}
\caption{Values of the LECs obtained in the several combined and global fits. For comparison purposes, in the case of $L_4$, $L_5$, $L_6$ and $L_8$ we also include the corresponding FLAG average value (last data point in pink).}
\label{fig:lecs}
\end{figure*}

To compare the LECs obtained in the different analyses done here with the Flag average~\cite{Aoki:2019cca}, we depict them in Fig.~\ref{fig:lecs}, where we show the results from Fit I (pseudoscalar meson mass and decay constant ratios), Fit II, (analysis of data over the $m_s=k$ trajectories), Fit III, ($\mathrm{Tr}{\cal M}=C$ trajectories), and  Fit IV (mean and standard deviation as a result of the study including both $m_s=k$ and $\mathrm{Tr}{\cal M}=C$ trajectories) together with the Flag average (pink color).
Indeed, we see that LECs from fits I and IV are very close, being also consistent with the FLAG average, which has larger errors. In general, fits I, II, IIIA and IV give closer results, while the LECs $L_6, L_7$ and $L_8$, from the analyses of PACS-CS and JL/TWQCD data, strongly disagree with other analyses.
Notice the precise values of the LECs provided by Fit IV.

\subsubsection*{The role of strangeness on the $\rho$-meson properties}

\begin{figure}
\begin{center}
\hspace{-0.5cm}\includegraphics[scale=0.34]{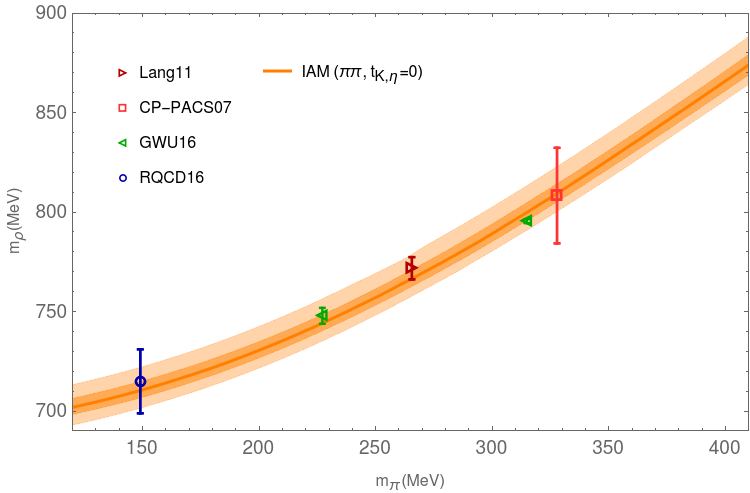}
\end{center}
\caption{Result for the $\rho$ meson mass when interacting terms involving kaons and etas, $t_{K,\eta}$, are set to zero as explained in the text, in comparison with $N_f=2$ lattice data.}
\label{fig:su2new}
\end{figure}

As we did before when discussing $f_\pi$, we can study the $\rho$-meson properties in a world where there are no kaons or etas by setting to zero their corresponding interacting terms, i.e., contact mass terms and diagrams involving loops with kaons and etas in Fig.~\ref{fig:diagrams} and pions in the initial and final state, (what we called $t_{K,\eta}$). In practice, this means that we solve the one-loop $\pi\pi$ IAM equation, see Eq.~\eqref{eq:iam}, taking the limits $f_{K,\eta}\to\infty$, and $m_{K,\eta}\to 0$. We refer the reader to a more extended explanation in Sect.~\ref{sec:nf2}. Then,  the pion mass and decay  constant are given now by $M_\pi$ and $F_\pi$ in Eqs.~(\ref{eq:mpip}) and (\ref{eq:fpip}). The result is shown in Fig.~\ref{fig:su2new}. Taking into account that we did not include $N_f=2$ data in our analysis but this result comes out as a prediction from our SU(3) IAM analysis, the agreement with the $N_f=2$ data from \cite{Aoki:2007rd,Lang:2011mn,Bali:2015gji,Guo:2016zos} is astonishing. 

In fact, starting from a three-flavor formulation, the $N_f=2$ formalism should be in principle obtained when one decouples the strange-quark contribution by sending  $m_s$ to infinity~\cite{Gasser:1984gg}. In this case the effect of kaons and etas is encoded into the bare pion decay constant, $f_0$, the constant $B_0$, and in terms proportional to $\nu_{K,\eta}=\left(\log(\hat{m}_{K,\eta}^2/\mu^2)+1\right)/32\pi^2$ ($\hat{m}_{K,\eta}$ refer to the limit where $m_{ud}\to 0$, which can be absorbed in a redefinition of the LECs. Here, though, we are simply studying the world where there are no kaons or etas relying on the fact that their interaction with pions comes from  terms where their masses and decay constants appear explicitly. 

Lattice $N_f = 2$ simulations typically use either $f_K$~\cite{Guo:2016zos,Fritzsch:2012wq} or the nucleon mass (via the QCD static potential)~\cite{Gockeler:2008kc,Feng:2010es,Lang:2011mn,Bali:2015gji} to fix the scale, which in general requires to perform a chiral extrapolation.  Nevertheless, given the observed dependence of $m_\pi$ and $f_\pi$ on the strange quark, and the fact that $f_\pi$ is correlated with $f_K$ and the nucleon mass, these methods seems to neglect this dependence. In fact, the agreement between $N_f=2$ lattice simulations and the IAM prediction without kaons and etas on the $\rho$-meson properties suggests that $N_f=2$ lattice simulations leave out  the contributions coming from the strange quark, and hence, they describe a world where the strange-quark is missing. 

\begin{figure}
\begin{center}
\hspace{-0.5cm}\includegraphics[scale=0.32]{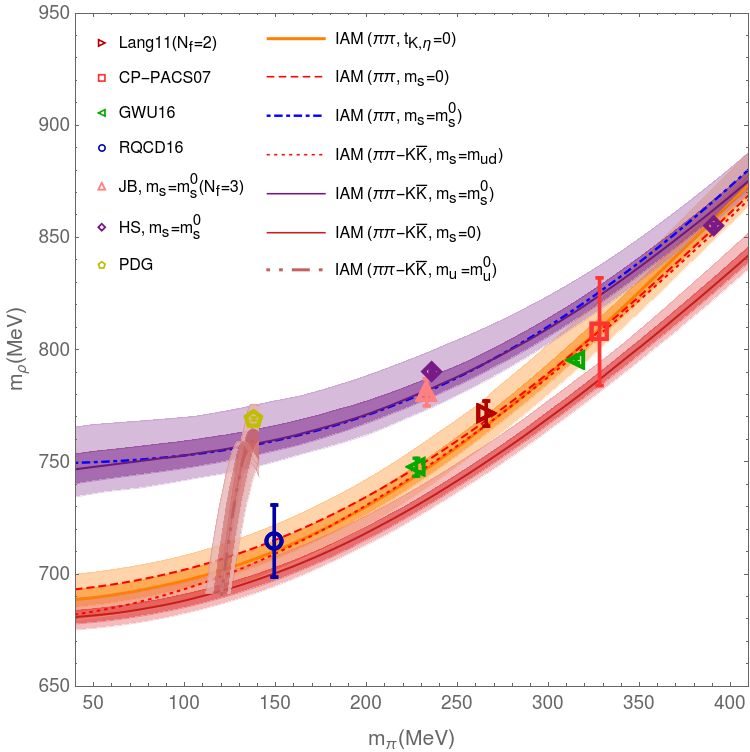}
\end{center}
\caption{Result for the $\rho$ meson mass when interacting terms involving kaons and etas, $t_{K,\eta}$, are set to zero as explained in the text, in comparison with $N_f=2$ lattice data, and with the result from SU(3) IAM in previous sections, $m_s=m_u,m_s^0,0$ and $m_s=ms^0$.}
\label{fig:compasu32}
\end{figure}

In Fig.~\ref{fig:compasu32} we compare this result with Fit IV over the chiral trajectories $m_s=\{m_s^0, m_{ud},0\}$ and $m_{u}=m_{u}^0$. Moreover, we also show the result of solving the one-channel $(\pi\pi)$ IAM equation with $m_s=\{0,m_s^0\}$, this is, keeping the $t_{k,\eta}$ terms in the $\pi\pi$ channel. Remarkably, the one-channel IAM result for $t_{K,\eta}=0$ (orange line), $m_s=0$ (dashed-red), and the two-coupled channel solution over $m_s=m_{ud}$ (dotted-red), provide very close results for the $\rho$ mass, which are also consistent with the $N_f=2$ lattice data. This is explained because the contribution of $t_{K,\eta}$ terms for light strange-quark masses is small. As explained before, these terms contribute in around $1-1.5$~MeV of the $f_\pi$ value when $m_s=0$, and around $6-7$~MeV for $m_s=m_s^0$ at the physical pion mass, see Fig.~\ref{fig:fpims}. This reduction on the value of $f_\pi$ for smaller strange-quark masses reflects also in smaller $\rho$-meson masses. Notice that in the $m_s=m_{ud}$ trajectories, pions and kaons are acting effectively in the $\rho$-meson mass as if only one flavor, the quark $u$, is present. 

It is also interesting to see what happens if one keeps the $t_{K,\eta}$ terms in $f_\pi$ and in the $\pi\pi$ scattering amplitude when solving the one-channel $\pi\pi$ IAM equation in the $m_s=m_s^0$ trajectory (dot-dashed blue). We see that this trajectory is consistent with the coupled-channel IAM solution for $m_s=m_s^0$, telling that the effect of the off-diagonal elements $t_{12}$ in Eq.~\eqref{eq:unicoupledel} is very small for physical strange-quark masses. Nevertheless, the coupled-channel effect becomes appreciable for lighter $m_s$, as as one can see by comparing the difference between the dashed and continuous red lines. The small contribution of the off-diagonal elements at the physical point is in contradiction with the results in ~\cite{Guo:2016zos,Hu:2016shf}, where the absence of these elements are found to be responsible for the dropping of the $\rho$ mass in the $N_f=2$ case. However, this is natural since in these works the same value of $f_\pi$ was used in the $N_f=2$ and $N_f=2+1$ predictions, and then, the effect of the kaon and eta contributions were absorbed in the off-diagonal elements. Indeed, we obtain here similar predictions for the $\rho$ meson mass in $N_f=2$ and $N_f=2+1$ simulations over $m_s=m_s^0$ than in \cite{Guo:2016zos,Hu:2016shf}. Nonetheless, we have gone through a deeper analysis in this work; by studying the $m_s$ dependence of $f_\pi$, we have obtained that $m_s$ regulates the contribution of the kaon and eta interacting terms and that the effect of these loops are absorbed in $f_\pi$ instead. Consistently with the predictions done in the works of~\cite{Guo:2016zos,Hu:2016shf}, we also obtain that the $\rho$ mass is reduced when these terms are omitted. In~\cite{Guo:2016zos}, the pion decay constant was determined and these values were used to make predictions for the $\rho$ mass using the UChPT model in~\cite{Oller:1998hw} with two and three flavors. Nevertheless, the pseudoscalar meson decay constants in~\cite{Guo:2016zos} were determined following the method of~\cite{Fritzsch:2012wq} where the kaon is introduced later in the quenched approximation and $m_K/f_K$ is fixed to the physical point. This leads to extrapolated values of $f_\pi$ in $N_f=2$ simulations consistent with the experimental value, where more than two flavors do exist. However, by doing this, one is missing the $m_s$ dependence and the effect of the kaon and eta loops in the pion decay constant, that we have shown here. These missing effects can lead to discrepancies between observables in $N_f=2$, such as the values of the $\rho$ mass and pion decay constant determinations. We have shown that a lower value of the $\rho$ mass than the physical one should be reflected also in lower values of $f_\pi$. In summary, assuming that the pion decay constant is the same in the two and three flavor simulations totally misses the effect of the strange quark and loops containing kaon and eta particles in pion observables. 

\subsection{Comparison with new $N_f=2$ and $N_f=2+1+1$ lattice results}

\begin{figure}
\begin{center}
\hspace{-0.5cm}\includegraphics[scale=0.39]{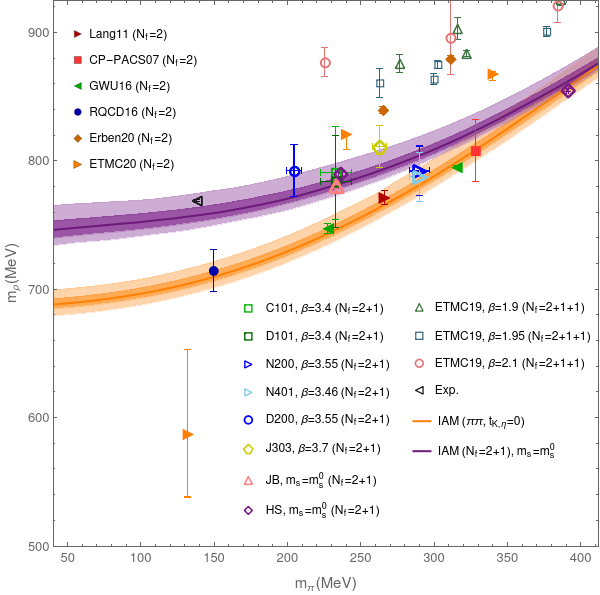}
\end{center}
\caption{Result for the $\rho$ meson mass obtained in the $N_f=2+1$ data analysis done here with the one quoted in the lattice papers for $N_f=2$ and $2+1+1$ simulations.}
\label{fig:compa234}
\end{figure}

After our analysis was completed, several $N_f=2$~\cite{Erben:2019nmx,Fischer:2020fvl} and $N_f=2+1+1$~\cite{Werner:2019hxc} new simulations have been performed using the pion decay constant to fix the scale. 
The comparison of these results with our $N_f=2+1$ analysis for the $\rho$-meson mass is depicted in Fig.~\ref{fig:compa234}. Looking at this figure it is clear that most data from Refs. \cite{Fischer:2020fvl,Erben:2019nmx} and \cite{Werner:2019hxc} follow a different trend than the  $N_f=2+1$ data analyzed in this article and the data from previous $N_f=2$ simulations~\cite{Aoki:2007rd,Lang:2011mn,Bali:2015gji,Guo:2016zos}. Moreover, the new $N_f=2$ simulation in~\cite{Fischer:2020fvl} includes results at the physical pion mass, which provide a $\rho$-meson mass from a IAM analysis around $600$ MeV, i.e., much lower than all other predictions in two flavors. In order to try to identify possible sources of discrepancies, we have summarized in Table~\ref{tab:va} the typical volumes and lattice spacings used in the simulations at different number of flavors. Comparing the data from Fig.~\ref{fig:compa234} and Table~\ref{tab:va}, we would like to make some final remarks regarding the differences observed among simulations with different number of flavors:
\begin{itemize}
 \item[1.] For $m_\pi\simeq230$ MeV, the ETMC $N_f=2+1+1$ simulation~\cite{Werner:2019hxc} obtains bigger values for the $\rho$-meson mass than in the $N_f=2$ simulation of the same collaboration for $m_\pi\simeq240$ MeV, even though the volume used in the $N_f=2$ simulation is bigger. This clearly points out that the dynamics involving the strange quark could be relevant in these simulations.
 \item[2.] The large discrepancy between the $N_f=2+1$ HS simulation~\cite{Wilson:2015dqa} and the $2+1+1$ ETMC19 result~\cite{Werner:2019hxc} at $m_\pi\simeq230$ MeV, cannot be explained because of the different volume or lattice spacing used. Namely, the HS data are compatible with the CLS simulation around $m_\pi\simeq 230$ when  the systematic error due to the lattice spacing  is included ($\simeq 4$ \% of the $\rho$ mass), even when CLS uses bigger volumes and smaller lattice spacing than HS.
   Similarly, these effects cannot explain the difference between the CLS simulation and the $N_f=2+1+1$ one at $m_\pi\simeq 265$ MeV.
   Then, these deviations between the $N_f=2+1$ and $2+1+1$ data can only be tight to the different methods employed by the collaborations in the simulations.
 \item[3.] The data for $N_f=2$ and $N_f=2+1+1$ simulations of the same ETMC collaboration show different trends of the $\rho$-meson mass dependence with the pion mass. Remarkably, an extrapolation by the eye of the $N_f=2+1+1$ simulation would lead to a much larger value of the $\rho$-meson mass than the simulation for physical pion mass of the $N_f=2$ data. 
 \item[4.] The two simulations done at physical pion masses for $N_f=2$, ETMC20 and RQCD16 disagree, even though the volumes and lattice spacings used are similar.
 \item[5.] The final analysis of the $N_f=2+1$ data done here, which includes the CLS data, indicates that the way the scale is set in the simulation can lead to a systematic source of error of around $4$ \% in the $\rho$-meson mass, still smaller than the differences observed in the data (which can be as large as $13$ \% for $m_\pi\simeq 230$ MeV). The collaborations could estimate this error by providing the average and typical deviation from different determinations of the lattice spacing. 
\item[6.] The differences observed between the $N_f=2$, $2+1$ and $2+1+1$ simulations can be understood partly by varying the mass of the strange quark while keeping the same scale setting in the simulations.
\end{itemize}

\begin{table}
\begin{center}
{\renewcommand{\arraystretch}{2}
\setlength\tabcolsep{0.001cm}
 \begin{tabular}{lcccc}
 \toprule
 Simulation&$N_f$&$m_\pi$&$L$ (fm)&$a$ (fm)\\
 \hline
 PACS-CS07~\cite{Aoki:2007rd}&$2$&$330$ &$2.4$&$0.2$\\
 Lang11~\cite{Lang:2011mn}&$2$&$270$&$1.9$&$0.12$\\
 GWU16~\cite{Guo:2016zos}&$2$&$230$&$2.9$ &$0.12$\\
 && &($L_z=2.9-5.8$)&\\
 RQCD16~\cite{Bali:2015gji}&$2$&$150$&$4.5$&$0.07$\\
 Erben20~\cite{Erben:2019nmx}&$2$&$265$&$3.2$&$0.07$\\
 ETMC20~\cite{Fischer:2020fvl}&$2$&$130$&$4.4$&$0.09$\\
  ETMC20~\cite{Fischer:2020fvl}&$2$&$240$&$4.4$&$0.09$\\\hline
 HS~\cite{Wilson:2015dqa}&$2+1$&$240$&$3.8$&$0.12$\\
 CLS~\cite{Andersen:2018mau}&$2+1$&$220-240$&$5-6$&$0.08-0.09$\\
 CLS~\cite{Andersen:2018mau}   &$2+1$ &$265$&$3.2$&$0.05$\\
 CLS~\cite{Andersen:2018mau}   &$2+1$&$205$&$4$&$0.06$\\\hline
 ETMC19~\cite{Werner:2019hxc}&$2+1+1$&$230$&$3.0$&$0.06$\\
 ETMC19~\cite{Werner:2019hxc}&$2+1+1$&$265$&$2.6$&$0.08$\\
 \hline
 \end{tabular}}
 \end{center}
 \caption{Typical values of the volumes and lattice spacings used for several collaborations in simulations for different flavors and the lightest pion masses.}
 \label{tab:va}
\end{table}
\section{Conclusions}\label{sec:con}

For the first time, we have studied simultaneously both the light- and strange-quark mass dependence of pseudoscalar meson masses, decay constants and $\rho$-meson properties, such as its mass, width and couplings to the pion and kaon channels. Our analysis is based on recent lattice data of these observables on the chiral trajectories $m_s=k$ and $\mathrm{Tr}{\cal M}=C$. In the analysis we resample pseudoscalar meson observables, energy levels (taking into account covariance matrices), and lattice spacings, providing a satisfactory solution at the $95\%$ confidence level. The IAM proves itself to be able to explain the pseudoscalar meson masses, decay constants and $\rho$-meson properties over different chiral trajectories. Therefore, the LECs obtained here are the most precise and the only ones up to now that are able to describe the strangeness dependence of these observables. The chiral extrapolation of $\rho$-meson phase shift data is also in remarkable agreement with experiment.

The dependence of the pion decay constant, $f_\pi$, with the strange-quark mass, $m_s$, is also studied for the first time. We have shown that, although to assume that the ratio $m_\pi/f_\pi$ is independent of $m_s$ can be a good approximation, the variation of $f_\pi$ with $m_s$ is  abrupt for light pion masses. Furthermore, this dependence is acting as a regulator of the size of the contribution of loops and contact terms involving kaons and etas. For instance, these terms contribute slightly to $f_\pi$ for $m_s=0$ but account for around $6-7$ MeV at $m_s=m_s^0$. This contribution to $f_\pi$ is sufficiently large, so that, their absence is able to explain successfully the lower values of the $\rho$-meson mass obtained in $N_f=2$ simulations. Even when we did not analyze here $N_f=2$ lattice data, but only $N_f=2+1$ simulations, the IAM has demonstrated to be able to describe simultaneously both, the $\rho$-meson mass over chiral trajectories in two and three flavor lattice simulations.  Regarding this last aspect, the results obtained here are consistent with the ones of~\cite{Guo:2016zos,Hu:2016shf}. However, we obtain here that the $\rho$-meson mass reduction in two-flavor calculations is due to the absence of the strange-quark mass and the contribution containing strange particles on the pion decay constant. Fixing the pseudoscalar decay constants to the physical point in two-flavor lattice simulations misses this dependence.

Some other interesting effects observed  involve the $K\bar{K}$ channel. First, as $m_s$ decreases the $\rho$-meson mass reduces; when $m_s$ approaches zero it drops around $70$ MeV respect to its values at the physical point. This is effectively more visible in the $m_u=
m_u^0$ trajectory.
Second, in the $m_s= m_s^0$ trajectory, as $m_\pi$ increases and the $\rho$-meson mass gets closer to the $K\bar{K}$ threshold, its coupling to $K\bar{K}$ increases, becoming eventually a bound state. Around $m_\pi=450$ MeV, it starts to decay into $K\bar{K}$  in the $\text{Tr}{\cal M}=\text{Tr}{\cal M}^0$ trajectory. For other trajectories, these transitions occur at different pion masses when the kaon becomes lighter than the pion. Third, the coupling ratio $g_{\pi\pi}/g_{K\bar{K}}=\sqrt{2}$ at the symmetric line, factor which comes from a SU(3) Clebsch-Gordan coefficient. Thus, SU(3) flavor symmetry is recovered in the symmetric line.

Our analysis also shows the operators that could be relevant in the energy region and for the light- and strange-quark masses considered in the lattice simulation . We hope that the results obtained here motivate the lattice community to investigate more on the hadron properties over different chiral trajectories, which indeed provide useful information to understand their dynamical nature.


\section{Appendix}\label{sec:app}
\subsection{Connecting ($\pi\pi-K\bar{K}$) coupled-channel IAM with $N_f=2$ lattice simulations}\label{sec:nf2}

In this section we analyze kaon and eta contributions into the $\rho$-meson properties, as well as we discuss how it is possible to disconnect their effect.
In the IAM coupled-channel formalism, kaons and etas contribute to pion-pion scattering through terms of the kind:
\begin{enumerate}
 \item The $\pi\pi\to K\bar{K}$ and $K\bar{K}\to K\bar{K}$ scattering amplitudes, which are named $t_{12}$ and $t_{22}$ in the coupled-channel IAM formulation, see Sect.~\ref{sec:iamcc}.\label{item1}
 \item Tadpoles and one-loop diagrams  involving kaons and etas in the $\pi\pi\to\pi\pi$ scattering amplitude, $t_{11}$, see Fig.~\ref{fig:diagrams}.\label{item2}
 \item Kaon and eta contact mass terms and tadpoles entering into the pion mass and decay constant, i.e., Eqs.~\eqref{eq:pimass}~and~\eqref{eq:fpis}.\label{item3}
\end{enumerate}

Regarding~(1), the amplitudes $t_{12}$ and $t_{22}$ are proportional to $1/f_K$ and $1/f_K^2$, respectively, since they involve diagrams with two and four external kaon legs.\footnote{Note that in the $\pi K$ and $K\bar K$ amplitudes in~\cite{GomezNicola:2001as} we have replaced $f_\pi^{1/2}$ by the decay constant of the corresponding external or internal GB leg. In addition, the NLO terms contains also higher powers of $1/f_P$.} It is clear that by sending $f_K\to \infty$ these contributions disappear. 
Note, though, that $f_K$ is related to $f_0$ through Eq.~\eqref{eq:fk}. Thus, in practice, taking this limit entails breaking the SU(3) symmetry in PCAC~\cite{Brooker:1970pe,Gounaris:1976ns,Oakes:1977vm}, i.e., to assume that the pion and kaon bare decay constants $f_{0,K}$ and $f_{0,\pi}$ do differ.
For the same reason, the terms in~\eqref{item2}, i.e., kaon and eta tadpoles and one-loop diagrams, are all proportional to $1/f_{K,\eta}^2$ and they also vanish when $f_{K,\eta}\to \infty$.

Finally, concerning~\eqref{item3}, while kaon and eta tadpoles entering in the pion mass and decay constant, Eqs.~\eqref{eq:pimass} and~\eqref{eq:fpis}, vanish when $f_{K,\eta}\to \infty$, there are still kaon mass contact terms, which can be removed only when one takes the limit $m_{K}\to0$. Thus, taking the limits $m^2_K=0$, $f_K,f_{\eta}\to \infty$, one obtains the pion mass and decay constant in a world where kaons are etas are both decoupled. Namely, 
\begin{align}
M_\pi^2=& M_{0\,\pi}^2\left[1+\mu_\pi+\frac{8 M_{0\,\pi}^2}{f_{0\pi}^2}\left(2L_6^r+2L_8^r-L_4^r-L_5^r\right)\right]\,,\label{eq:mpis1}
\end{align}
and 
\begin{align}
F_\pi=& f_{0\pi}\left[1-2\mu_\pi+\frac{4 M_{0\,\pi}^2}{f_{0\pi}^2}\left(L_4^r+L_5^r\right)\right]\ .\label{eq:fpis1}
\end{align}
 
The above relations are the same given in Eqs.~\eqref{eq:mpip} and~\eqref{eq:fpip}, which are depicted in Figs.~\ref{fig:mpiza} and~\ref{fig:fpims}, and discussed in the paragraphs around these figures.

Then, once all the contributions in \eqref{item1}-\eqref{item3}, which are called $t_{K,\eta}$ in the text, are removed, one can solve the one-channel $\pi\pi$ IAM, Eq. \eqref{eq:iam}, 
\begin{equation}
t(s)^{\mathrm{IAM}}=\frac{t_2(s)^2}{ t_2(s)-t_4(s)}\nonumber,
\end{equation}
using as input the pion mass and decay constant given in Eqs. \eqref{eq:mpis1}, \eqref{eq:fpis1}, to obtain pion-pion scattering in a world where kaons and etas are absent. This is plotted in the orange line in Fig.~\eqref{fig:compasu32} referred to as $t_{K\eta}=0$, and discussed in the Sect.~\ref{sec:uni}.
\newline

\subsection{Breit-Wigner reanalyses of IAM solutions}\label{app:BWIAM}

\begin{figure*}
 \begin{center}
  \includegraphics[scale=0.5]{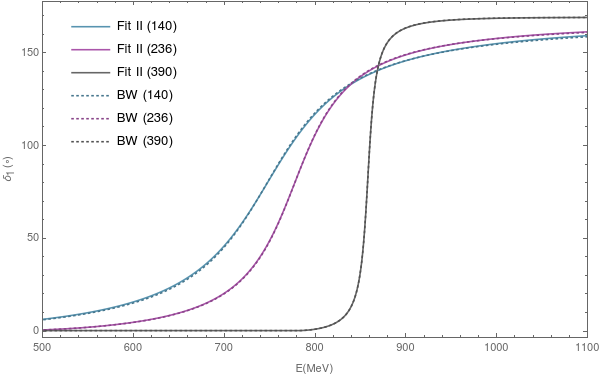}
 \end{center}
\caption{Breit-Wigner (BW) reanalyses of the phase shifts obtained in Fit II with IAM for the HadSpec masses and the extrapolation to the physical point. The figure shows that the behavior is compatible with a BW parameterizations and BW resonance parameters can be extracted. The BW results of the individual fits of sections~\ref{sec:ms} and~\ref{sec:trm} are comparable with the values given in the lattice articles.}
\label{fig:bw}
\end{figure*}

In Fig.~\ref{fig:bw} we show the refit of the IAM solution for the $m_s=m_s^0$ lattice data analyzed in section~\ref{sec:ms}. The Breit-Wigner parameterization used in these fits for the phase shift is
\begin{eqnarray}
 \mathrm{tan}\,\delta(E)=\frac{E\,\Gamma(E)}{m^2_\rho-E^2}\quad\mathrm{with}\quad \Gamma(E)=\frac{g^2_{\rho\pi\pi} p^3}{6\pi E^2}\ .\nonumber\\\label{eq:bw}
\end{eqnarray}

\subsection{Covariance matrix}\label{app:cov}

In Eq.~\ref{eq:cor} we provide the correlation matrix of our fitting parameters.

\begin{widetext}
  \begin{eqnarray}
  \left(\begin{array}{ccccccccccc}
             1.&- 0.71 &- 0.12 &-0.34 &- 0.16 &- 0.27 &- 0.25 &0.20&0.26 &0.29 &0.33\\
             - 0.71 &1.&0.0056 &0.33 &0.24  &0.22 &0.095 &- 0.21 &- 0.36 &- 0.35 &- 0.32 \\ 
             - 0.12 &0.0056 &1.&- 0.091 &0.73 &0.66 &- 0.53 &0.063 &- 0.082 &- 0.18 &- 0.14 \\
             - 0.34 &0.33 & - 0.091 & 1.&0.26  &- 0.46 &0.60 & - 0.65 &- 0.58  &- 0.59  &- 0.71\\ 
             - 0.16  &0.24 & 0.73 &0.26 &1.&0.32 &- 0.44  &- 0.53 &- 0.65  &- 0.70 & - 0.60\\
             - 0.27 & 0.22  &0.66 &- 0.46 &0.32&1. &-0.76& 0.54 & 0.36 & 0.29 &0.41 \\
             - 0.25 &0.095  &- 0.53 &0.60 &- 0.44  &- 0.75  &1.&- 0.36 &- 0.25 &- 0.21  &- 0.41 \\
             0.20& - 0.21 & 0.063 &- 0.65 &- 0.53& 0.54&- 0.36 &1.&0.95& 0.93 &0.94\\
             0.26 &- 0.36 &- 0.082  &- 0.58 &- 0.65  &0.36  &- 0.25  &0.95  &1.&0.96 &0.95 \\
             0.29 &- 0.35 &- 0.18  & - 0.59 & - 0.70  & 0.29  &- 0.21  & 0.93 &0.96 &1.&0.95\\
             0.33 &- 0.32 &- 0.14 &- 0.71 &- 0.6 &0.41 &- 0.41 &0.94 &0.95 &0.95 &1.\\
           \end{array}
            \right)\nonumber\\\label{eq:cor}
           \end{eqnarray}
         \end{widetext}
         
\section*{Acknowledgments}
We acknowledge discussions with C. Bernard, J. Bulava, S. Schaefer, M. Bruno, G. Colangelo, R. Brice\~no, J. Dudek, M. Niehus, M. Mai and the MILC Collaboration.
We also thank J. Bulava for providing the CLS phase-shift lattice data. R.M. acknowledges financial support from the Fundac\~ao de amparo \`a pesquisa do estado de S\~ao Paulo (FAPESP), the Talento Program of the Community of Madrid and the Complutense University of Madrid, under the project with Ref.~2018-T1/TIC-11167, and the CIDEGENT program with Ref. CIDEGENT/2019/015 and from the spanish national grant PID2019-106080GB-C21. This project has received funding from the European Union’s Horizon 2020 research and innovation programme under grant agreement No. 824093 for the STRONG-2020 project 
JRE is supported by the Swiss National Science Foundation, project No. PZ00P2 174228. 

\bibliography{biblio}

\begin{thebibliography}{168}
\expandafter\ifx\csname natexlab\endcsname\relax\def\natexlab#1{#1}\fi
\expandafter\ifx\csname bibnamefont\endcsname\relax
  \def\bibnamefont#1{#1}\fi
\expandafter\ifx\csname bibfnamefont\endcsname\relax
  \def\bibfnamefont#1{#1}\fi
\expandafter\ifx\csname citenamefont\endcsname\relax
  \def\citenamefont#1{#1}\fi
\expandafter\ifx\csname url\endcsname\relax
  \def\url#1{\texttt{#1}}\fi
\expandafter\ifx\csname urlprefix\endcsname\relax\def\urlprefix{URL }\fi
\providecommand{\bibinfo}[2]{#2}
\providecommand{\eprint}[2][]{\url{#2}}

\bibitem[{\citenamefont{Pisut and Roos}(1968)}]{Pisut:1968zza}
\bibinfo{author}{\bibfnamefont{J.}~\bibnamefont{Pisut}} \bibnamefont{and}
  \bibinfo{author}{\bibfnamefont{M.}~\bibnamefont{Roos}},
  \bibinfo{journal}{Nucl. Phys.} \textbf{\bibinfo{volume}{B6}},
  \bibinfo{pages}{325} (\bibinfo{year}{1968}).

\bibitem[{\citenamefont{Lafferty}(1993)}]{Lafferty:1993sx}
\bibinfo{author}{\bibfnamefont{G.~D.} \bibnamefont{Lafferty}},
  \bibinfo{journal}{Z. Phys.} \textbf{\bibinfo{volume}{C60}},
  \bibinfo{pages}{659} (\bibinfo{year}{1993}).

\bibitem[{\citenamefont{Tanabashi et~al.}(2018)}]{Tanabashi:2018oca}
\bibinfo{author}{\bibfnamefont{M.}~\bibnamefont{Tanabashi}}
  \bibnamefont{et~al.} (\bibinfo{collaboration}{Particle Data Group}),
  \bibinfo{journal}{Phys. Rev.} \textbf{\bibinfo{volume}{D98}},
  \bibinfo{pages}{030001} (\bibinfo{year}{2018}).

\bibitem[{\citenamefont{Ananthanarayan
  et~al.}(2001)\citenamefont{Ananthanarayan, Colangelo, Gasser, and
  Leutwyler}}]{Ananthanarayan:2000ht}
\bibinfo{author}{\bibfnamefont{B.}~\bibnamefont{Ananthanarayan}},
  \bibinfo{author}{\bibfnamefont{G.}~\bibnamefont{Colangelo}},
  \bibinfo{author}{\bibfnamefont{J.}~\bibnamefont{Gasser}}, \bibnamefont{and}
  \bibinfo{author}{\bibfnamefont{H.}~\bibnamefont{Leutwyler}},
  \bibinfo{journal}{Phys. Rept.} \textbf{\bibinfo{volume}{353}},
  \bibinfo{pages}{207} (\bibinfo{year}{2001}), \eprint{hep-ph/0005297}.

\bibitem[{\citenamefont{Colangelo et~al.}(2001)\citenamefont{Colangelo, Gasser,
  and Leutwyler}}]{Colangelo:2001df}
\bibinfo{author}{\bibfnamefont{G.}~\bibnamefont{Colangelo}},
  \bibinfo{author}{\bibfnamefont{J.}~\bibnamefont{Gasser}}, \bibnamefont{and}
  \bibinfo{author}{\bibfnamefont{H.}~\bibnamefont{Leutwyler}},
  \bibinfo{journal}{Nucl. Phys.} \textbf{\bibinfo{volume}{B603}},
  \bibinfo{pages}{125} (\bibinfo{year}{2001}), \eprint{hep-ph/0103088}.

\bibitem[{\citenamefont{Garcia-Martin
  et~al.}(2011{\natexlab{a}})\citenamefont{Garcia-Martin, Kaminski, Pelaez,
  Ruiz~de Elvira, and Yndurain}}]{GarciaMartin:2011cn}
\bibinfo{author}{\bibfnamefont{R.}~\bibnamefont{Garcia-Martin}},
  \bibinfo{author}{\bibfnamefont{R.}~\bibnamefont{Kaminski}},
  \bibinfo{author}{\bibfnamefont{J.~R.} \bibnamefont{Pelaez}},
  \bibinfo{author}{\bibfnamefont{J.}~\bibnamefont{Ruiz~de Elvira}},
  \bibnamefont{and} \bibinfo{author}{\bibfnamefont{F.~J.}
  \bibnamefont{Yndurain}}, \bibinfo{journal}{Phys. Rev.}
  \textbf{\bibinfo{volume}{D83}}, \bibinfo{pages}{074004}
  (\bibinfo{year}{2011}{\natexlab{a}}), \eprint{1102.2183}.

\bibitem[{\citenamefont{Garcia-Martin
  et~al.}(2011{\natexlab{b}})\citenamefont{Garcia-Martin, Kaminski, Pelaez, and
  Ruiz~de Elvira}}]{GarciaMartin:2011jx}
\bibinfo{author}{\bibfnamefont{R.}~\bibnamefont{Garcia-Martin}},
  \bibinfo{author}{\bibfnamefont{R.}~\bibnamefont{Kaminski}},
  \bibinfo{author}{\bibfnamefont{J.~R.} \bibnamefont{Pelaez}},
  \bibnamefont{and} \bibinfo{author}{\bibfnamefont{J.}~\bibnamefont{Ruiz~de
  Elvira}}, \bibinfo{journal}{Phys. Rev. Lett.} \textbf{\bibinfo{volume}{107}},
  \bibinfo{pages}{072001} (\bibinfo{year}{2011}{\natexlab{b}}),
  \eprint{1107.1635}.

\bibitem[{\citenamefont{Pelaez et~al.}(2019)\citenamefont{Pelaez, Rodas, and
  Ruiz De~Elvira}}]{Pelaez:2019eqa}
\bibinfo{author}{\bibfnamefont{J.}~\bibnamefont{Pelaez}},
  \bibinfo{author}{\bibfnamefont{A.}~\bibnamefont{Rodas}}, \bibnamefont{and}
  \bibinfo{author}{\bibfnamefont{J.}~\bibnamefont{Ruiz De~Elvira}},
  \bibinfo{journal}{Eur. Phys. J. C} \textbf{\bibinfo{volume}{79}},
  \bibinfo{pages}{1008} (\bibinfo{year}{2019}), \eprint{1907.13162}.

\bibitem[{\citenamefont{Aubert et~al.}(2009)}]{Aubert:2009ad}
\bibinfo{author}{\bibfnamefont{B.}~\bibnamefont{Aubert}} \bibnamefont{et~al.}
  (\bibinfo{collaboration}{BaBar}), \bibinfo{journal}{Phys. Rev. Lett.}
  \textbf{\bibinfo{volume}{103}}, \bibinfo{pages}{231801}
  (\bibinfo{year}{2009}), \eprint{0908.3589}.

\bibitem[{\citenamefont{Babusci et~al.}(2013)}]{Babusci:2012rp}
\bibinfo{author}{\bibfnamefont{D.}~\bibnamefont{Babusci}} \bibnamefont{et~al.}
  (\bibinfo{collaboration}{KLOE}), \bibinfo{journal}{Phys. Lett.}
  \textbf{\bibinfo{volume}{B720}}, \bibinfo{pages}{336} (\bibinfo{year}{2013}),
  \eprint{1212.4524}.

\bibitem[{\citenamefont{Ablikim et~al.}(2016)}]{Ablikim:2015orh}
\bibinfo{author}{\bibfnamefont{M.}~\bibnamefont{Ablikim}} \bibnamefont{et~al.}
  (\bibinfo{collaboration}{BESIII}), \bibinfo{journal}{Phys. Lett.}
  \textbf{\bibinfo{volume}{B753}}, \bibinfo{pages}{629} (\bibinfo{year}{2016}),
  \eprint{1507.08188}.

\bibitem[{\citenamefont{Eidelman and Jegerlehner}(1995)}]{Eidelman:1995ny}
\bibinfo{author}{\bibfnamefont{S.}~\bibnamefont{Eidelman}} \bibnamefont{and}
  \bibinfo{author}{\bibfnamefont{F.}~\bibnamefont{Jegerlehner}},
  \bibinfo{journal}{Z. Phys.} \textbf{\bibinfo{volume}{C67}},
  \bibinfo{pages}{585} (\bibinfo{year}{1995}), \eprint{hep-ph/9502298}.

\bibitem[{\citenamefont{Jegerlehner and Nyffeler}(2009)}]{Jegerlehner:2009ry}
\bibinfo{author}{\bibfnamefont{F.}~\bibnamefont{Jegerlehner}} \bibnamefont{and}
  \bibinfo{author}{\bibfnamefont{A.}~\bibnamefont{Nyffeler}},
  \bibinfo{journal}{Phys. Rept.} \textbf{\bibinfo{volume}{477}},
  \bibinfo{pages}{1} (\bibinfo{year}{2009}), \eprint{0902.3360}.

\bibitem[{\citenamefont{Colangelo et~al.}(2017)\citenamefont{Colangelo,
  Hoferichter, Procura, and Stoffer}}]{Colangelo:2017qdm}
\bibinfo{author}{\bibfnamefont{G.}~\bibnamefont{Colangelo}},
  \bibinfo{author}{\bibfnamefont{M.}~\bibnamefont{Hoferichter}},
  \bibinfo{author}{\bibfnamefont{M.}~\bibnamefont{Procura}}, \bibnamefont{and}
  \bibinfo{author}{\bibfnamefont{P.}~\bibnamefont{Stoffer}},
  \bibinfo{journal}{Phys. Rev. Lett.} \textbf{\bibinfo{volume}{118}},
  \bibinfo{pages}{232001} (\bibinfo{year}{2017}), \eprint{1701.06554}.

\bibitem[{\citenamefont{Colangelo et~al.}(2019)\citenamefont{Colangelo,
  Hoferichter, and Stoffer}}]{Colangelo:2018mtw}
\bibinfo{author}{\bibfnamefont{G.}~\bibnamefont{Colangelo}},
  \bibinfo{author}{\bibfnamefont{M.}~\bibnamefont{Hoferichter}},
  \bibnamefont{and} \bibinfo{author}{\bibfnamefont{P.}~\bibnamefont{Stoffer}},
  \bibinfo{journal}{JHEP} \textbf{\bibinfo{volume}{02}}, \bibinfo{pages}{006}
  (\bibinfo{year}{2019}), \eprint{1810.00007}.

\bibitem[{\citenamefont{Belushkin et~al.}(2007)\citenamefont{Belushkin, Hammer,
  and Mei{\ss}ner}}]{Belushkin:2006qa}
\bibinfo{author}{\bibfnamefont{M.~A.} \bibnamefont{Belushkin}},
  \bibinfo{author}{\bibfnamefont{H.~W.} \bibnamefont{Hammer}},
  \bibnamefont{and} \bibinfo{author}{\bibfnamefont{U.~G.}
  \bibnamefont{Mei{\ss}ner}}, \bibinfo{journal}{Phys. Rev.}
  \textbf{\bibinfo{volume}{C75}}, \bibinfo{pages}{035202}
  (\bibinfo{year}{2007}), \eprint{hep-ph/0608337}.

\bibitem[{\citenamefont{Lorenz et~al.}(2015)\citenamefont{Lorenz, Mei{\ss}ner,
  Hammer, and Dong}}]{Lorenz:2014yda}
\bibinfo{author}{\bibfnamefont{I.~T.} \bibnamefont{Lorenz}},
  \bibinfo{author}{\bibfnamefont{U.-G.} \bibnamefont{Mei{\ss}ner}},
  \bibinfo{author}{\bibfnamefont{H.~W.} \bibnamefont{Hammer}},
  \bibnamefont{and} \bibinfo{author}{\bibfnamefont{Y.~B.} \bibnamefont{Dong}},
  \bibinfo{journal}{Phys. Rev.} \textbf{\bibinfo{volume}{D91}},
  \bibinfo{pages}{014023} (\bibinfo{year}{2015}), \eprint{1411.1704}.

\bibitem[{\citenamefont{Hoferichter
  et~al.}(2016{\natexlab{a}})\citenamefont{Hoferichter, Kubis, Ruiz~de Elvira,
  Hammer, and Mei{\ss}ner}}]{Hoferichter:2016duk}
\bibinfo{author}{\bibfnamefont{M.}~\bibnamefont{Hoferichter}},
  \bibinfo{author}{\bibfnamefont{B.}~\bibnamefont{Kubis}},
  \bibinfo{author}{\bibfnamefont{J.}~\bibnamefont{Ruiz~de Elvira}},
  \bibinfo{author}{\bibfnamefont{H.~W.} \bibnamefont{Hammer}},
  \bibnamefont{and} \bibinfo{author}{\bibfnamefont{U.~G.}
  \bibnamefont{Mei{\ss}ner}}, \bibinfo{journal}{Eur. Phys. J.}
  \textbf{\bibinfo{volume}{A52}}, \bibinfo{pages}{331}
  (\bibinfo{year}{2016}{\natexlab{a}}), \eprint{1609.06722}.

\bibitem[{\citenamefont{Hoferichter et~al.}(2019)\citenamefont{Hoferichter,
  Kubis, Ruiz~de Elvira, and Stoffer}}]{Hoferichter:2018zwu}
\bibinfo{author}{\bibfnamefont{M.}~\bibnamefont{Hoferichter}},
  \bibinfo{author}{\bibfnamefont{B.}~\bibnamefont{Kubis}},
  \bibinfo{author}{\bibfnamefont{J.}~\bibnamefont{Ruiz~de Elvira}},
  \bibnamefont{and} \bibinfo{author}{\bibfnamefont{P.}~\bibnamefont{Stoffer}},
  \bibinfo{journal}{Phys. Rev. Lett.} \textbf{\bibinfo{volume}{122}},
  \bibinfo{pages}{122001} (\bibinfo{year}{2019}), \eprint{1811.11181}.

\bibitem[{\citenamefont{Kang et~al.}(2014)\citenamefont{Kang, Kubis, Hanhart,
  and Mei{\ss}ner}}]{Kang:2013jaa}
\bibinfo{author}{\bibfnamefont{X.-W.} \bibnamefont{Kang}},
  \bibinfo{author}{\bibfnamefont{B.}~\bibnamefont{Kubis}},
  \bibinfo{author}{\bibfnamefont{C.}~\bibnamefont{Hanhart}}, \bibnamefont{and}
  \bibinfo{author}{\bibfnamefont{U.-G.} \bibnamefont{Mei{\ss}ner}},
  \bibinfo{journal}{Phys. Rev.} \textbf{\bibinfo{volume}{D89}},
  \bibinfo{pages}{053015} (\bibinfo{year}{2014}), \eprint{1312.1193}.

\bibitem[{\citenamefont{Niecknig and Kubis}(2015)}]{Niecknig:2015ija}
\bibinfo{author}{\bibfnamefont{F.}~\bibnamefont{Niecknig}} \bibnamefont{and}
  \bibinfo{author}{\bibfnamefont{B.}~\bibnamefont{Kubis}},
  \bibinfo{journal}{JHEP} \textbf{\bibinfo{volume}{10}}, \bibinfo{pages}{142}
  (\bibinfo{year}{2015}), \eprint{1509.03188}.

\bibitem[{\citenamefont{Pisarski}(1995)}]{Pisarski:1995xu}
\bibinfo{author}{\bibfnamefont{R.~D.} \bibnamefont{Pisarski}},
  \bibinfo{journal}{Phys. Rev.} \textbf{\bibinfo{volume}{D52}},
  \bibinfo{pages}{R3773} (\bibinfo{year}{1995}), \eprint{hep-ph/9503328}.

\bibitem[{\citenamefont{Harada and Yamawaki}(2001)}]{Harada:2000kb}
\bibinfo{author}{\bibfnamefont{M.}~\bibnamefont{Harada}} \bibnamefont{and}
  \bibinfo{author}{\bibfnamefont{K.}~\bibnamefont{Yamawaki}},
  \bibinfo{journal}{Phys. Rev. Lett.} \textbf{\bibinfo{volume}{86}},
  \bibinfo{pages}{757} (\bibinfo{year}{2001}), \eprint{hep-ph/0010207}.

\bibitem[{\citenamefont{Rapp et~al.}(2010)\citenamefont{Rapp, Wambach, and van
  Hees}}]{Rapp:2009yu}
\bibinfo{author}{\bibfnamefont{R.}~\bibnamefont{Rapp}},
  \bibinfo{author}{\bibfnamefont{J.}~\bibnamefont{Wambach}}, \bibnamefont{and}
  \bibinfo{author}{\bibfnamefont{H.}~\bibnamefont{van Hees}},
  \bibinfo{journal}{Landolt-Bornstein} \textbf{\bibinfo{volume}{23}},
  \bibinfo{pages}{134} (\bibinfo{year}{2010}), \eprint{0901.3289}.

\bibitem[{\citenamefont{Gomez~Nicola
  et~al.}(2013{\natexlab{a}})\citenamefont{Gomez~Nicola, Pelaez, and Ruiz~de
  Elvira}}]{GomezNicola:2012uc}
\bibinfo{author}{\bibfnamefont{A.}~\bibnamefont{Gomez~Nicola}},
  \bibinfo{author}{\bibfnamefont{J.~R.} \bibnamefont{Pelaez}},
  \bibnamefont{and} \bibinfo{author}{\bibfnamefont{J.}~\bibnamefont{Ruiz~de
  Elvira}}, \bibinfo{journal}{Phys. Rev.} \textbf{\bibinfo{volume}{D87}},
  \bibinfo{pages}{016001} (\bibinfo{year}{2013}{\natexlab{a}}),
  \eprint{1210.7977}.

\bibitem[{\citenamefont{Gomez~Nicola
  et~al.}(2013{\natexlab{b}})\citenamefont{Gomez~Nicola, Ruiz~de Elvira, and
  Torres~Andres}}]{Nicola:2013vma}
\bibinfo{author}{\bibfnamefont{A.}~\bibnamefont{Gomez~Nicola}},
  \bibinfo{author}{\bibfnamefont{J.}~\bibnamefont{Ruiz~de Elvira}},
  \bibnamefont{and}
  \bibinfo{author}{\bibfnamefont{R.}~\bibnamefont{Torres~Andres}},
  \bibinfo{journal}{Phys. Rev.} \textbf{\bibinfo{volume}{D88}},
  \bibinfo{pages}{076007} (\bibinfo{year}{2013}{\natexlab{b}}),
  \eprint{1304.3356}.

\bibitem[{\citenamefont{G\'omez~Nicola and Ruiz~de
  Elvira}(2016)}]{Nicola:2016jlj}
\bibinfo{author}{\bibfnamefont{A.}~\bibnamefont{G\'omez~Nicola}}
  \bibnamefont{and} \bibinfo{author}{\bibfnamefont{J.}~\bibnamefont{Ruiz~de
  Elvira}}, \bibinfo{journal}{JHEP} \textbf{\bibinfo{volume}{03}},
  \bibinfo{pages}{186} (\bibinfo{year}{2016}), \eprint{1602.01476}.

\bibitem[{\citenamefont{Gomez~Nicola and Ruiz~de
  Elvira}(2018)}]{GomezNicola:2017bhm}
\bibinfo{author}{\bibfnamefont{A.}~\bibnamefont{Gomez~Nicola}}
  \bibnamefont{and} \bibinfo{author}{\bibfnamefont{J.}~\bibnamefont{Ruiz~de
  Elvira}}, \bibinfo{journal}{Phys. Rev.} \textbf{\bibinfo{volume}{D97}},
  \bibinfo{pages}{074016} (\bibinfo{year}{2018}), \eprint{1704.05036}.

\bibitem[{\citenamefont{G\'omez~Nicola and Ruiz
  De~Elvira}(2018)}]{Nicola:2018vug}
\bibinfo{author}{\bibfnamefont{A.}~\bibnamefont{G\'omez~Nicola}}
  \bibnamefont{and} \bibinfo{author}{\bibfnamefont{J.}~\bibnamefont{Ruiz
  De~Elvira}}, \bibinfo{journal}{Phys. Rev.} \textbf{\bibinfo{volume}{D98}},
  \bibinfo{pages}{014020} (\bibinfo{year}{2018}), \eprint{1803.08517}.

\bibitem[{\citenamefont{G\'omez~Nicola
  et~al.}(2019)\citenamefont{G\'omez~Nicola, Ruiz De~Elvira, and
  Vioque-Rodr\'iguez}}]{Nicola:2019ohb}
\bibinfo{author}{\bibfnamefont{A.}~\bibnamefont{G\'omez~Nicola}},
  \bibinfo{author}{\bibfnamefont{J.}~\bibnamefont{Ruiz De~Elvira}},
  \bibnamefont{and}
  \bibinfo{author}{\bibfnamefont{A.}~\bibnamefont{Vioque-Rodr\'iguez}},
  \bibinfo{journal}{JHEP} \textbf{\bibinfo{volume}{11}}, \bibinfo{pages}{086}
  (\bibinfo{year}{2019}), \eprint{1907.11734}.

\bibitem[{\citenamefont{Hu et~al.}(2016)\citenamefont{Hu, Molina, D{\"o}ring,
  and Alexandru}}]{Hu:2016shf}
\bibinfo{author}{\bibfnamefont{B.}~\bibnamefont{Hu}},
  \bibinfo{author}{\bibfnamefont{R.}~\bibnamefont{Molina}},
  \bibinfo{author}{\bibfnamefont{M.}~\bibnamefont{D{\"o}ring}},
  \bibnamefont{and}
  \bibinfo{author}{\bibfnamefont{A.}~\bibnamefont{Alexandru}},
  \bibinfo{journal}{Phys. Rev. Lett.} \textbf{\bibinfo{volume}{117}},
  \bibinfo{pages}{122001} (\bibinfo{year}{2016}), \eprint{1605.04823}.

\bibitem[{\citenamefont{Brice\~no et~al.}(2018)\citenamefont{Brice\~no, Dudek,
  and Young}}]{Briceno:2017max}
\bibinfo{author}{\bibfnamefont{R.~A.} \bibnamefont{Brice\~no}},
  \bibinfo{author}{\bibfnamefont{J.~J.} \bibnamefont{Dudek}}, \bibnamefont{and}
  \bibinfo{author}{\bibfnamefont{R.~D.} \bibnamefont{Young}},
  \bibinfo{journal}{Rev. Mod. Phys.} \textbf{\bibinfo{volume}{90}},
  \bibinfo{pages}{025001} (\bibinfo{year}{2018}), \eprint{1706.06223}.

\bibitem[{\citenamefont{Mohler}(2015)}]{Mohler:2015zsa}
\bibinfo{author}{\bibfnamefont{D.}~\bibnamefont{Mohler}}, in
  \emph{\bibinfo{booktitle}{{Proceedings, CHARM 2015}}} (\bibinfo{year}{2015}),
  \eprint{1508.02753},
  \urlprefix\url{http://lss.fnal.gov/archive/2015/conf/fermilab-conf-15-342-t.pdf}.

\bibitem[{\citenamefont{'t~Hooft}(1974)}]{tHooft:1973alw}
\bibinfo{author}{\bibfnamefont{G.}~\bibnamefont{'t~Hooft}},
  \bibinfo{journal}{Nucl. Phys.} \textbf{\bibinfo{volume}{B72}},
  \bibinfo{pages}{461} (\bibinfo{year}{1974}), \bibinfo{note}{[,337(1973)]}.

\bibitem[{\citenamefont{Witten}(1979)}]{Witten:1979kh}
\bibinfo{author}{\bibfnamefont{E.}~\bibnamefont{Witten}},
  \bibinfo{journal}{Nucl. Phys.} \textbf{\bibinfo{volume}{B160}},
  \bibinfo{pages}{57} (\bibinfo{year}{1979}).

\bibitem[{\citenamefont{Cohen et~al.}(2014)\citenamefont{Cohen, Llanes-Estrada,
  Pelaez, and Ruiz~de Elvira}}]{Cohen:2014vta}
\bibinfo{author}{\bibfnamefont{T.}~\bibnamefont{Cohen}},
  \bibinfo{author}{\bibfnamefont{F.~J.} \bibnamefont{Llanes-Estrada}},
  \bibinfo{author}{\bibfnamefont{J.~R.} \bibnamefont{Pelaez}},
  \bibnamefont{and} \bibinfo{author}{\bibfnamefont{J.}~\bibnamefont{Ruiz~de
  Elvira}}, \bibinfo{journal}{Phys. Rev.} \textbf{\bibinfo{volume}{D90}},
  \bibinfo{pages}{036003} (\bibinfo{year}{2014}), \eprint{1405.4831}.

\bibitem[{\citenamefont{Pelaez and Rios}(2006)}]{Pelaez:2006nj}
\bibinfo{author}{\bibfnamefont{J.~R.} \bibnamefont{Pelaez}} \bibnamefont{and}
  \bibinfo{author}{\bibfnamefont{G.}~\bibnamefont{Rios}},
  \bibinfo{journal}{Phys. Rev. Lett.} \textbf{\bibinfo{volume}{97}},
  \bibinfo{pages}{242002} (\bibinfo{year}{2006}), \eprint{hep-ph/0610397}.

\bibitem[{\citenamefont{Ruiz~de Elvira et~al.}(2011)\citenamefont{Ruiz~de
  Elvira, Pelaez, Pennington, and Wilson}}]{RuizdeElvira:2010cs}
\bibinfo{author}{\bibfnamefont{J.}~\bibnamefont{Ruiz~de Elvira}},
  \bibinfo{author}{\bibfnamefont{J.~R.} \bibnamefont{Pelaez}},
  \bibinfo{author}{\bibfnamefont{M.~R.} \bibnamefont{Pennington}},
  \bibnamefont{and} \bibinfo{author}{\bibfnamefont{D.~J.}
  \bibnamefont{Wilson}}, \bibinfo{journal}{Phys. Rev.}
  \textbf{\bibinfo{volume}{D84}}, \bibinfo{pages}{096006}
  (\bibinfo{year}{2011}), \eprint{1009.6204}.

\bibitem[{\citenamefont{Guo et~al.}(2012{\natexlab{a}})\citenamefont{Guo,
  Oller, and Ruiz~de Elvira}}]{Guo:2012ym}
\bibinfo{author}{\bibfnamefont{Z.-H.} \bibnamefont{Guo}},
  \bibinfo{author}{\bibfnamefont{J.~A.} \bibnamefont{Oller}}, \bibnamefont{and}
  \bibinfo{author}{\bibfnamefont{J.}~\bibnamefont{Ruiz~de Elvira}},
  \bibinfo{journal}{Phys. Lett.} \textbf{\bibinfo{volume}{B712}},
  \bibinfo{pages}{407} (\bibinfo{year}{2012}{\natexlab{a}}),
  \eprint{1203.4381}.

\bibitem[{\citenamefont{Guo et~al.}(2012{\natexlab{b}})\citenamefont{Guo,
  Oller, and Ruiz~de Elvira}}]{Guo:2012yt}
\bibinfo{author}{\bibfnamefont{Z.-H.} \bibnamefont{Guo}},
  \bibinfo{author}{\bibfnamefont{J.~A.} \bibnamefont{Oller}}, \bibnamefont{and}
  \bibinfo{author}{\bibfnamefont{J.}~\bibnamefont{Ruiz~de Elvira}},
  \bibinfo{journal}{Phys. Rev.} \textbf{\bibinfo{volume}{D86}},
  \bibinfo{pages}{054006} (\bibinfo{year}{2012}{\natexlab{b}}),
  \eprint{1206.4163}.

\bibitem[{\citenamefont{Ledwig et~al.}(2014)\citenamefont{Ledwig, Nieves, Pich,
  Ruiz~Arriola, and Ruiz~de Elvira}}]{Ledwig:2014cla}
\bibinfo{author}{\bibfnamefont{T.}~\bibnamefont{Ledwig}},
  \bibinfo{author}{\bibfnamefont{J.}~\bibnamefont{Nieves}},
  \bibinfo{author}{\bibfnamefont{A.}~\bibnamefont{Pich}},
  \bibinfo{author}{\bibfnamefont{E.}~\bibnamefont{Ruiz~Arriola}},
  \bibnamefont{and} \bibinfo{author}{\bibfnamefont{J.}~\bibnamefont{Ruiz~de
  Elvira}}, \bibinfo{journal}{Phys. Rev.} \textbf{\bibinfo{volume}{D90}},
  \bibinfo{pages}{114020} (\bibinfo{year}{2014}), \eprint{1407.3750}.

\bibitem[{\citenamefont{Ruiz~de Elvira et~al.}(2017)\citenamefont{Ruiz~de
  Elvira, Mei{\ss}ner, Rusetsky, and Schierholz}}]{RuizdeElvira:2017aet}
\bibinfo{author}{\bibfnamefont{J.}~\bibnamefont{Ruiz~de Elvira}},
  \bibinfo{author}{\bibfnamefont{U.~G.} \bibnamefont{Mei{\ss}ner}},
  \bibinfo{author}{\bibfnamefont{A.}~\bibnamefont{Rusetsky}}, \bibnamefont{and}
  \bibinfo{author}{\bibfnamefont{G.}~\bibnamefont{Schierholz}},
  \bibinfo{journal}{Eur. Phys. J.} \textbf{\bibinfo{volume}{C77}},
  \bibinfo{pages}{659} (\bibinfo{year}{2017}), \eprint{1706.09015}.

\bibitem[{\citenamefont{Weinberg}(1979)}]{Weinberg:1978kz}
\bibinfo{author}{\bibfnamefont{S.}~\bibnamefont{Weinberg}},
  \bibinfo{journal}{Physica} \textbf{\bibinfo{volume}{A96}},
  \bibinfo{pages}{327} (\bibinfo{year}{1979}).

\bibitem[{\citenamefont{Gasser and Leutwyler}(1984)}]{Gasser:1983yg}
\bibinfo{author}{\bibfnamefont{J.}~\bibnamefont{Gasser}} \bibnamefont{and}
  \bibinfo{author}{\bibfnamefont{H.}~\bibnamefont{Leutwyler}},
  \bibinfo{journal}{Annals Phys.} \textbf{\bibinfo{volume}{158}},
  \bibinfo{pages}{142} (\bibinfo{year}{1984}).

\bibitem[{\citenamefont{Gasser and Leutwyler}(1985)}]{Gasser:1984gg}
\bibinfo{author}{\bibfnamefont{J.}~\bibnamefont{Gasser}} \bibnamefont{and}
  \bibinfo{author}{\bibfnamefont{H.}~\bibnamefont{Leutwyler}},
  \bibinfo{journal}{Nucl. Phys.} \textbf{\bibinfo{volume}{B250}},
  \bibinfo{pages}{465} (\bibinfo{year}{1985}).

\bibitem[{\citenamefont{Truong}(1988)}]{Truong:1988zp}
\bibinfo{author}{\bibfnamefont{T.~N.} \bibnamefont{Truong}},
  \bibinfo{journal}{Phys. Rev. Lett.} \textbf{\bibinfo{volume}{61}},
  \bibinfo{pages}{2526} (\bibinfo{year}{1988}).

\bibitem[{\citenamefont{Dobado et~al.}(1990)\citenamefont{Dobado, Herrero, and
  Truong}}]{Dobado:1989qm}
\bibinfo{author}{\bibfnamefont{A.}~\bibnamefont{Dobado}},
  \bibinfo{author}{\bibfnamefont{M.~J.} \bibnamefont{Herrero}},
  \bibnamefont{and} \bibinfo{author}{\bibfnamefont{T.~N.}
  \bibnamefont{Truong}}, \bibinfo{journal}{Phys. Lett.}
  \textbf{\bibinfo{volume}{B235}}, \bibinfo{pages}{134} (\bibinfo{year}{1990}).

\bibitem[{\citenamefont{Dobado and Pelaez}(1993)}]{Dobado:1992ha}
\bibinfo{author}{\bibfnamefont{A.}~\bibnamefont{Dobado}} \bibnamefont{and}
  \bibinfo{author}{\bibfnamefont{J.~R.} \bibnamefont{Pelaez}},
  \bibinfo{journal}{Phys. Rev.} \textbf{\bibinfo{volume}{D47}},
  \bibinfo{pages}{4883} (\bibinfo{year}{1993}), \eprint{hep-ph/9301276}.

\bibitem[{\citenamefont{Dobado and Pelaez}(1997)}]{Dobado:1996ps}
\bibinfo{author}{\bibfnamefont{A.}~\bibnamefont{Dobado}} \bibnamefont{and}
  \bibinfo{author}{\bibfnamefont{J.~R.} \bibnamefont{Pelaez}},
  \bibinfo{journal}{Phys. Rev.} \textbf{\bibinfo{volume}{D56}},
  \bibinfo{pages}{3057} (\bibinfo{year}{1997}), \eprint{hep-ph/9604416}.

\bibitem[{\citenamefont{Nieves and Ruiz~Arriola}(1999)}]{Nieves:1998hp}
\bibinfo{author}{\bibfnamefont{J.}~\bibnamefont{Nieves}} \bibnamefont{and}
  \bibinfo{author}{\bibfnamefont{E.}~\bibnamefont{Ruiz~Arriola}},
  \bibinfo{journal}{Phys. Lett.} \textbf{\bibinfo{volume}{B455}},
  \bibinfo{pages}{30} (\bibinfo{year}{1999}), \eprint{nucl-th/9807035}.

\bibitem[{\citenamefont{Oller et~al.}(1999)\citenamefont{Oller, Oset, and
  Pelaez}}]{Oller:1998hw}
\bibinfo{author}{\bibfnamefont{J.~A.} \bibnamefont{Oller}},
  \bibinfo{author}{\bibfnamefont{E.}~\bibnamefont{Oset}}, \bibnamefont{and}
  \bibinfo{author}{\bibfnamefont{J.~R.} \bibnamefont{Pelaez}},
  \bibinfo{journal}{Phys. Rev.} \textbf{\bibinfo{volume}{D59}},
  \bibinfo{pages}{074001} (\bibinfo{year}{1999}), \bibinfo{note}{[Erratum:
  Phys. Rev.D75,099903(2007)]}, \eprint{hep-ph/9804209}.

\bibitem[{\citenamefont{Nieves and Ruiz~Arriola}(2000)}]{Nieves:1999bx}
\bibinfo{author}{\bibfnamefont{J.}~\bibnamefont{Nieves}} \bibnamefont{and}
  \bibinfo{author}{\bibfnamefont{E.}~\bibnamefont{Ruiz~Arriola}},
  \bibinfo{journal}{Nucl. Phys.} \textbf{\bibinfo{volume}{A679}},
  \bibinfo{pages}{57} (\bibinfo{year}{2000}), \eprint{hep-ph/9907469}.

\bibitem[{\citenamefont{Gomez~Nicola and Pelaez}(2002)}]{GomezNicola:2001as}
\bibinfo{author}{\bibfnamefont{A.}~\bibnamefont{Gomez~Nicola}}
  \bibnamefont{and} \bibinfo{author}{\bibfnamefont{J.~R.}
  \bibnamefont{Pelaez}}, \bibinfo{journal}{Phys. Rev.}
  \textbf{\bibinfo{volume}{D65}}, \bibinfo{pages}{054009}
  (\bibinfo{year}{2002}), \eprint{hep-ph/0109056}.

\bibitem[{\citenamefont{Aoki et~al.}(2007)}]{Aoki:2007rd}
\bibinfo{author}{\bibfnamefont{S.}~\bibnamefont{Aoki}} \bibnamefont{et~al.}
  (\bibinfo{collaboration}{CP-PACS}), \bibinfo{journal}{Phys. Rev.}
  \textbf{\bibinfo{volume}{D76}}, \bibinfo{pages}{094506}
  (\bibinfo{year}{2007}), \eprint{0708.3705}.

\bibitem[{\citenamefont{Gockeler et~al.}(2008)\citenamefont{Gockeler, Horsley,
  Nakamura, Pleiter, Rakow, Schierholz, and Zanotti}}]{Gockeler:2008kc}
\bibinfo{author}{\bibfnamefont{M.}~\bibnamefont{Gockeler}},
  \bibinfo{author}{\bibfnamefont{R.}~\bibnamefont{Horsley}},
  \bibinfo{author}{\bibfnamefont{Y.}~\bibnamefont{Nakamura}},
  \bibinfo{author}{\bibfnamefont{D.}~\bibnamefont{Pleiter}},
  \bibinfo{author}{\bibfnamefont{P.~E.~L.} \bibnamefont{Rakow}},
  \bibinfo{author}{\bibfnamefont{G.}~\bibnamefont{Schierholz}},
  \bibnamefont{and} \bibinfo{author}{\bibfnamefont{J.}~\bibnamefont{Zanotti}}
  (\bibinfo{collaboration}{QCDSF}), \bibinfo{journal}{PoS}
  \textbf{\bibinfo{volume}{LATTICE2008}}, \bibinfo{pages}{136}
  (\bibinfo{year}{2008}), \eprint{0810.5337}.

\bibitem[{\citenamefont{Feng et~al.}(2011)\citenamefont{Feng, Jansen, and
  Renner}}]{Feng:2010es}
\bibinfo{author}{\bibfnamefont{X.}~\bibnamefont{Feng}},
  \bibinfo{author}{\bibfnamefont{K.}~\bibnamefont{Jansen}}, \bibnamefont{and}
  \bibinfo{author}{\bibfnamefont{D.~B.} \bibnamefont{Renner}},
  \bibinfo{journal}{Phys. Rev.} \textbf{\bibinfo{volume}{D83}},
  \bibinfo{pages}{094505} (\bibinfo{year}{2011}), \eprint{1011.5288}.

\bibitem[{\citenamefont{Lang et~al.}(2011)\citenamefont{Lang, Mohler,
  Prelovsek, and Vidmar}}]{Lang:2011mn}
\bibinfo{author}{\bibfnamefont{C.~B.} \bibnamefont{Lang}},
  \bibinfo{author}{\bibfnamefont{D.}~\bibnamefont{Mohler}},
  \bibinfo{author}{\bibfnamefont{S.}~\bibnamefont{Prelovsek}},
  \bibnamefont{and} \bibinfo{author}{\bibfnamefont{M.}~\bibnamefont{Vidmar}},
  \bibinfo{journal}{Phys. Rev.} \textbf{\bibinfo{volume}{D84}},
  \bibinfo{pages}{054503} (\bibinfo{year}{2011}), \bibinfo{note}{[Erratum:
  Phys. Rev.D89,no.5,059903(2014)]}, \eprint{1105.5636}.

\bibitem[{\citenamefont{Pelissier and Alexandru}(2013)}]{Pelissier:2012pi}
\bibinfo{author}{\bibfnamefont{C.}~\bibnamefont{Pelissier}} \bibnamefont{and}
  \bibinfo{author}{\bibfnamefont{A.}~\bibnamefont{Alexandru}},
  \bibinfo{journal}{Phys. Rev.} \textbf{\bibinfo{volume}{D87}},
  \bibinfo{pages}{014503} (\bibinfo{year}{2013}), \eprint{1211.0092}.

\bibitem[{\citenamefont{Bali et~al.}(2016)\citenamefont{Bali, Collins, Cox,
  Donald, G{\"o}ckeler, Lang, and Schäfer}}]{Bali:2015gji}
\bibinfo{author}{\bibfnamefont{G.~S.} \bibnamefont{Bali}},
  \bibinfo{author}{\bibfnamefont{S.}~\bibnamefont{Collins}},
  \bibinfo{author}{\bibfnamefont{A.}~\bibnamefont{Cox}},
  \bibinfo{author}{\bibfnamefont{G.}~\bibnamefont{Donald}},
  \bibinfo{author}{\bibfnamefont{M.}~\bibnamefont{G{\"o}ckeler}},
  \bibinfo{author}{\bibfnamefont{C.~B.} \bibnamefont{Lang}}, \bibnamefont{and}
  \bibinfo{author}{\bibfnamefont{A.}~\bibnamefont{Schäfer}}
  (\bibinfo{collaboration}{RQCD}), \bibinfo{journal}{Phys. Rev.}
  \textbf{\bibinfo{volume}{D93}}, \bibinfo{pages}{054509}
  (\bibinfo{year}{2016}), \eprint{1512.08678}.

\bibitem[{\citenamefont{Guo et~al.}(2016)\citenamefont{Guo, Alexandru, Molina,
  and D{\"o}ring}}]{Guo:2016zos}
\bibinfo{author}{\bibfnamefont{D.}~\bibnamefont{Guo}},
  \bibinfo{author}{\bibfnamefont{A.}~\bibnamefont{Alexandru}},
  \bibinfo{author}{\bibfnamefont{R.}~\bibnamefont{Molina}}, \bibnamefont{and}
  \bibinfo{author}{\bibfnamefont{M.}~\bibnamefont{D{\"o}ring}},
  \bibinfo{journal}{Phys. Rev.} \textbf{\bibinfo{volume}{D94}},
  \bibinfo{pages}{034501} (\bibinfo{year}{2016}), \eprint{1605.03993}.

\bibitem[{\citenamefont{Erben et~al.}(2020)\citenamefont{Erben, Green, Mohler,
  and Wittig}}]{Erben:2019nmx}
\bibinfo{author}{\bibfnamefont{F.}~\bibnamefont{Erben}},
  \bibinfo{author}{\bibfnamefont{J.~R.} \bibnamefont{Green}},
  \bibinfo{author}{\bibfnamefont{D.}~\bibnamefont{Mohler}}, \bibnamefont{and}
  \bibinfo{author}{\bibfnamefont{H.}~\bibnamefont{Wittig}},
  \bibinfo{journal}{Phys. Rev. D} \textbf{\bibinfo{volume}{101}},
  \bibinfo{pages}{054504} (\bibinfo{year}{2020}), \eprint{1910.01083}.

\bibitem[{\citenamefont{Fischer et~al.}(2020)\citenamefont{Fischer, Kostrzewa,
  Mai, Petschlies, Pittler, Ueding, Urbach, and Werner}}]{Fischer:2020fvl}
\bibinfo{author}{\bibfnamefont{M.}~\bibnamefont{Fischer}},
  \bibinfo{author}{\bibfnamefont{B.}~\bibnamefont{Kostrzewa}},
  \bibinfo{author}{\bibfnamefont{M.}~\bibnamefont{Mai}},
  \bibinfo{author}{\bibfnamefont{M.}~\bibnamefont{Petschlies}},
  \bibinfo{author}{\bibfnamefont{F.}~\bibnamefont{Pittler}},
  \bibinfo{author}{\bibfnamefont{M.}~\bibnamefont{Ueding}},
  \bibinfo{author}{\bibfnamefont{C.}~\bibnamefont{Urbach}}, \bibnamefont{and}
  \bibinfo{author}{\bibfnamefont{M.}~\bibnamefont{Werner}}
  (\bibinfo{collaboration}{ETM}) (\bibinfo{year}{2020}), \eprint{2006.13805}.

\bibitem[{\citenamefont{Wilson et~al.}(2015)\citenamefont{Wilson, Brice\~no,
  Dudek, Edwards, and Thomas}}]{Wilson:2015dqa}
\bibinfo{author}{\bibfnamefont{D.~J.} \bibnamefont{Wilson}},
  \bibinfo{author}{\bibfnamefont{R.~A.} \bibnamefont{Brice\~no}},
  \bibinfo{author}{\bibfnamefont{J.~J.} \bibnamefont{Dudek}},
  \bibinfo{author}{\bibfnamefont{R.~G.} \bibnamefont{Edwards}},
  \bibnamefont{and} \bibinfo{author}{\bibfnamefont{C.~E.}
  \bibnamefont{Thomas}}, \bibinfo{journal}{Phys. Rev.}
  \textbf{\bibinfo{volume}{D92}}, \bibinfo{pages}{094502}
  (\bibinfo{year}{2015}), \eprint{1507.02599}.

\bibitem[{\citenamefont{Dudek et~al.}(2013)\citenamefont{Dudek, Edwards, and
  Thomas}}]{Dudek:2012xn}
\bibinfo{author}{\bibfnamefont{J.~J.} \bibnamefont{Dudek}},
  \bibinfo{author}{\bibfnamefont{R.~G.} \bibnamefont{Edwards}},
  \bibnamefont{and} \bibinfo{author}{\bibfnamefont{C.~E.} \bibnamefont{Thomas}}
  (\bibinfo{collaboration}{Hadron Spectrum}), \bibinfo{journal}{Phys. Rev.}
  \textbf{\bibinfo{volume}{D87}}, \bibinfo{pages}{034505}
  (\bibinfo{year}{2013}), \bibinfo{note}{[Erratum: Phys.
  Rev.D90,no.9,099902(2014)]}, \eprint{1212.0830}.

\bibitem[{\citenamefont{Bulava et~al.}(2016)\citenamefont{Bulava, Fahy, Horz,
  Juge, Morningstar, and Wong}}]{Bulava:2016mks}
\bibinfo{author}{\bibfnamefont{J.}~\bibnamefont{Bulava}},
  \bibinfo{author}{\bibfnamefont{B.}~\bibnamefont{Fahy}},
  \bibinfo{author}{\bibfnamefont{B.}~\bibnamefont{Horz}},
  \bibinfo{author}{\bibfnamefont{K.~J.} \bibnamefont{Juge}},
  \bibinfo{author}{\bibfnamefont{C.}~\bibnamefont{Morningstar}},
  \bibnamefont{and} \bibinfo{author}{\bibfnamefont{C.~H.} \bibnamefont{Wong}},
  \bibinfo{journal}{Nucl. Phys.} \textbf{\bibinfo{volume}{B910}},
  \bibinfo{pages}{842} (\bibinfo{year}{2016}), \eprint{1604.05593}.

\bibitem[{\citenamefont{Feng et~al.}(2015)\citenamefont{Feng, Aoki, Hashimoto,
  and Kaneko}}]{Feng:2014gba}
\bibinfo{author}{\bibfnamefont{X.}~\bibnamefont{Feng}},
  \bibinfo{author}{\bibfnamefont{S.}~\bibnamefont{Aoki}},
  \bibinfo{author}{\bibfnamefont{S.}~\bibnamefont{Hashimoto}},
  \bibnamefont{and} \bibinfo{author}{\bibfnamefont{T.}~\bibnamefont{Kaneko}},
  \bibinfo{journal}{Phys. Rev.} \textbf{\bibinfo{volume}{D91}},
  \bibinfo{pages}{054504} (\bibinfo{year}{2015}), \eprint{1412.6319}.

\bibitem[{\citenamefont{Alexandrou et~al.}(2017)\citenamefont{Alexandrou,
  Leskovec, Meinel, Negele, Paul, Petschlies, Pochinsky, Rendon, and
  Syritsyn}}]{Alexandrou:2017mpi}
\bibinfo{author}{\bibfnamefont{C.}~\bibnamefont{Alexandrou}},
  \bibinfo{author}{\bibfnamefont{L.}~\bibnamefont{Leskovec}},
  \bibinfo{author}{\bibfnamefont{S.}~\bibnamefont{Meinel}},
  \bibinfo{author}{\bibfnamefont{J.}~\bibnamefont{Negele}},
  \bibinfo{author}{\bibfnamefont{S.}~\bibnamefont{Paul}},
  \bibinfo{author}{\bibfnamefont{M.}~\bibnamefont{Petschlies}},
  \bibinfo{author}{\bibfnamefont{A.}~\bibnamefont{Pochinsky}},
  \bibinfo{author}{\bibfnamefont{G.}~\bibnamefont{Rendon}}, \bibnamefont{and}
  \bibinfo{author}{\bibfnamefont{S.}~\bibnamefont{Syritsyn}},
  \bibinfo{journal}{Phys. Rev.} \textbf{\bibinfo{volume}{D96}},
  \bibinfo{pages}{034525} (\bibinfo{year}{2017}), \eprint{1704.05439}.

\bibitem[{\citenamefont{Fu and Wang}(2016)}]{Fu:2016}
\bibinfo{author}{\bibfnamefont{Z.}~\bibnamefont{Fu}} \bibnamefont{and}
  \bibinfo{author}{\bibfnamefont{L.}~\bibnamefont{Wang}},
  \bibinfo{journal}{Phys. Rev.} \textbf{\bibinfo{volume}{D94}},
  \bibinfo{pages}{034505} (\bibinfo{year}{2016}), \eprint{1608.07478}.

\bibitem[{\citenamefont{Metivet}(2015)}]{metivet}
\bibinfo{author}{\bibfnamefont{T.}~\bibnamefont{Metivet}}
  (\bibinfo{collaboration}{Budapest-Marseille-Wuppertal}),
  \bibinfo{journal}{PoS} \textbf{\bibinfo{volume}{LATTICE2014}},
  \bibinfo{pages}{079} (\bibinfo{year}{2015}), \eprint{1410.8447}.

\bibitem[{\citenamefont{Werner et~al.}(2020)}]{Werner:2019hxc}
\bibinfo{author}{\bibfnamefont{M.}~\bibnamefont{Werner}} \bibnamefont{et~al.},
  \bibinfo{journal}{Eur. Phys. J.} \textbf{\bibinfo{volume}{A56}},
  \bibinfo{pages}{61} (\bibinfo{year}{2020}), \eprint{1907.01237}.

\bibitem[{\citenamefont{Miller et~al.}(2020)}]{Miller:2020xhy}
\bibinfo{author}{\bibfnamefont{N.}~\bibnamefont{Miller}} \bibnamefont{et~al.},
  \bibinfo{journal}{Phys. Rev. D} \textbf{\bibinfo{volume}{102}},
  \bibinfo{pages}{034507} (\bibinfo{year}{2020}), \eprint{2005.04795}.

\bibitem[{\citenamefont{Bruno et~al.}(2017)\citenamefont{Bruno, Korzec, and
  Schaefer}}]{Bruno:2016plf}
\bibinfo{author}{\bibfnamefont{M.}~\bibnamefont{Bruno}},
  \bibinfo{author}{\bibfnamefont{T.}~\bibnamefont{Korzec}}, \bibnamefont{and}
  \bibinfo{author}{\bibfnamefont{S.}~\bibnamefont{Schaefer}},
  \bibinfo{journal}{Phys. Rev.} \textbf{\bibinfo{volume}{D95}},
  \bibinfo{pages}{074504} (\bibinfo{year}{2017}), \eprint{1608.08900}.

\bibitem[{\citenamefont{Blum et~al.}(2016)}]{Blum:2014tka}
\bibinfo{author}{\bibfnamefont{T.}~\bibnamefont{Blum}} \bibnamefont{et~al.}
  (\bibinfo{collaboration}{RBC, UKQCD}), \bibinfo{journal}{Phys. Rev.}
  \textbf{\bibinfo{volume}{D93}}, \bibinfo{pages}{074505}
  (\bibinfo{year}{2016}), \eprint{1411.7017}.

\bibitem[{\citenamefont{Bazavov et~al.}(2010{\natexlab{a}})}]{Bazavov:2010hj}
\bibinfo{author}{\bibfnamefont{A.}~\bibnamefont{Bazavov}} \bibnamefont{et~al.}
  (\bibinfo{collaboration}{MILC}), \bibinfo{journal}{PoS}
  \textbf{\bibinfo{volume}{LATTICE2010}}, \bibinfo{pages}{074}
  (\bibinfo{year}{2010}{\natexlab{a}}), \eprint{1012.0868}.

\bibitem[{\citenamefont{Bazavov et~al.}(2010{\natexlab{b}})}]{Bazavov:2009bb}
\bibinfo{author}{\bibfnamefont{A.}~\bibnamefont{Bazavov}} \bibnamefont{et~al.}
  (\bibinfo{collaboration}{MILC}), \bibinfo{journal}{Rev. Mod. Phys.}
  \textbf{\bibinfo{volume}{82}}, \bibinfo{pages}{1349}
  (\bibinfo{year}{2010}{\natexlab{b}}), \eprint{0903.3598}.

\bibitem[{\citenamefont{Aubin et~al.}(2008)\citenamefont{Aubin, Laiho, and
  Van~de Water}}]{Aubin:2008ie}
\bibinfo{author}{\bibfnamefont{C.}~\bibnamefont{Aubin}},
  \bibinfo{author}{\bibfnamefont{J.}~\bibnamefont{Laiho}}, \bibnamefont{and}
  \bibinfo{author}{\bibfnamefont{R.~S.} \bibnamefont{Van~de Water}},
  \bibinfo{journal}{PoS} \textbf{\bibinfo{volume}{LATTICE2008}},
  \bibinfo{pages}{105} (\bibinfo{year}{2008}), \eprint{0810.4328}.

\bibitem[{\citenamefont{Noaki et~al.}(2009)}]{Noaki:2009sk}
\bibinfo{author}{\bibfnamefont{J.}~\bibnamefont{Noaki}} \bibnamefont{et~al.}
  (\bibinfo{collaboration}{TWQCD, JLQCD}), \bibinfo{journal}{PoS}
  \textbf{\bibinfo{volume}{LAT2009}}, \bibinfo{pages}{096}
  (\bibinfo{year}{2009}), \eprint{0910.5532}.

\bibitem[{\citenamefont{Aoki et~al.}(2009)}]{Aoki:2008sm}
\bibinfo{author}{\bibfnamefont{S.}~\bibnamefont{Aoki}} \bibnamefont{et~al.}
  (\bibinfo{collaboration}{PACS-CS}), \bibinfo{journal}{Phys. Rev.}
  \textbf{\bibinfo{volume}{D79}}, \bibinfo{pages}{034503}
  (\bibinfo{year}{2009}), \eprint{0807.1661}.

\bibitem[{\citenamefont{Baron et~al.}(2010)}]{Baron:2010bv}
\bibinfo{author}{\bibfnamefont{R.}~\bibnamefont{Baron}} \bibnamefont{et~al.},
  \bibinfo{journal}{JHEP} \textbf{\bibinfo{volume}{06}}, \bibinfo{pages}{111}
  (\bibinfo{year}{2010}), \eprint{1004.5284}.

\bibitem[{\citenamefont{Aoki et~al.}(2019)}]{Aoki:2019cca}
\bibinfo{author}{\bibfnamefont{S.}~\bibnamefont{Aoki}} \bibnamefont{et~al.}
  (\bibinfo{collaboration}{Flavour Lattice Averaging Group})
  (\bibinfo{year}{2019}), \eprint{1902.08191}.

\bibitem[{\citenamefont{Andersen et~al.}(2019)\citenamefont{Andersen, Bulava,
  Hörz, and Morningstar}}]{Andersen:2018mau}
\bibinfo{author}{\bibfnamefont{C.}~\bibnamefont{Andersen}},
  \bibinfo{author}{\bibfnamefont{J.}~\bibnamefont{Bulava}},
  \bibinfo{author}{\bibfnamefont{B.}~\bibnamefont{Hörz}}, \bibnamefont{and}
  \bibinfo{author}{\bibfnamefont{C.}~\bibnamefont{Morningstar}},
  \bibinfo{journal}{Nucl. Phys. B} \textbf{\bibinfo{volume}{939}},
  \bibinfo{pages}{145} (\bibinfo{year}{2019}), \eprint{1808.05007}.

\bibitem[{\citenamefont{Sakurai}(1969)}]{sakurai}
\bibinfo{author}{\bibfnamefont{J.}~\bibnamefont{Sakurai}},
  \bibinfo{journal}{Currents and mesons, University of Chicago Press, Chicago
  II,}  (\bibinfo{year}{1969}).

\bibitem[{\citenamefont{Birse}(1996)}]{Birse:1996hd}
\bibinfo{author}{\bibfnamefont{M.~C.} \bibnamefont{Birse}},
  \bibinfo{journal}{Z. Phys.} \textbf{\bibinfo{volume}{A355}},
  \bibinfo{pages}{231} (\bibinfo{year}{1996}), \eprint{hep-ph/9603251}.

\bibitem[{\citenamefont{Riazuddin and Fayyazuddin}(1966)}]{Riazuddin:1966sw}
\bibinfo{author}{\bibnamefont{Riazuddin}} \bibnamefont{and}
  \bibinfo{author}{\bibnamefont{Fayyazuddin}}, \bibinfo{journal}{Phys. Rev.}
  \textbf{\bibinfo{volume}{147}}, \bibinfo{pages}{1071} (\bibinfo{year}{1966}).

\bibitem[{\citenamefont{Hanhart et~al.}(2008)\citenamefont{Hanhart, Pelaez, and
  Rios}}]{Hanhart:2008mx}
\bibinfo{author}{\bibfnamefont{C.}~\bibnamefont{Hanhart}},
  \bibinfo{author}{\bibfnamefont{J.}~\bibnamefont{Pelaez}}, \bibnamefont{and}
  \bibinfo{author}{\bibfnamefont{G.}~\bibnamefont{Rios}},
  \bibinfo{journal}{Phys. Rev. Lett.} \textbf{\bibinfo{volume}{100}},
  \bibinfo{pages}{152001} (\bibinfo{year}{2008}), \eprint{0801.2871}.

\bibitem[{\citenamefont{Pelaez and Rios}(2010)}]{Pelaez:2010fj}
\bibinfo{author}{\bibfnamefont{J.~R.} \bibnamefont{Pelaez}} \bibnamefont{and}
  \bibinfo{author}{\bibfnamefont{G.}~\bibnamefont{Rios}},
  \bibinfo{journal}{Phys. Rev.} \textbf{\bibinfo{volume}{D82}},
  \bibinfo{pages}{114002} (\bibinfo{year}{2010}), \eprint{1010.6008}.

\bibitem[{\citenamefont{Nebreda and Pelaez.}(2010)}]{Nebreda:2010wv}
\bibinfo{author}{\bibfnamefont{J.}~\bibnamefont{Nebreda}} \bibnamefont{and}
  \bibinfo{author}{\bibfnamefont{J.~R.} \bibnamefont{Pelaez.}},
  \bibinfo{journal}{Phys. Rev.} \textbf{\bibinfo{volume}{D81}},
  \bibinfo{pages}{054035} (\bibinfo{year}{2010}), \eprint{1001.5237}.

\bibitem[{\citenamefont{Hu et~al.}(2017)\citenamefont{Hu, Molina, D{\"o}ring,
  Mai, and Alexandru}}]{Hu:2017wli}
\bibinfo{author}{\bibfnamefont{B.}~\bibnamefont{Hu}},
  \bibinfo{author}{\bibfnamefont{R.}~\bibnamefont{Molina}},
  \bibinfo{author}{\bibfnamefont{M.}~\bibnamefont{D{\"o}ring}},
  \bibinfo{author}{\bibfnamefont{M.}~\bibnamefont{Mai}}, \bibnamefont{and}
  \bibinfo{author}{\bibfnamefont{A.}~\bibnamefont{Alexandru}},
  \bibinfo{journal}{Phys. Rev.} \textbf{\bibinfo{volume}{D96}},
  \bibinfo{pages}{034520} (\bibinfo{year}{2017}), \eprint{1704.06248}.

\bibitem[{\citenamefont{Niehus et~al.}(2020)\citenamefont{Niehus, Hoferichter,
  Kubis, and Ruiz~de Elvira}}]{Niehus:2020gmf}
\bibinfo{author}{\bibfnamefont{M.}~\bibnamefont{Niehus}},
  \bibinfo{author}{\bibfnamefont{M.}~\bibnamefont{Hoferichter}},
  \bibinfo{author}{\bibfnamefont{B.}~\bibnamefont{Kubis}}, \bibnamefont{and}
  \bibinfo{author}{\bibfnamefont{J.}~\bibnamefont{Ruiz~de Elvira}}
  (\bibinfo{year}{2020}), \eprint{2009.04479}.

\bibitem[{\citenamefont{Gell-Mann et~al.}(1968)\citenamefont{Gell-Mann, Oakes,
  and Renner}}]{GellMann:1968rz}
\bibinfo{author}{\bibfnamefont{M.}~\bibnamefont{Gell-Mann}},
  \bibinfo{author}{\bibfnamefont{R.~J.} \bibnamefont{Oakes}}, \bibnamefont{and}
  \bibinfo{author}{\bibfnamefont{B.}~\bibnamefont{Renner}},
  \bibinfo{journal}{Phys. Rev.} \textbf{\bibinfo{volume}{175}},
  \bibinfo{pages}{2195} (\bibinfo{year}{1968}).

\bibitem[{\citenamefont{Bijnens and Jemos}(2012)}]{Bijnens:2011tb}
\bibinfo{author}{\bibfnamefont{J.}~\bibnamefont{Bijnens}} \bibnamefont{and}
  \bibinfo{author}{\bibfnamefont{I.}~\bibnamefont{Jemos}},
  \bibinfo{journal}{Nucl. Phys.} \textbf{\bibinfo{volume}{B854}},
  \bibinfo{pages}{631} (\bibinfo{year}{2012}), \eprint{1103.5945}.

\bibitem[{\citenamefont{Bijnens and Ecker}(2014)}]{Bijnens:2014lea}
\bibinfo{author}{\bibfnamefont{J.}~\bibnamefont{Bijnens}} \bibnamefont{and}
  \bibinfo{author}{\bibfnamefont{G.}~\bibnamefont{Ecker}},
  \bibinfo{journal}{Ann. Rev. Nucl. Part. Sci.} \textbf{\bibinfo{volume}{64}},
  \bibinfo{pages}{149} (\bibinfo{year}{2014}), \eprint{1405.6488}.

\bibitem[{\citenamefont{Hoferichter
  et~al.}(2015{\natexlab{a}})\citenamefont{Hoferichter, Ruiz~de Elvira, Kubis,
  and Mei{\ss}ner}}]{Hoferichter:2015tha}
\bibinfo{author}{\bibfnamefont{M.}~\bibnamefont{Hoferichter}},
  \bibinfo{author}{\bibfnamefont{J.}~\bibnamefont{Ruiz~de Elvira}},
  \bibinfo{author}{\bibfnamefont{B.}~\bibnamefont{Kubis}}, \bibnamefont{and}
  \bibinfo{author}{\bibfnamefont{U.-G.} \bibnamefont{Mei{\ss}ner}},
  \bibinfo{journal}{Phys. Rev. Lett.} \textbf{\bibinfo{volume}{115}},
  \bibinfo{pages}{192301} (\bibinfo{year}{2015}{\natexlab{a}}),
  \eprint{1507.07552}.

\bibitem[{\citenamefont{Siemens et~al.}(2017)\citenamefont{Siemens, Ruiz~de
  Elvira, Epelbaum, Hoferichter, Krebs, Kubis, and
  Mei{\ss}ner}}]{Siemens:2016jwj}
\bibinfo{author}{\bibfnamefont{D.}~\bibnamefont{Siemens}},
  \bibinfo{author}{\bibfnamefont{J.}~\bibnamefont{Ruiz~de Elvira}},
  \bibinfo{author}{\bibfnamefont{E.}~\bibnamefont{Epelbaum}},
  \bibinfo{author}{\bibfnamefont{M.}~\bibnamefont{Hoferichter}},
  \bibinfo{author}{\bibfnamefont{H.}~\bibnamefont{Krebs}},
  \bibinfo{author}{\bibfnamefont{B.}~\bibnamefont{Kubis}}, \bibnamefont{and}
  \bibinfo{author}{\bibfnamefont{U.~G.} \bibnamefont{Mei{\ss}ner}},
  \bibinfo{journal}{Phys. Lett.} \textbf{\bibinfo{volume}{B770}},
  \bibinfo{pages}{27} (\bibinfo{year}{2017}), \eprint{1610.08978}.

\bibitem[{\citenamefont{Leutwyler}(2015)}]{Leutwyler:2015jga}
\bibinfo{author}{\bibfnamefont{H.}~\bibnamefont{Leutwyler}},
  \bibinfo{journal}{PoS} \textbf{\bibinfo{volume}{CD15}}, \bibinfo{pages}{022}
  (\bibinfo{year}{2015}), \eprint{1510.07511}.

\bibitem[{\citenamefont{Bernard
  et~al.}(1991{\natexlab{a}})\citenamefont{Bernard, Kaiser, and
  Mei{\ss}ner}}]{Bernard:1990kx}
\bibinfo{author}{\bibfnamefont{V.}~\bibnamefont{Bernard}},
  \bibinfo{author}{\bibfnamefont{N.}~\bibnamefont{Kaiser}}, \bibnamefont{and}
  \bibinfo{author}{\bibfnamefont{U.~G.} \bibnamefont{Mei{\ss}ner}},
  \bibinfo{journal}{Phys. Rev.} \textbf{\bibinfo{volume}{D43}},
  \bibinfo{pages}{2757} (\bibinfo{year}{1991}{\natexlab{a}}).

\bibitem[{\citenamefont{Bernard
  et~al.}(1991{\natexlab{b}})\citenamefont{Bernard, Kaiser, and
  Mei{\ss}ner}}]{Bernard:1990kw}
\bibinfo{author}{\bibfnamefont{V.}~\bibnamefont{Bernard}},
  \bibinfo{author}{\bibfnamefont{N.}~\bibnamefont{Kaiser}}, \bibnamefont{and}
  \bibinfo{author}{\bibfnamefont{U.~G.} \bibnamefont{Mei{\ss}ner}},
  \bibinfo{journal}{Nucl. Phys.} \textbf{\bibinfo{volume}{B357}},
  \bibinfo{pages}{129} (\bibinfo{year}{1991}{\natexlab{b}}).

\bibitem[{\citenamefont{Bernard
  et~al.}(1991{\natexlab{c}})\citenamefont{Bernard, Kaiser, and
  Mei{\ss}ner}}]{Bernard:1991xb}
\bibinfo{author}{\bibfnamefont{V.}~\bibnamefont{Bernard}},
  \bibinfo{author}{\bibfnamefont{N.}~\bibnamefont{Kaiser}}, \bibnamefont{and}
  \bibinfo{author}{\bibfnamefont{U.~G.} \bibnamefont{Mei{\ss}ner}},
  \bibinfo{journal}{Phys. Rev.} \textbf{\bibinfo{volume}{D44}},
  \bibinfo{pages}{3698} (\bibinfo{year}{1991}{\natexlab{c}}).

\bibitem[{\citenamefont{Knecht et~al.}(1995)\citenamefont{Knecht, Moussallam,
  Stern, and Fuchs}}]{Knecht:1995tr}
\bibinfo{author}{\bibfnamefont{M.}~\bibnamefont{Knecht}},
  \bibinfo{author}{\bibfnamefont{B.}~\bibnamefont{Moussallam}},
  \bibinfo{author}{\bibfnamefont{J.}~\bibnamefont{Stern}}, \bibnamefont{and}
  \bibinfo{author}{\bibfnamefont{N.~H.} \bibnamefont{Fuchs}},
  \bibinfo{journal}{Nucl. Phys.} \textbf{\bibinfo{volume}{B457}},
  \bibinfo{pages}{513} (\bibinfo{year}{1995}), \eprint{hep-ph/9507319}.

\bibitem[{\citenamefont{Bijnens et~al.}(1996)\citenamefont{Bijnens, Colangelo,
  Ecker, Gasser, and Sainio}}]{Bijnens:1995yn}
\bibinfo{author}{\bibfnamefont{J.}~\bibnamefont{Bijnens}},
  \bibinfo{author}{\bibfnamefont{G.}~\bibnamefont{Colangelo}},
  \bibinfo{author}{\bibfnamefont{G.}~\bibnamefont{Ecker}},
  \bibinfo{author}{\bibfnamefont{J.}~\bibnamefont{Gasser}}, \bibnamefont{and}
  \bibinfo{author}{\bibfnamefont{M.~E.} \bibnamefont{Sainio}},
  \bibinfo{journal}{Phys. Lett.} \textbf{\bibinfo{volume}{B374}},
  \bibinfo{pages}{210} (\bibinfo{year}{1996}), \eprint{hep-ph/9511397}.

\bibitem[{\citenamefont{Bijnens
  et~al.}(2004{\natexlab{a}})\citenamefont{Bijnens, Dhonte, and
  Talavera}}]{Bijnens:2004eu}
\bibinfo{author}{\bibfnamefont{J.}~\bibnamefont{Bijnens}},
  \bibinfo{author}{\bibfnamefont{P.}~\bibnamefont{Dhonte}}, \bibnamefont{and}
  \bibinfo{author}{\bibfnamefont{P.}~\bibnamefont{Talavera}},
  \bibinfo{journal}{JHEP} \textbf{\bibinfo{volume}{01}}, \bibinfo{pages}{050}
  (\bibinfo{year}{2004}{\natexlab{a}}), \eprint{hep-ph/0401039}.

\bibitem[{\citenamefont{Bijnens
  et~al.}(2004{\natexlab{b}})\citenamefont{Bijnens, Dhonte, and
  Talavera}}]{Bijnens:2004bu}
\bibinfo{author}{\bibfnamefont{J.}~\bibnamefont{Bijnens}},
  \bibinfo{author}{\bibfnamefont{P.}~\bibnamefont{Dhonte}}, \bibnamefont{and}
  \bibinfo{author}{\bibfnamefont{P.}~\bibnamefont{Talavera}},
  \bibinfo{journal}{JHEP} \textbf{\bibinfo{volume}{05}}, \bibinfo{pages}{036}
  (\bibinfo{year}{2004}{\natexlab{b}}), \eprint{hep-ph/0404150}.

\bibitem[{\citenamefont{Bijnens et~al.}(2019)\citenamefont{Bijnens,
  Hermansson-Truedsson, and Wang}}]{Bijnens:2018lez}
\bibinfo{author}{\bibfnamefont{J.}~\bibnamefont{Bijnens}},
  \bibinfo{author}{\bibfnamefont{N.}~\bibnamefont{Hermansson-Truedsson}},
  \bibnamefont{and} \bibinfo{author}{\bibfnamefont{S.}~\bibnamefont{Wang}},
  \bibinfo{journal}{JHEP} \textbf{\bibinfo{volume}{01}}, \bibinfo{pages}{102}
  (\bibinfo{year}{2019}), \eprint{1810.06834}.

\bibitem[{\citenamefont{Gomez~Nicola et~al.}(2010)\citenamefont{Gomez~Nicola,
  Pelaez, and Ruiz~de Elvira}}]{GomezNicola:2010tb}
\bibinfo{author}{\bibfnamefont{A.}~\bibnamefont{Gomez~Nicola}},
  \bibinfo{author}{\bibfnamefont{J.~R.} \bibnamefont{Pelaez}},
  \bibnamefont{and} \bibinfo{author}{\bibfnamefont{J.}~\bibnamefont{Ruiz~de
  Elvira}}, \bibinfo{journal}{Phys. Rev.} \textbf{\bibinfo{volume}{D82}},
  \bibinfo{pages}{074012} (\bibinfo{year}{2010}), \eprint{1005.4370}.

\bibitem[{\citenamefont{Ruiz~de Elvira and
  Ruiz~Arriola}(2018)}]{RuizdeElvira:2018hsv}
\bibinfo{author}{\bibfnamefont{J.}~\bibnamefont{Ruiz~de Elvira}}
  \bibnamefont{and}
  \bibinfo{author}{\bibfnamefont{E.}~\bibnamefont{Ruiz~Arriola}},
  \bibinfo{journal}{Eur. Phys. J.} \textbf{\bibinfo{volume}{C78}},
  \bibinfo{pages}{878} (\bibinfo{year}{2018}), \eprint{1807.10837}.

\bibitem[{\citenamefont{Mei{\ss}ner}(1991)}]{Meissner:1990kz}
\bibinfo{author}{\bibfnamefont{U.~G.} \bibnamefont{Mei{\ss}ner}},
  \bibinfo{journal}{Comments Nucl. Part. Phys.} \textbf{\bibinfo{volume}{20}},
  \bibinfo{pages}{119} (\bibinfo{year}{1991}).

\bibitem[{\citenamefont{Protopopescu et~al.}(1973)\citenamefont{Protopopescu,
  Alston-Garnjost, Barbaro-Galtieri, Flatte, Friedman, Lasinski, Lynch, Rabin,
  and Solmitz}}]{Protopopescu:1973sh}
\bibinfo{author}{\bibfnamefont{S.~D.} \bibnamefont{Protopopescu}},
  \bibinfo{author}{\bibfnamefont{M.}~\bibnamefont{Alston-Garnjost}},
  \bibinfo{author}{\bibfnamefont{A.}~\bibnamefont{Barbaro-Galtieri}},
  \bibinfo{author}{\bibfnamefont{S.~M.} \bibnamefont{Flatte}},
  \bibinfo{author}{\bibfnamefont{J.~H.} \bibnamefont{Friedman}},
  \bibinfo{author}{\bibfnamefont{T.~A.} \bibnamefont{Lasinski}},
  \bibinfo{author}{\bibfnamefont{G.~R.} \bibnamefont{Lynch}},
  \bibinfo{author}{\bibfnamefont{M.~S.} \bibnamefont{Rabin}}, \bibnamefont{and}
  \bibinfo{author}{\bibfnamefont{F.~T.} \bibnamefont{Solmitz}},
  \bibinfo{journal}{Phys. Rev.} \textbf{\bibinfo{volume}{D7}},
  \bibinfo{pages}{1279} (\bibinfo{year}{1973}).

\bibitem[{\citenamefont{Hyams et~al.}(1973)}]{Hyams:1973zf}
\bibinfo{author}{\bibfnamefont{B.}~\bibnamefont{Hyams}} \bibnamefont{et~al.},
  \bibinfo{journal}{Nucl. Phys.} \textbf{\bibinfo{volume}{B64}},
  \bibinfo{pages}{134} (\bibinfo{year}{1973}).

\bibitem[{\citenamefont{Grayer et~al.}(1974)}]{Grayer:1974cr}
\bibinfo{author}{\bibfnamefont{G.}~\bibnamefont{Grayer}} \bibnamefont{et~al.},
  \bibinfo{journal}{Nucl. Phys.} \textbf{\bibinfo{volume}{B75}},
  \bibinfo{pages}{189} (\bibinfo{year}{1974}).

\bibitem[{\citenamefont{Estabrooks and Martin}(1974)}]{Estabrooks:1974vu}
\bibinfo{author}{\bibfnamefont{P.}~\bibnamefont{Estabrooks}} \bibnamefont{and}
  \bibinfo{author}{\bibfnamefont{A.~D.} \bibnamefont{Martin}},
  \bibinfo{journal}{Nucl. Phys.} \textbf{\bibinfo{volume}{B79}},
  \bibinfo{pages}{301} (\bibinfo{year}{1974}).

\bibitem[{\citenamefont{Navarro~P\'erez
  et~al.}(2015)\citenamefont{Navarro~P\'erez, Ruiz~Arriola, and Ruiz~de
  Elvira}}]{Perez:2015pea}
\bibinfo{author}{\bibfnamefont{R.}~\bibnamefont{Navarro~P\'erez}},
  \bibinfo{author}{\bibfnamefont{E.}~\bibnamefont{Ruiz~Arriola}},
  \bibnamefont{and} \bibinfo{author}{\bibfnamefont{J.}~\bibnamefont{Ruiz~de
  Elvira}}, \bibinfo{journal}{Phys. Rev.} \textbf{\bibinfo{volume}{D91}},
  \bibinfo{pages}{074014} (\bibinfo{year}{2015}), \eprint{1502.03361}.

\bibitem[{\citenamefont{Gupta}(1977)}]{Gupta:1977pd}
\bibinfo{author}{\bibfnamefont{S.~N.} \bibnamefont{Gupta}},
  \emph{\bibinfo{title}{{Quantum Electrodynamics}}} (\bibinfo{year}{1977}).

\bibitem[{\citenamefont{Oller and Oset}(1997)}]{Oller:1997ti}
\bibinfo{author}{\bibfnamefont{J.~A.} \bibnamefont{Oller}} \bibnamefont{and}
  \bibinfo{author}{\bibfnamefont{E.}~\bibnamefont{Oset}},
  \bibinfo{journal}{Nucl. Phys.} \textbf{\bibinfo{volume}{A620}},
  \bibinfo{pages}{438} (\bibinfo{year}{1997}), \bibinfo{note}{[Erratum: Nucl.
  Phys.A652,407(1999)]}, \eprint{hep-ph/9702314}.

\bibitem[{\citenamefont{Oller and Oset}(1999)}]{Oller:1998zr}
\bibinfo{author}{\bibfnamefont{J.~A.} \bibnamefont{Oller}} \bibnamefont{and}
  \bibinfo{author}{\bibfnamefont{E.}~\bibnamefont{Oset}},
  \bibinfo{journal}{Phys. Rev.} \textbf{\bibinfo{volume}{D60}},
  \bibinfo{pages}{074023} (\bibinfo{year}{1999}), \eprint{hep-ph/9809337}.

\bibitem[{\citenamefont{Mandelstam}(1959)}]{Mandelstam:1959bc}
\bibinfo{author}{\bibfnamefont{S.}~\bibnamefont{Mandelstam}},
  \bibinfo{journal}{Phys. Rev.} \textbf{\bibinfo{volume}{115}},
  \bibinfo{pages}{1741} (\bibinfo{year}{1959}).

\bibitem[{\citenamefont{Froissart}(1961)}]{Froissart:1961ux}
\bibinfo{author}{\bibfnamefont{M.}~\bibnamefont{Froissart}},
  \bibinfo{journal}{Phys. Rev.} \textbf{\bibinfo{volume}{123}},
  \bibinfo{pages}{1053} (\bibinfo{year}{1961}).

\bibitem[{\citenamefont{Martin}(1963)}]{Martin:1962rt}
\bibinfo{author}{\bibfnamefont{A.}~\bibnamefont{Martin}},
  \bibinfo{journal}{Phys. Rev.} \textbf{\bibinfo{volume}{129}},
  \bibinfo{pages}{1432} (\bibinfo{year}{1963}).

\bibitem[{\citenamefont{Roy}(1971)}]{Roy:1971tc}
\bibinfo{author}{\bibfnamefont{S.~M.} \bibnamefont{Roy}},
  \bibinfo{journal}{Phys. Lett.} \textbf{\bibinfo{volume}{36B}},
  \bibinfo{pages}{353} (\bibinfo{year}{1971}).

\bibitem[{\citenamefont{Hite and Steiner}(1973)}]{Hite:1973pm}
\bibinfo{author}{\bibfnamefont{G.~E.} \bibnamefont{Hite}} \bibnamefont{and}
  \bibinfo{author}{\bibfnamefont{F.}~\bibnamefont{Steiner}},
  \bibinfo{journal}{Nuovo Cim.} \textbf{\bibinfo{volume}{A18}},
  \bibinfo{pages}{237} (\bibinfo{year}{1973}).

\bibitem[{\citenamefont{Buettiker et~al.}(2004)\citenamefont{Buettiker,
  Descotes-Genon, and Moussallam}}]{Buettiker:2003pp}
\bibinfo{author}{\bibfnamefont{P.}~\bibnamefont{Buettiker}},
  \bibinfo{author}{\bibfnamefont{S.}~\bibnamefont{Descotes-Genon}},
  \bibnamefont{and}
  \bibinfo{author}{\bibfnamefont{B.}~\bibnamefont{Moussallam}},
  \bibinfo{journal}{Eur. Phys. J.} \textbf{\bibinfo{volume}{C33}},
  \bibinfo{pages}{409} (\bibinfo{year}{2004}), \eprint{hep-ph/0310283}.

\bibitem[{\citenamefont{Hoferichter
  et~al.}(2016{\natexlab{b}})\citenamefont{Hoferichter, Ruiz~de Elvira, Kubis,
  and Mei{\ss}ner}}]{Hoferichter:2015hva}
\bibinfo{author}{\bibfnamefont{M.}~\bibnamefont{Hoferichter}},
  \bibinfo{author}{\bibfnamefont{J.}~\bibnamefont{Ruiz~de Elvira}},
  \bibinfo{author}{\bibfnamefont{B.}~\bibnamefont{Kubis}}, \bibnamefont{and}
  \bibinfo{author}{\bibfnamefont{U.-G.} \bibnamefont{Mei{\ss}ner}},
  \bibinfo{journal}{Phys. Rept.} \textbf{\bibinfo{volume}{625}},
  \bibinfo{pages}{1} (\bibinfo{year}{2016}{\natexlab{b}}), \eprint{1510.06039}.

\bibitem[{\citenamefont{Caprini et~al.}(2006)\citenamefont{Caprini, Colangelo,
  and Leutwyler}}]{Caprini:2005zr}
\bibinfo{author}{\bibfnamefont{I.}~\bibnamefont{Caprini}},
  \bibinfo{author}{\bibfnamefont{G.}~\bibnamefont{Colangelo}},
  \bibnamefont{and}
  \bibinfo{author}{\bibfnamefont{H.}~\bibnamefont{Leutwyler}},
  \bibinfo{journal}{Phys. Rev. Lett.} \textbf{\bibinfo{volume}{96}},
  \bibinfo{pages}{132001} (\bibinfo{year}{2006}), \eprint{hep-ph/0512364}.

\bibitem[{\citenamefont{Descotes-Genon and
  Moussallam}(2006)}]{DescotesGenon:2006uk}
\bibinfo{author}{\bibfnamefont{S.}~\bibnamefont{Descotes-Genon}}
  \bibnamefont{and}
  \bibinfo{author}{\bibfnamefont{B.}~\bibnamefont{Moussallam}},
  \bibinfo{journal}{Eur. Phys. J.} \textbf{\bibinfo{volume}{C48}},
  \bibinfo{pages}{553} (\bibinfo{year}{2006}), \eprint{hep-ph/0607133}.

\bibitem[{\citenamefont{Masjuan et~al.}(2014)\citenamefont{Masjuan, Ruiz~de
  Elvira, and Sanz-Cillero}}]{Masjuan:2014psa}
\bibinfo{author}{\bibfnamefont{P.}~\bibnamefont{Masjuan}},
  \bibinfo{author}{\bibfnamefont{J.}~\bibnamefont{Ruiz~de Elvira}},
  \bibnamefont{and} \bibinfo{author}{\bibfnamefont{J.~J.}
  \bibnamefont{Sanz-Cillero}}, \bibinfo{journal}{Phys. Rev.}
  \textbf{\bibinfo{volume}{D90}}, \bibinfo{pages}{097901}
  (\bibinfo{year}{2014}), \eprint{1410.2397}.

\bibitem[{\citenamefont{Caprini et~al.}(2016)\citenamefont{Caprini, Masjuan,
  Ruiz~de Elvira, and Sanz-Cillero}}]{Caprini:2016uxy}
\bibinfo{author}{\bibfnamefont{I.}~\bibnamefont{Caprini}},
  \bibinfo{author}{\bibfnamefont{P.}~\bibnamefont{Masjuan}},
  \bibinfo{author}{\bibfnamefont{J.}~\bibnamefont{Ruiz~de Elvira}},
  \bibnamefont{and} \bibinfo{author}{\bibfnamefont{J.~J.}
  \bibnamefont{Sanz-Cillero}}, \bibinfo{journal}{Phys. Rev.}
  \textbf{\bibinfo{volume}{D93}}, \bibinfo{pages}{076004}
  (\bibinfo{year}{2016}), \eprint{1602.02062}.

\bibitem[{\citenamefont{Pel\'aez et~al.}(2017)\citenamefont{Pel\'aez, Rodas,
  and Ruiz~de Elvira}}]{Pelaez:2016klv}
\bibinfo{author}{\bibfnamefont{J.~R.} \bibnamefont{Pel\'aez}},
  \bibinfo{author}{\bibfnamefont{A.}~\bibnamefont{Rodas}}, \bibnamefont{and}
  \bibinfo{author}{\bibfnamefont{J.}~\bibnamefont{Ruiz~de Elvira}},
  \bibinfo{journal}{Eur. Phys. J.} \textbf{\bibinfo{volume}{C77}},
  \bibinfo{pages}{91} (\bibinfo{year}{2017}), \eprint{1612.07966}.

\bibitem[{\citenamefont{Hoferichter
  et~al.}(2015{\natexlab{b}})\citenamefont{Hoferichter, Ruiz~de Elvira, Kubis,
  and Mei{\ss}ner}}]{Hoferichter:2015dsa}
\bibinfo{author}{\bibfnamefont{M.}~\bibnamefont{Hoferichter}},
  \bibinfo{author}{\bibfnamefont{J.}~\bibnamefont{Ruiz~de Elvira}},
  \bibinfo{author}{\bibfnamefont{B.}~\bibnamefont{Kubis}}, \bibnamefont{and}
  \bibinfo{author}{\bibfnamefont{U.-G.} \bibnamefont{Mei{\ss}ner}},
  \bibinfo{journal}{Phys. Rev. Lett.} \textbf{\bibinfo{volume}{115}},
  \bibinfo{pages}{092301} (\bibinfo{year}{2015}{\natexlab{b}}),
  \eprint{1506.04142}.

\bibitem[{\citenamefont{Hoferichter
  et~al.}(2016{\natexlab{c}})\citenamefont{Hoferichter, Ruiz~de Elvira, Kubis,
  and Mei{\ss}ner}}]{Hoferichter:2016ocj}
\bibinfo{author}{\bibfnamefont{M.}~\bibnamefont{Hoferichter}},
  \bibinfo{author}{\bibfnamefont{J.}~\bibnamefont{Ruiz~de Elvira}},
  \bibinfo{author}{\bibfnamefont{B.}~\bibnamefont{Kubis}}, \bibnamefont{and}
  \bibinfo{author}{\bibfnamefont{U.-G.} \bibnamefont{Mei{\ss}ner}},
  \bibinfo{journal}{Phys. Lett.} \textbf{\bibinfo{volume}{B760}},
  \bibinfo{pages}{74} (\bibinfo{year}{2016}{\natexlab{c}}),
  \eprint{1602.07688}.

\bibitem[{\citenamefont{Ruiz~de Elvira et~al.}(2018)\citenamefont{Ruiz~de
  Elvira, Hoferichter, Kubis, and Mei{\ss}ner}}]{RuizdeElvira:2017stg}
\bibinfo{author}{\bibfnamefont{J.}~\bibnamefont{Ruiz~de Elvira}},
  \bibinfo{author}{\bibfnamefont{M.}~\bibnamefont{Hoferichter}},
  \bibinfo{author}{\bibfnamefont{B.}~\bibnamefont{Kubis}}, \bibnamefont{and}
  \bibinfo{author}{\bibfnamefont{U.-G.} \bibnamefont{Mei{\ss}ner}},
  \bibinfo{journal}{J. Phys.} \textbf{\bibinfo{volume}{G45}},
  \bibinfo{pages}{024001} (\bibinfo{year}{2018}), \eprint{1706.01465}.

\bibitem[{\citenamefont{Nieves et~al.}(2002)\citenamefont{Nieves,
  Pavon~Valderrama, and Ruiz~Arriola}}]{Nieves:2001de}
\bibinfo{author}{\bibfnamefont{J.}~\bibnamefont{Nieves}},
  \bibinfo{author}{\bibfnamefont{M.}~\bibnamefont{Pavon~Valderrama}},
  \bibnamefont{and}
  \bibinfo{author}{\bibfnamefont{E.}~\bibnamefont{Ruiz~Arriola}},
  \bibinfo{journal}{Phys. Rev.} \textbf{\bibinfo{volume}{D65}},
  \bibinfo{pages}{036002} (\bibinfo{year}{2002}), \eprint{hep-ph/0109077}.

\bibitem[{\citenamefont{Gomez~Nicola et~al.}(2008)\citenamefont{Gomez~Nicola,
  Pelaez, and Rios}}]{GomezNicola:2007qj}
\bibinfo{author}{\bibfnamefont{A.}~\bibnamefont{Gomez~Nicola}},
  \bibinfo{author}{\bibfnamefont{J.~R.} \bibnamefont{Pelaez}},
  \bibnamefont{and} \bibinfo{author}{\bibfnamefont{G.}~\bibnamefont{Rios}},
  \bibinfo{journal}{Phys. Rev.} \textbf{\bibinfo{volume}{D77}},
  \bibinfo{pages}{056006} (\bibinfo{year}{2008}), \eprint{0712.2763}.

\bibitem[{\citenamefont{Gasser and Mei{\ss}ner}(1991)}]{Gasser:1990bv}
\bibinfo{author}{\bibfnamefont{J.}~\bibnamefont{Gasser}} \bibnamefont{and}
  \bibinfo{author}{\bibfnamefont{U.~G.} \bibnamefont{Mei{\ss}ner}},
  \bibinfo{journal}{Nucl. Phys.} \textbf{\bibinfo{volume}{B357}},
  \bibinfo{pages}{90} (\bibinfo{year}{1991}).

\bibitem[{\citenamefont{Pelaez}(2004)}]{Pelaez:2003dy}
\bibinfo{author}{\bibfnamefont{J.~R.} \bibnamefont{Pelaez}},
  \bibinfo{journal}{Phys. Rev. Lett.} \textbf{\bibinfo{volume}{92}},
  \bibinfo{pages}{102001} (\bibinfo{year}{2004}), \eprint{hep-ph/0309292}.

\bibitem[{\citenamefont{Iagolnitzer et~al.}(1973)\citenamefont{Iagolnitzer,
  Zinn-Justin, and Zuber}}]{Iagolnitzer:1973fq}
\bibinfo{author}{\bibfnamefont{D.}~\bibnamefont{Iagolnitzer}},
  \bibinfo{author}{\bibfnamefont{J.}~\bibnamefont{Zinn-Justin}},
  \bibnamefont{and} \bibinfo{author}{\bibfnamefont{J.~B.} \bibnamefont{Zuber}},
  \bibinfo{journal}{Nucl. Phys.} \textbf{\bibinfo{volume}{B60}},
  \bibinfo{pages}{233} (\bibinfo{year}{1973}).

\bibitem[{\citenamefont{Badalian et~al.}(1982)\citenamefont{Badalian, Kok,
  Polikarpov, and Simonov}}]{Badalian:1981xj}
\bibinfo{author}{\bibfnamefont{A.~M.} \bibnamefont{Badalian}},
  \bibinfo{author}{\bibfnamefont{L.~P.} \bibnamefont{Kok}},
  \bibinfo{author}{\bibfnamefont{M.~I.} \bibnamefont{Polikarpov}},
  \bibnamefont{and} \bibinfo{author}{\bibfnamefont{{\relax Yu}.~A.}
  \bibnamefont{Simonov}}, \bibinfo{journal}{Phys. Rept.}
  \textbf{\bibinfo{volume}{82}}, \bibinfo{pages}{31} (\bibinfo{year}{1982}).

\bibitem[{\citenamefont{Guerrero and Oller}(1999)}]{Guerrero:1998ei}
\bibinfo{author}{\bibfnamefont{F.}~\bibnamefont{Guerrero}} \bibnamefont{and}
  \bibinfo{author}{\bibfnamefont{J.~A.} \bibnamefont{Oller}},
  \bibinfo{journal}{Nucl. Phys.} \textbf{\bibinfo{volume}{B537}},
  \bibinfo{pages}{459} (\bibinfo{year}{1999}), \bibinfo{note}{[Erratum: Nucl.
  Phys.B602,641(2001)]}, \eprint{hep-ph/9805334}.

\bibitem[{\citenamefont{Pelaez and Gomez~Nicola}(2003)}]{Pelaez:2003xd}
\bibinfo{author}{\bibfnamefont{J.~R.} \bibnamefont{Pelaez}} \bibnamefont{and}
  \bibinfo{author}{\bibfnamefont{A.}~\bibnamefont{Gomez~Nicola}},
  \bibinfo{journal}{AIP Conf. Proc.} \textbf{\bibinfo{volume}{660}},
  \bibinfo{pages}{102} (\bibinfo{year}{2003}), \eprint{hep-ph/0301049}.

\bibitem[{\citenamefont{Luscher}(1986)}]{Luscher:1986pf}
\bibinfo{author}{\bibfnamefont{M.}~\bibnamefont{Luscher}},
  \bibinfo{journal}{Commun. Math. Phys.} \textbf{\bibinfo{volume}{105}},
  \bibinfo{pages}{153} (\bibinfo{year}{1986}).

\bibitem[{\citenamefont{Luscher}(1991)}]{Luscher:1990ux}
\bibinfo{author}{\bibfnamefont{M.}~\bibnamefont{Luscher}},
  \bibinfo{journal}{Nucl. Phys.} \textbf{\bibinfo{volume}{B354}},
  \bibinfo{pages}{531} (\bibinfo{year}{1991}).

\bibitem[{\citenamefont{Liu et~al.}(2006)\citenamefont{Liu, Feng, and
  He}}]{Liu:2005kr}
\bibinfo{author}{\bibfnamefont{C.}~\bibnamefont{Liu}},
  \bibinfo{author}{\bibfnamefont{X.}~\bibnamefont{Feng}}, \bibnamefont{and}
  \bibinfo{author}{\bibfnamefont{S.}~\bibnamefont{He}}, \bibinfo{journal}{Int.
  J. Mod. Phys.} \textbf{\bibinfo{volume}{A21}}, \bibinfo{pages}{847}
  (\bibinfo{year}{2006}), \eprint{hep-lat/0508022}.

\bibitem[{\citenamefont{Bernard et~al.}(2008)\citenamefont{Bernard, Lage,
  Mei{\ss}ner, and Rusetsky}}]{Bernard:2008ax}
\bibinfo{author}{\bibfnamefont{V.}~\bibnamefont{Bernard}},
  \bibinfo{author}{\bibfnamefont{M.}~\bibnamefont{Lage}},
  \bibinfo{author}{\bibfnamefont{U.-G.} \bibnamefont{Mei{\ss}ner}},
  \bibnamefont{and} \bibinfo{author}{\bibfnamefont{A.}~\bibnamefont{Rusetsky}},
  \bibinfo{journal}{JHEP} \textbf{\bibinfo{volume}{08}}, \bibinfo{pages}{024}
  (\bibinfo{year}{2008}), \eprint{0806.4495}.

\bibitem[{\citenamefont{Lage et~al.}(2009)\citenamefont{Lage, Mei{\ss}ner, and
  Rusetsky}}]{Lage:2009zv}
\bibinfo{author}{\bibfnamefont{M.}~\bibnamefont{Lage}},
  \bibinfo{author}{\bibfnamefont{U.-G.} \bibnamefont{Mei{\ss}ner}},
  \bibnamefont{and} \bibinfo{author}{\bibfnamefont{A.}~\bibnamefont{Rusetsky}},
  \bibinfo{journal}{Phys. Lett.} \textbf{\bibinfo{volume}{B681}},
  \bibinfo{pages}{439} (\bibinfo{year}{2009}), \eprint{0905.0069}.

\bibitem[{\citenamefont{Hansen and Sharpe}(2012)}]{Hansen:2012tf}
\bibinfo{author}{\bibfnamefont{M.~T.} \bibnamefont{Hansen}} \bibnamefont{and}
  \bibinfo{author}{\bibfnamefont{S.~R.} \bibnamefont{Sharpe}},
  \bibinfo{journal}{Phys. Rev.} \textbf{\bibinfo{volume}{D86}},
  \bibinfo{pages}{016007} (\bibinfo{year}{2012}), \eprint{1204.0826}.

\bibitem[{\citenamefont{Brice\~no and Davoudi}(2013)}]{Briceno:2012yi}
\bibinfo{author}{\bibfnamefont{R.~A.} \bibnamefont{Brice\~no}}
  \bibnamefont{and} \bibinfo{author}{\bibfnamefont{Z.}~\bibnamefont{Davoudi}},
  \bibinfo{journal}{Phys. Rev.} \textbf{\bibinfo{volume}{D88}},
  \bibinfo{pages}{094507} (\bibinfo{year}{2013}), \eprint{1204.1110}.

\bibitem[{\citenamefont{Li and Liu}(2013)}]{Li:2012bi}
\bibinfo{author}{\bibfnamefont{N.}~\bibnamefont{Li}} \bibnamefont{and}
  \bibinfo{author}{\bibfnamefont{C.}~\bibnamefont{Liu}},
  \bibinfo{journal}{Phys. Rev.} \textbf{\bibinfo{volume}{D87}},
  \bibinfo{pages}{014502} (\bibinfo{year}{2013}), \eprint{1209.2201}.

\bibitem[{\citenamefont{Guo et~al.}(2013)\citenamefont{Guo, Dudek, Edwards, and
  Szczepaniak}}]{Guo:2012hv}
\bibinfo{author}{\bibfnamefont{P.}~\bibnamefont{Guo}},
  \bibinfo{author}{\bibfnamefont{J.}~\bibnamefont{Dudek}},
  \bibinfo{author}{\bibfnamefont{R.}~\bibnamefont{Edwards}}, \bibnamefont{and}
  \bibinfo{author}{\bibfnamefont{A.~P.} \bibnamefont{Szczepaniak}},
  \bibinfo{journal}{Phys. Rev.} \textbf{\bibinfo{volume}{D88}},
  \bibinfo{pages}{014501} (\bibinfo{year}{2013}), \eprint{1211.0929}.

\bibitem[{\citenamefont{Polejaeva and Rusetsky}(2012)}]{Polejaeva:2012ut}
\bibinfo{author}{\bibfnamefont{K.}~\bibnamefont{Polejaeva}} \bibnamefont{and}
  \bibinfo{author}{\bibfnamefont{A.}~\bibnamefont{Rusetsky}},
  \bibinfo{journal}{Eur. Phys. J. A} \textbf{\bibinfo{volume}{48}},
  \bibinfo{pages}{67} (\bibinfo{year}{2012}), \eprint{1203.1241}.

\bibitem[{\citenamefont{Hansen and Sharpe}(2014)}]{Hansen:2014eka}
\bibinfo{author}{\bibfnamefont{M.~T.} \bibnamefont{Hansen}} \bibnamefont{and}
  \bibinfo{author}{\bibfnamefont{S.~R.} \bibnamefont{Sharpe}},
  \bibinfo{journal}{Phys. Rev. D} \textbf{\bibinfo{volume}{90}},
  \bibinfo{pages}{116003} (\bibinfo{year}{2014}), \eprint{1408.5933}.

\bibitem[{\citenamefont{Briceño et~al.}(2017)\citenamefont{Briceño, Hansen,
  and Sharpe}}]{Briceno:2017tce}
\bibinfo{author}{\bibfnamefont{R.~A.} \bibnamefont{Briceño}},
  \bibinfo{author}{\bibfnamefont{M.~T.} \bibnamefont{Hansen}},
  \bibnamefont{and} \bibinfo{author}{\bibfnamefont{S.~R.}
  \bibnamefont{Sharpe}}, \bibinfo{journal}{Phys. Rev. D}
  \textbf{\bibinfo{volume}{95}}, \bibinfo{pages}{074510}
  (\bibinfo{year}{2017}), \eprint{1701.07465}.

\bibitem[{\citenamefont{Mai and D{\"o}ring}(2017)}]{Mai:2017bge}
\bibinfo{author}{\bibfnamefont{M.}~\bibnamefont{Mai}} \bibnamefont{and}
  \bibinfo{author}{\bibfnamefont{M.}~\bibnamefont{D{\"o}ring}},
  \bibinfo{journal}{Eur. Phys. J. A} \textbf{\bibinfo{volume}{53}},
  \bibinfo{pages}{240} (\bibinfo{year}{2017}), \eprint{1709.08222}.

\bibitem[{\citenamefont{D{\"o}ring et~al.}(2018)\citenamefont{D{\"o}ring,
  Hammer, Mai, Pang, Rusetsky, and Wu}}]{Doring:2018xxx}
\bibinfo{author}{\bibfnamefont{M.}~\bibnamefont{D{\"o}ring}},
  \bibinfo{author}{\bibfnamefont{H.-W.} \bibnamefont{Hammer}},
  \bibinfo{author}{\bibfnamefont{M.}~\bibnamefont{Mai}},
  \bibinfo{author}{\bibfnamefont{J.-Y.} \bibnamefont{Pang}},
  \bibinfo{author}{\bibfnamefont{A.}~\bibnamefont{Rusetsky}}, \bibnamefont{and}
  \bibinfo{author}{\bibfnamefont{J.}~\bibnamefont{Wu}}, \bibinfo{journal}{Phys.
  Rev. D} \textbf{\bibinfo{volume}{97}}, \bibinfo{pages}{114508}
  (\bibinfo{year}{2018}), \eprint{1802.03362}.

\bibitem[{\citenamefont{Hansen and Sharpe}(2019)}]{Hansen:2019nir}
\bibinfo{author}{\bibfnamefont{M.~T.} \bibnamefont{Hansen}} \bibnamefont{and}
  \bibinfo{author}{\bibfnamefont{S.~R.} \bibnamefont{Sharpe}},
  \bibinfo{journal}{Ann. Rev. Nucl. Part. Sci.} \textbf{\bibinfo{volume}{69}},
  \bibinfo{pages}{65} (\bibinfo{year}{2019}), \eprint{1901.00483}.

\bibitem[{\citenamefont{Blanton et~al.}(2019)\citenamefont{Blanton,
  Romero-López, and Sharpe}}]{Blanton:2019igq}
\bibinfo{author}{\bibfnamefont{T.~D.} \bibnamefont{Blanton}},
  \bibinfo{author}{\bibfnamefont{F.}~\bibnamefont{Romero-López}},
  \bibnamefont{and} \bibinfo{author}{\bibfnamefont{S.~R.}
  \bibnamefont{Sharpe}}, \bibinfo{journal}{JHEP} \textbf{\bibinfo{volume}{03}},
  \bibinfo{pages}{106} (\bibinfo{year}{2019}), \eprint{1901.07095}.

\bibitem[{\citenamefont{Pang et~al.}(2019)\citenamefont{Pang, Wu, Hammer,
  Meißner, and Rusetsky}}]{Pang:2019dfe}
\bibinfo{author}{\bibfnamefont{J.-Y.} \bibnamefont{Pang}},
  \bibinfo{author}{\bibfnamefont{J.-J.} \bibnamefont{Wu}},
  \bibinfo{author}{\bibfnamefont{H.-W.} \bibnamefont{Hammer}},
  \bibinfo{author}{\bibfnamefont{U.-G.} \bibnamefont{Meißner}},
  \bibnamefont{and} \bibinfo{author}{\bibfnamefont{A.}~\bibnamefont{Rusetsky}},
  \bibinfo{journal}{Phys. Rev. D} \textbf{\bibinfo{volume}{99}},
  \bibinfo{pages}{074513} (\bibinfo{year}{2019}), \eprint{1902.01111}.

\bibitem[{\citenamefont{Briceño et~al.}(2019)\citenamefont{Briceño, Hansen,
  Sharpe, and Szczepaniak}}]{Briceno:2019muc}
\bibinfo{author}{\bibfnamefont{R.~A.} \bibnamefont{Briceño}},
  \bibinfo{author}{\bibfnamefont{M.~T.} \bibnamefont{Hansen}},
  \bibinfo{author}{\bibfnamefont{S.~R.} \bibnamefont{Sharpe}},
  \bibnamefont{and} \bibinfo{author}{\bibfnamefont{A.~P.}
  \bibnamefont{Szczepaniak}}, \bibinfo{journal}{Phys. Rev. D}
  \textbf{\bibinfo{volume}{100}}, \bibinfo{pages}{054508}
  (\bibinfo{year}{2019}), \eprint{1905.11188}.

\bibitem[{\citenamefont{Romero-López et~al.}(2019)\citenamefont{Romero-López,
  Sharpe, Blanton, Briceño, and Hansen}}]{Romero-Lopez:2019qrt}
\bibinfo{author}{\bibfnamefont{F.}~\bibnamefont{Romero-López}},
  \bibinfo{author}{\bibfnamefont{S.~R.} \bibnamefont{Sharpe}},
  \bibinfo{author}{\bibfnamefont{T.~D.} \bibnamefont{Blanton}},
  \bibinfo{author}{\bibfnamefont{R.~A.} \bibnamefont{Briceño}},
  \bibnamefont{and} \bibinfo{author}{\bibfnamefont{M.~T.}
  \bibnamefont{Hansen}}, \bibinfo{journal}{JHEP} \textbf{\bibinfo{volume}{10}},
  \bibinfo{pages}{007} (\bibinfo{year}{2019}), \eprint{1908.02411}.

\bibitem[{\citenamefont{Hansen et~al.}(2020)\citenamefont{Hansen,
  Romero-L\'opez, and Sharpe}}]{Hansen:2020zhy}
\bibinfo{author}{\bibfnamefont{M.~T.} \bibnamefont{Hansen}},
  \bibinfo{author}{\bibfnamefont{F.}~\bibnamefont{Romero-L\'opez}},
  \bibnamefont{and} \bibinfo{author}{\bibfnamefont{S.~R.}
  \bibnamefont{Sharpe}}, \bibinfo{journal}{JHEP} \textbf{\bibinfo{volume}{07}},
  \bibinfo{pages}{047} (\bibinfo{year}{2020}), \eprint{2003.10974}.

\bibitem[{\citenamefont{Chen and Oset}(2013)}]{Chen:2012rp}
\bibinfo{author}{\bibfnamefont{H.-X.} \bibnamefont{Chen}} \bibnamefont{and}
  \bibinfo{author}{\bibfnamefont{E.}~\bibnamefont{Oset}},
  \bibinfo{journal}{Phys. Rev.} \textbf{\bibinfo{volume}{D87}},
  \bibinfo{pages}{016014} (\bibinfo{year}{2013}), \eprint{1202.2787}.

\bibitem[{\citenamefont{D{\"o}ring et~al.}(2011)\citenamefont{D{\"o}ring,
  Mei{\ss}ner, Oset, and Rusetsky}}]{Doring:2011vk}
\bibinfo{author}{\bibfnamefont{M.}~\bibnamefont{D{\"o}ring}},
  \bibinfo{author}{\bibfnamefont{U.-G.} \bibnamefont{Mei{\ss}ner}},
  \bibinfo{author}{\bibfnamefont{E.}~\bibnamefont{Oset}}, \bibnamefont{and}
  \bibinfo{author}{\bibfnamefont{A.}~\bibnamefont{Rusetsky}},
  \bibinfo{journal}{Eur. Phys. J.} \textbf{\bibinfo{volume}{A47}},
  \bibinfo{pages}{139} (\bibinfo{year}{2011}), \eprint{1107.3988}.

\bibitem[{\citenamefont{D{\"o}ring et~al.}(2012)\citenamefont{D{\"o}ring,
  Mei{\ss}ner, Oset, and Rusetsky}}]{Doring:2012eu}
\bibinfo{author}{\bibfnamefont{M.}~\bibnamefont{D{\"o}ring}},
  \bibinfo{author}{\bibfnamefont{U.~G.} \bibnamefont{Mei{\ss}ner}},
  \bibinfo{author}{\bibfnamefont{E.}~\bibnamefont{Oset}}, \bibnamefont{and}
  \bibinfo{author}{\bibfnamefont{A.}~\bibnamefont{Rusetsky}},
  \bibinfo{journal}{Eur. Phys. J.} \textbf{\bibinfo{volume}{A48}},
  \bibinfo{pages}{114} (\bibinfo{year}{2012}), \eprint{1205.4838}.

\bibitem[{\citenamefont{Albaladejo et~al.}(2012)\citenamefont{Albaladejo,
  Oller, Oset, Rios, and Roca}}]{Albaladejo:2012jr}
\bibinfo{author}{\bibfnamefont{M.}~\bibnamefont{Albaladejo}},
  \bibinfo{author}{\bibfnamefont{J.~A.} \bibnamefont{Oller}},
  \bibinfo{author}{\bibfnamefont{E.}~\bibnamefont{Oset}},
  \bibinfo{author}{\bibfnamefont{G.}~\bibnamefont{Rios}}, \bibnamefont{and}
  \bibinfo{author}{\bibfnamefont{L.}~\bibnamefont{Roca}},
  \bibinfo{journal}{JHEP} \textbf{\bibinfo{volume}{08}}, \bibinfo{pages}{071}
  (\bibinfo{year}{2012}), \eprint{1205.3582}.

\bibitem[{\citenamefont{Albaladejo et~al.}(2013)\citenamefont{Albaladejo, Rios,
  Oller, and Roca}}]{Albaladejo:2013bra}
\bibinfo{author}{\bibfnamefont{M.}~\bibnamefont{Albaladejo}},
  \bibinfo{author}{\bibfnamefont{G.}~\bibnamefont{Rios}},
  \bibinfo{author}{\bibfnamefont{J.~A.} \bibnamefont{Oller}}, \bibnamefont{and}
  \bibinfo{author}{\bibfnamefont{L.}~\bibnamefont{Roca}}
  (\bibinfo{year}{2013}), \eprint{1307.5169}.

\bibitem[{\citenamefont{Bolton et~al.}(2016)\citenamefont{Bolton, Briceno, and
  Wilson}}]{Bolton:2015psa}
\bibinfo{author}{\bibfnamefont{D.~R.} \bibnamefont{Bolton}},
  \bibinfo{author}{\bibfnamefont{R.~A.} \bibnamefont{Briceno}},
  \bibnamefont{and} \bibinfo{author}{\bibfnamefont{D.~J.}
  \bibnamefont{Wilson}}, \bibinfo{journal}{Phys. Lett.}
  \textbf{\bibinfo{volume}{B757}}, \bibinfo{pages}{50} (\bibinfo{year}{2016}),
  \eprint{1507.07928}.

\bibitem[{\citenamefont{Bazavov et~al.}(2009)}]{Bazavov:2009fk}
\bibinfo{author}{\bibfnamefont{A.}~\bibnamefont{Bazavov}} \bibnamefont{et~al.}
  (\bibinfo{collaboration}{MILC}), \bibinfo{journal}{PoS}
  \textbf{\bibinfo{volume}{CD09}}, \bibinfo{pages}{007} (\bibinfo{year}{2009}),
  \eprint{0910.2966}.

\bibitem[{\citenamefont{Fritzsch et~al.}(2012)\citenamefont{Fritzsch, Knechtli,
  Leder, Marinkovic, Schaefer, Sommer, and Virotta}}]{Fritzsch:2012wq}
\bibinfo{author}{\bibfnamefont{P.}~\bibnamefont{Fritzsch}},
  \bibinfo{author}{\bibfnamefont{F.}~\bibnamefont{Knechtli}},
  \bibinfo{author}{\bibfnamefont{B.}~\bibnamefont{Leder}},
  \bibinfo{author}{\bibfnamefont{M.}~\bibnamefont{Marinkovic}},
  \bibinfo{author}{\bibfnamefont{S.}~\bibnamefont{Schaefer}},
  \bibinfo{author}{\bibfnamefont{R.}~\bibnamefont{Sommer}}, \bibnamefont{and}
  \bibinfo{author}{\bibfnamefont{F.}~\bibnamefont{Virotta}},
  \bibinfo{journal}{Nucl. Phys.} \textbf{\bibinfo{volume}{B865}},
  \bibinfo{pages}{397} (\bibinfo{year}{2012}), \eprint{1205.5380}.

\bibitem[{\citenamefont{Brooker and Taylor}(1970)}]{Brooker:1970pe}
\bibinfo{author}{\bibfnamefont{P.}~\bibnamefont{Brooker}} \bibnamefont{and}
  \bibinfo{author}{\bibfnamefont{J.}~\bibnamefont{Taylor}},
  \bibinfo{journal}{Nucl. Phys. B} \textbf{\bibinfo{volume}{17}},
  \bibinfo{pages}{461} (\bibinfo{year}{1970}).

\bibitem[{\citenamefont{Gounaris and Sarantakos}(1977)}]{Gounaris:1976ns}
\bibinfo{author}{\bibfnamefont{G.}~\bibnamefont{Gounaris}} \bibnamefont{and}
  \bibinfo{author}{\bibfnamefont{S.}~\bibnamefont{Sarantakos}},
  \bibinfo{journal}{Nuovo Cim. A} \textbf{\bibinfo{volume}{39}},
  \bibinfo{pages}{554} (\bibinfo{year}{1977}).

\bibitem[{\citenamefont{Oakes and Sorba}(1979)}]{Oakes:1977vm}
\bibinfo{author}{\bibfnamefont{R.}~\bibnamefont{Oakes}} \bibnamefont{and}
  \bibinfo{author}{\bibfnamefont{P.}~\bibnamefont{Sorba}},
  \bibinfo{journal}{Nuovo Cim. A} \textbf{\bibinfo{volume}{50}},
  \bibinfo{pages}{291} (\bibinfo{year}{1979}).

\end{thebibliography}

\end{document}